\documentclass[aps, prd, amsmath, floats, floatfix, twocolumn, nofootinbib, superscriptaddress, showpacs]{revtex4-1}
\usepackage{graphicx}
\usepackage{multirow}
\usepackage{color}
\usepackage{latexsym}
\usepackage{amsfonts}
\usepackage[colorlinks]{hyperref}

\newcommand{\be}{\begin{eqnarray}}
\newcommand{\ee}{\end{eqnarray}}

\begin{document}

	\title{Quasi-harmonic oscillations of charged particles in static axially symmetric space-times immersed in a uniform magnetic field}

	\author{Carlos~A.~Benavides-Gallego}
	\email{abgcarlos17@fudan.edu.cn}
	\affiliation{Center for Field Theory and Particle Physics and Department of Physics, Fudan University, 200438 Shanghai, China}
	
	\author{Ahmadjon Abdujabbarov}
	\email{ahmadjon@astrin.uz}
	\affiliation{Tashkent Institute of Irrigation and Agricultural Mechanization Engineers, Kori Niyoziy, 39, Tashkent 100000, Uzbekistan}
	\affiliation{Shanghai Astronomical Observatory, 80 Nandan Road, Shanghai 
	200030, China}
	\affiliation{Ulugh Beg Astronomical Institute, Astronomicheskaya 33, 
	Tashkent 100052, Uzbekistan}
	\affiliation{National University of Uzbekistan, Tashkent 100174, 
	Uzbekistan}
	\affiliation{Institute of Nuclear Physics, Tashkent 100214, Uzbekistan}
	
	\author{Daniele~Malafarina}
	\email{daniele.malafarina@nu.edu.kz}
	\affiliation{Department of Physics, Nazarbayev University, 53 Kabanbay Batyr avenue, 010000 Nur-Sultan, Kazakhstan }
	
	\author{Cosimo~Bambi}
	\email[Corresponding author: ]{bambi@fudan.edu.cn}
	\affiliation{Center for Field Theory and Particle Physics and Department of Physics, Fudan University, 200438 Shanghai, China}

	\date{\today}

	\begin{abstract}
		In continuation of our previous study [Phys.~Rev.~D 99 (2019) 4, 044012], we investigate the motion of charged particles in the $\gamma$-metric. We provide some examples of curled trajectories in the equatorial plane and escape trajectories outside the equatorial plane. Finally, we consider harmonic oscillations due to small displacements from stable circular orbits on the equatorial plane and compute the epicyclic frequencies for different values of deformation parameter and magnetic field.
	\end{abstract}

	\maketitle

	\section{Introduction}
	
	Black holes in the general theory of relativity are entirely described by three parameters, namely mass, angular momentum, and charge
	\cite{Israel:1967wq,Carter:1971zc,Gurlebeck:2015xpa}.
	We know that classical black hole solutions contain curvature singularities which are expected to be resolved by a theory of quantum-gravity
	\cite{Penrose:1964wq,Hawking:1969sw}.
	This naturally leads to the question of whether such quantum-gravity effects remain confined within the horizon or may affect the exterior geometry, thus separating (astro)physical black holes from mathematical black holes.
	A hint toward the answer to this question comes from the study of non-singular gravitational collapse (see for example \cite{Bambi:2013caa,Bambi:2016uda,Chakrabarty:2019omm,Malafarina:2017csn,Carballo-Rubio:2019fnb,Carballo-Rubio:2019nel,Giddings:2019ujs,Giddings:2017jts,Barcelo:2015noa}), which suggests that the resolution of the singularity must affect the trapped surfaces in the space-time.
	Another hint comes from the investigation of exact solutions of Einstein's equations that do not describe black holes. For example, it is well known that static axially symmetric vacuum solutions generically posses naked curvature singularities when multipole moments of higher order are present~\cite{Weyl,WeylI,Quevedo1989,Quevedo1990,Bonnor1992,Pastora1994}.
	This fact can be interpreted in two ways: 
	\begin{itemize}
		\item[(i)] A collapsing body must shed away all higher multipole moments before crossing the horizon threshold.
		\item[(ii)] The space-time resulting from the collapse of a non-spherical body requires quantum-gravity modification already at the horizon scale to account for the non-vanishing multipole moments.
	\end{itemize}
	The issue will likely be resolved once we will be able to precisely probe experimentally the geometry around astrophysical black hole candidates.
	
	As of today, the nature of the geometry around extreme astrophysical compact objects is still unknown.
	However, recent results such as X-ray reflection spectroscopy~\cite{Bambi:2016sac,Cao:2017kdq,Tripathi:2018lhx}
	or the image of the `shadow' of the super-massive black hole candidate at the center of the galaxy M87~\cite{Akiyama:2019bqs} 
	suggest that experimental tests of the so-called Kerr hypothesis~\cite{Bambi:2015kza,Berti:2015itd}, 
	namely the hypothesis that all astrophysical black hole candidates are described by the Kerr metric, may be possible in the near future.
	
	In this work, we focus our attention on the Zipoy-Voorhees space-time, also known as the $\gamma$-metric, which is a static, axially symmetric line element describing the gravitational field outside a prolate or oblate object
	\cite{Voorhees:1971wh,Zipoy}. 
	The space-time is continuously linked to the Schwarzschild metric through the value of one parameter $\gamma$, which is related to the non-vanishing multipole moments and in the limit of $\gamma=1$ it reduces to Schwarzschild. As mentioned before, for $\gamma\neq 1$ the line-element presents a curvature singularity at the surface $r=2m$, making the manifold geodesically incomplete
	\cite{Kodama:2003ch,Herrera:1998rj,Herrera:1998eq}.
	The properties of the $\gamma$-metric, its geodesics, interior solutions and geometrical aspects, have been studied by many authors 
	\cite{Herrera:2000tz,LukesGerakopoulos:2012pq,Boshkayev:2015jaa,Virbhadra:1996cz,Papadopoulos:1981wr,Stewart1982,Herrera:2004ra}.
	In a series of previous articles, some of us investigated the properties of this space-time in connection with the possibility of observing departures from the Schwarzschild or Kerr line-elements in astrophysical observations
	\cite{Abdikamalov:2019ztb,Toshmatov:2019qih,Narzilloev:2020qdc}.

    In practice, to probe experimentally the geometry of extreme compact objects one can rely on different methods. Shadow and orbits are limited to only two super-massive black hole candidates and therefore to obtain a larger sample of observations one must look at X-ray binaries.
    
    The Rossi X-ray Timing Explorer (RXTE) mission has provided thousands of observations of black holes transients~\cite{Belloni:2012sv} and the study of X-ray binaries has been considered with great interest because it opens the possibility to probe fundamental physics~\cite{Rezzolla:2003zx}. The X-rays emitted by binary systems are produced by matter falling from the donor (usually a normal star) to the accretor: a neutron star or a black hole. Thanks to the RXTE mission, complex variability patterns were discovered, including the detection of Quasi-Periodic Oscillation (QPO) at frequencies higher than 40Hz~\cite{Belloni:2012sv}. According to Belloni et al., the QPOs “\textit{open a new window onto fast phenomena in the innermost regions of an accretion disk.}” \cite{Belloni:2012sv}. The high-frequency QPOs give us information about the masses and radii of neutron stars~\cite{Kluzniak:1997sh,Miller:1997if}, the masses and spin of the central objects~\cite{Wagoner:2001uj,Abramowicz:2001bi}. Furthermore, since the high-frequency QPOs are observed close to the orbital frequencies of the marginally stable circular orbits (MSO), the effects from strong gravity must be essential to explain their behavior~\cite{Abramowicz:2004je, Stella:1999sj, Stuchlik:2008fy, Kotrlova:2008xs}. Finally, from the analysis of the frequencies, it is also possible to obtain information about the electromagnetic field in the vicinity of black hole candidates. 
		
	Many of the observed black hole candidates have accretion discs formed by plasma whose dynamics can generate a regular magnetic field~\cite{Kolos:2015iva}. In this sense, it is essential to consider the role of the electromagnetic field in the processes taking place in the surroundings of a black hole. Recent observations have suggested that the center of the Milky Way galaxy has a strong magnetic field that is not related to the accretion disc of the black hole~\cite{Eatough:2013nva}. Hence, black holes can also be immersed in an external, large-scale electromagnetic field that can have a complicated structure in the vicinity of field source, but asymptotically (at large distances) its character can be close to being homogeneous~\cite{Stuchlik:2015nlt}. In this sense, the idea of a black hole immersed in a uniform magnetic field has been used in Ref.~\cite{Kolos:2015iva} as a model to explain the frequencies of the 3:2 high-frequency QPOs observed in the three Galactic microquasars GRS 1915+105, XTE 1550–564 and GRO1655–40, which cannot be explained by a model based on the frequencies of the geodesic epicyclic motion, if the accepted limits on the mass and spin of the black holes are taken into account~\cite{Torok:2011qy,Bambi:2012pa}.  	
	
	In the present work, we shall focus on the motion of charged test particles in the equatorial plane of the Zipoy-Voorhees space-time immersed in an external magnetic field to characterize the effects of the non-vanishing quadrupole moment on the QPOs (the case of QPOs for neutral particles was considered in \cite{Toshmatov:2019qih}). The motion of test particles in the magnetized Schwarzschild space-time was first considered in Ref.~\cite{Wald:1974np}.  Since then, solutions of Einstein's equations in external magnetic fields have been widely studied (see for example \cite{Petterson:1975sg,Kolos:2015iva,Frolov:2010mi,Narzilloev:2019hep}). 
 	
	The article is organized as follows. In section~\ref{III}, we review the $\gamma$-metric and its main properties together with the basic ideas of the description of a space-time immersed in an external magnetic field~\cite{Wald:1974np}. 
	In section~\ref{IV}, following the Hamiltonian formalism presented in Ref.~\cite{Kolos:2015iva}, we discuss the charged particle motion in the magnetized Zipoy-Voorhees space-time and obtain the effective potential for charged test particles. Finally section~\ref{V} is devoted to the study of harmonic oscillations about the circular orbits of charged particles.
	Throughout the manuscript, we use the signature $(-,+,+,+)$, and use geometrized units thus setting $G=c=1$. Greeks indices run from $0$ to $3$.


	\section{Zipoy-Voorhees space-time immersed in a uniform magnetic field. \label{III}}
	
	The Zipoy-Voorhees space-time, also known as the $\gamma$-metric \cite{Voorhees:1971wh,Zipoy} is a well known asymptotically flat vacuum solution of Einstein's equations that belongs to the Weyl class of static, axially symmetric space-times~\cite{Weyl}. 
	In Erez-Rosen coordinates, the line element is given by~\cite{Herrera:1998eq} 
	\begin{equation}
	\label{II.1}
	ds^2=-Fdt^2+F^{-1}\left[Gdr^2+Hd\theta^2+(r^2-2mr)\sin^2\theta d\phi^2\right],
	\end{equation}
	where 
	\begin{equation}
	\label{II.2}
	\begin{aligned}
	F(r)&=\left(1-\frac{2m}{r}\right)^\gamma,\\
	G(r,\theta)&=\left(\frac{r^2-2mr}{r^2-2mr+m^2\sin^2\theta}\right)^{\gamma^2-1},\\
	H(r,\theta)&=\frac{(r^2-2mr)^{\gamma^2}}{\left(r^2-2mr+m^2\sin^2\theta\right)^{\gamma^2-1}}.\\
	\end{aligned}
	\end{equation}
	When $\gamma=1$ the line element reduces to Schwarzschild in Schwarzschild coordinates. From the expansion of the gravitational potential it is easy to evaluate the total mass $M$ of the source as~\cite{Herrera:1998eq} 
	\begin{equation}
	\label{II.3}
	M=\gamma m,
	\end{equation} 
	while the quadrupole moment $Q$ is given by
	\begin{equation}
	\label{II.4}
	Q=\frac{\gamma}{3}\left(1-\gamma^2\right)M^3.\\
	\end{equation}
	From Eq. (\ref{II.4}) we see that $\gamma>1$ correspond to oblate spheroids, while $\gamma<1$ correspond to prolate spheroids.
	
	
	Symmetries in the space-time can be studied by means of Killing vectors. In the case of vacuum, stationary and axial symmetric space-times, we have two Killing vectors: one related to time translations and the other related to spatial rotations about the symmetry axis. Such vectors are related to conserved quantities, energy and angular momentum, and satisfy the Killing equation~\cite{Wald:1984rg,Wald:1974np}
	\begin{equation}
	\label{III.1}
	\mathsterling_\xi g_{\mu\nu}=\nabla_\mu\xi_\nu+\nabla_\nu\xi_\mu=0,
	\end{equation}  
	where $\mathsterling_\xi g_{\mu\nu}$ is the Lie derivative of the metric tensor and $\nabla$ is the covariant derivative. One interesting property that follows form Eq.(\ref{III.1}) is the relation
	\begin{equation}
	\label{III.9}
	\nabla^\nu\nabla_\nu\xi_\mu=-R^{\;\;\;\lambda}_{\mu}\xi_\lambda=-R^\lambda_{\;\;\;\mu}\xi_\lambda.
	\end{equation}
	
On the other hand, in curved space-times, Maxwell's equations for the vector potential $A_\mu$ in the Lorentz gauge ($\nabla_\mu A^\mu=0$) are given by~\cite{Wald:1984rg,MTW}
	\begin{equation}
	\label{III.10}
	\nabla^\nu\nabla_\nu A_\mu-R^\lambda_{\;\;\mu}A_\lambda=-4\pi J_\mu.
	\end{equation}
	In this sense, if we consider $J_\mu=0$, Eq.~(\ref{III.10}) reduces to 
	\begin{equation}
	\label{III.11}
	\nabla^\nu\nabla_\nu A_\mu=R^\lambda_{\;\;\mu}A_\lambda.
	\end{equation} 
Note that there is a sign difference between Eqs.~(\ref{III.9}) and (\ref{III.11}), so that the two equations coincide in the case of vacuum space-times. This means that, when $R^\lambda_{\;\;\;\mu}=0$, the Killing vector $\xi_\mu$ must satisfy the source-free Maxwell's equations for a vector potential in the Lorentz gauge. In this sense, the Killing vector $\xi_\mu$ in vacuum is endowed with the property of being proportional to some vector potential $A_\mu$ and can be used to derive a solution for the electromagnetic field occurring when a stationary, axisymmetric space-time is placed in an external magnetic field aligned along the axis of symmetry~\cite{Azreg-Ainou:2016tkt}. Hence, the Faraday tensor of the electromagnetic field $\mathrm{F}_{\mu\nu}$ can be expressed as~\cite{Wald:1974np}
	\begin{equation}
	\label{III.2}
	\mathrm{F}_{\mu\nu}=\nabla_\mu\xi_\nu-\nabla_\nu\xi_\mu=-2\nabla_\nu\xi_\mu,
	\end{equation} 
	and the vector potential $A_{\mu}$ can be written as

	\begin{equation}
	\label{III.12}
	A_\mu=C_1\xi^{(t)}_\mu+C_2\xi^{(\phi)}_\mu,
	\end{equation}
	where $\xi^{(t)}_\mu$ and $\xi^{(\phi)}_\mu$ are the Killing vector fields associated with time translations and rotations about the symmetry axis.
	In this work, we consider the case of a magnetic field that is uniform, with magnitude $B$ at spatial infinity. If the field is oriented perpendicularly to the equatorial plane which is orthogonal to the symmetry axis, then the four-vector potential takes the form
	\begin{equation}
	\label{III.13}
	A_\mu=\frac{B}{2}\xi^{(\phi)}_\mu.
	\end{equation}
	Consequently, the only nonzero component of the potential of the electromagnetic field is~\cite{Wald:1984rg}
	\begin{equation}
	\label{III.14}
	A_\phi=\frac{B}{2}g_{\phi\phi}.
	\end{equation}
	When applied to the $\gamma$-metric, given in Eq.~(\ref{II.1}), the vector potential is given by~\cite{Benavides-Gallego:2018htf}
	\begin{equation}
	\label{III.15}
	A_\phi=\frac{B}{2}\left(1-\frac{2m}{r}\right)^{1-\gamma}r^2\sin^2\theta,
	\end{equation}
	and it automatically reduces to the well-known vector potential in the Schwarzschild space-time for $\gamma$=1. We shall now move to consider the motion of test particles in the Zipoy-Voorhees geometry immersed in the magnetic field discussed above.
	
	\begin{center}
		\begin{figure*}[hhh]
			\includegraphics[scale=0.25]{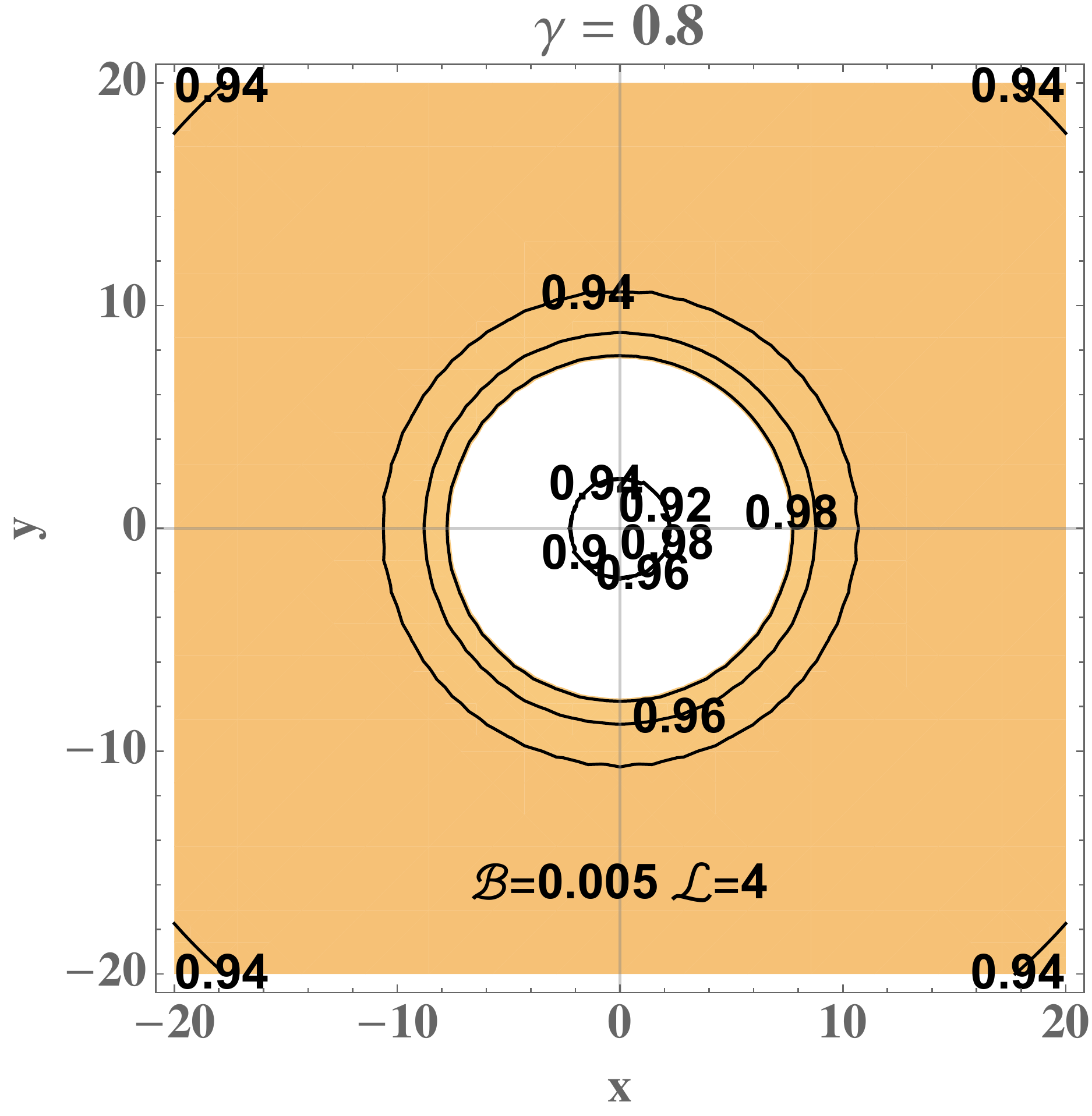}
			\hspace{0.5cm}
			\includegraphics[scale=0.25]{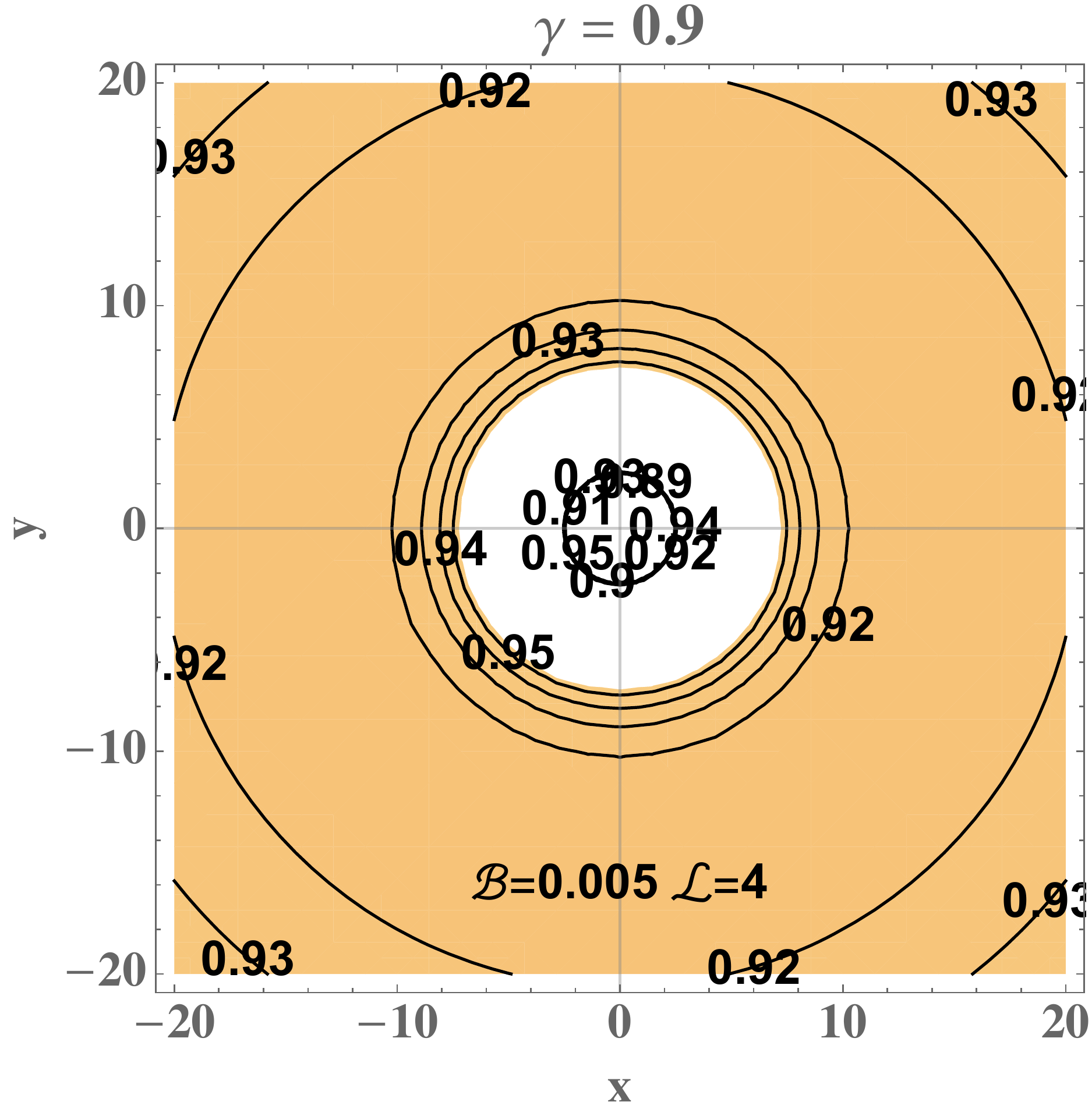}
			\hspace{0.5cm}
			\includegraphics[scale=0.25]{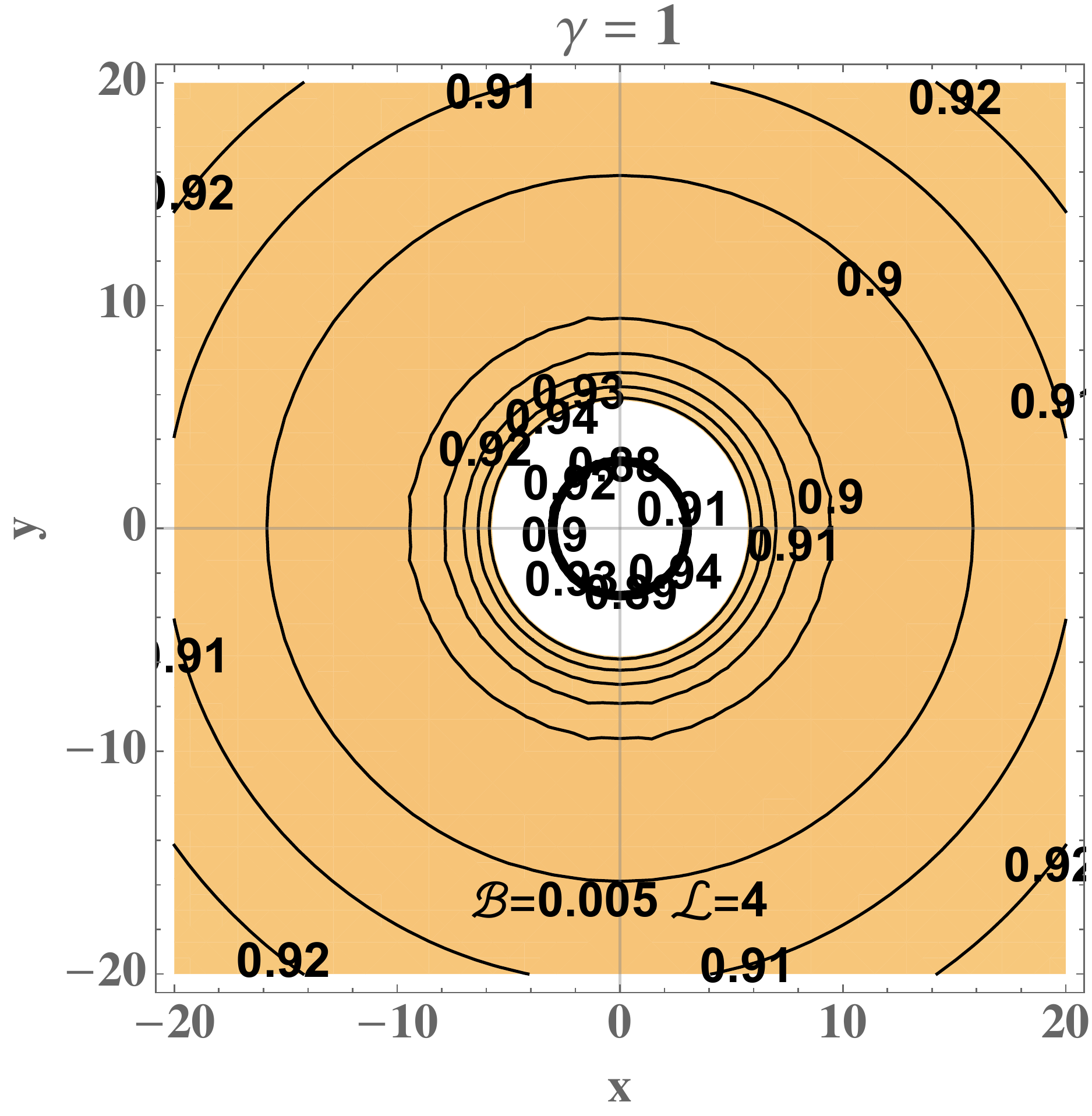}
			\hspace{0.5cm}
			\includegraphics[scale=0.25]{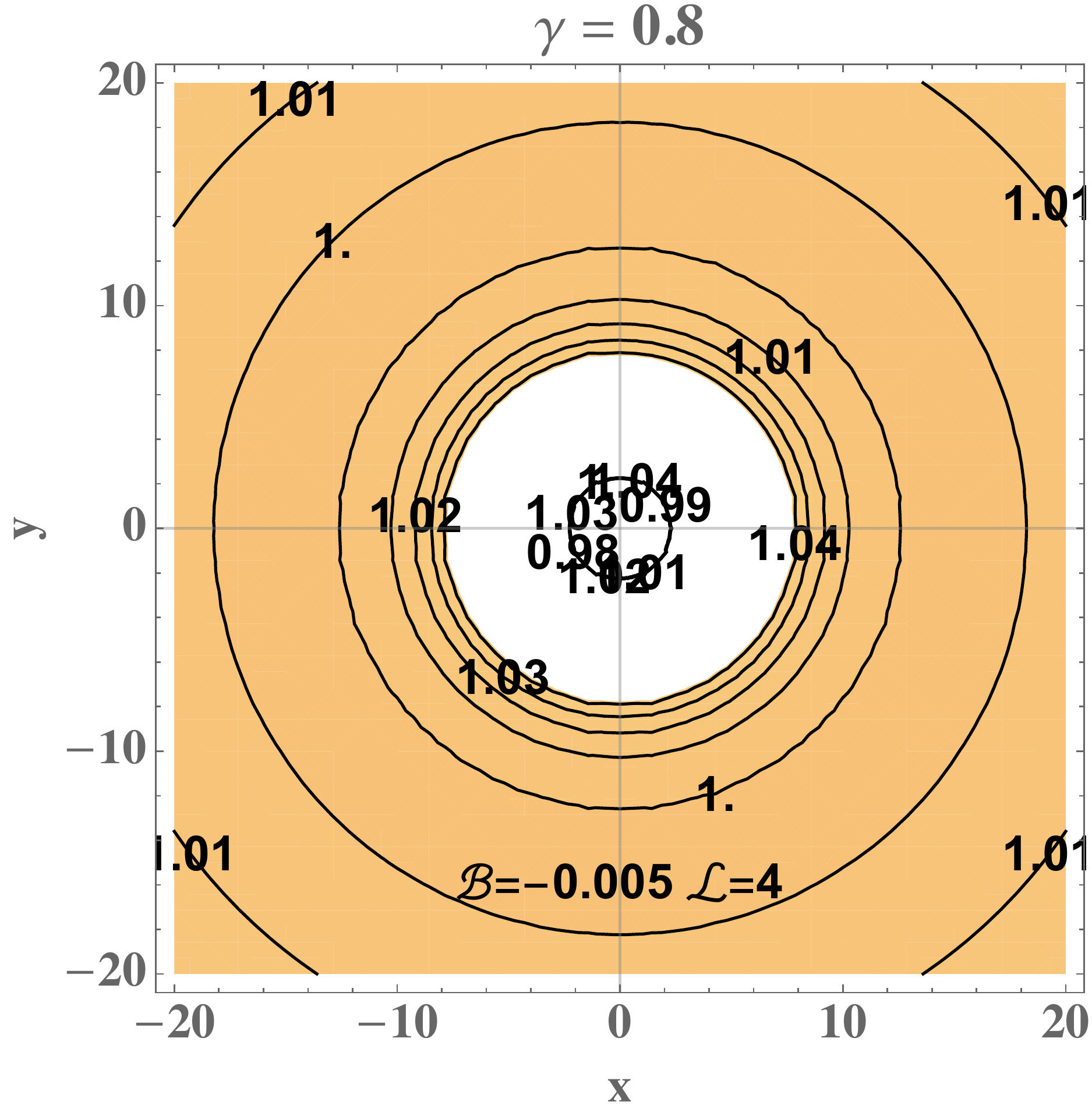}
			\hspace{0.5cm}
			\includegraphics[scale=0.25]{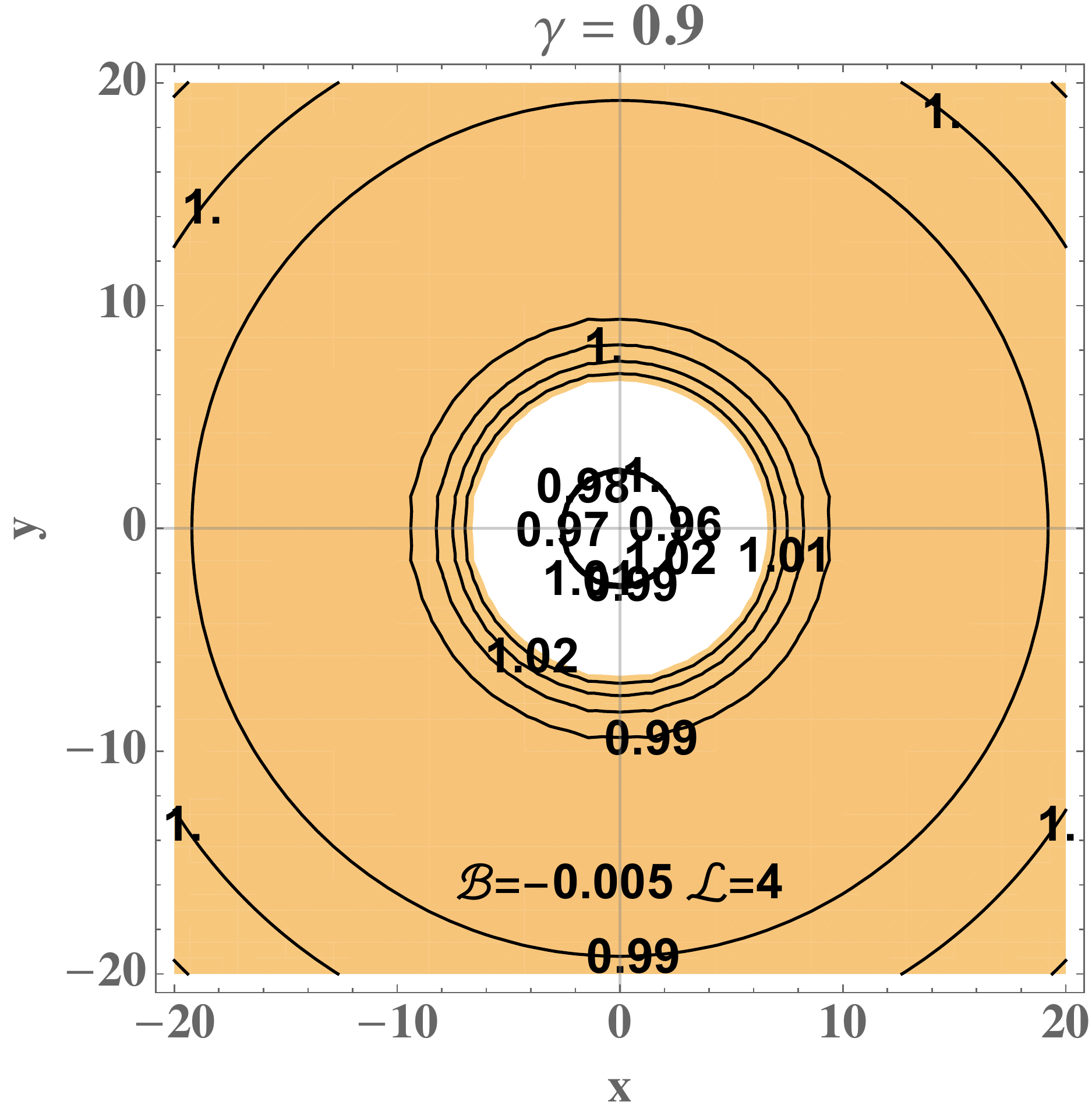}
			\hspace{0.5cm}
			\includegraphics[scale=0.25]{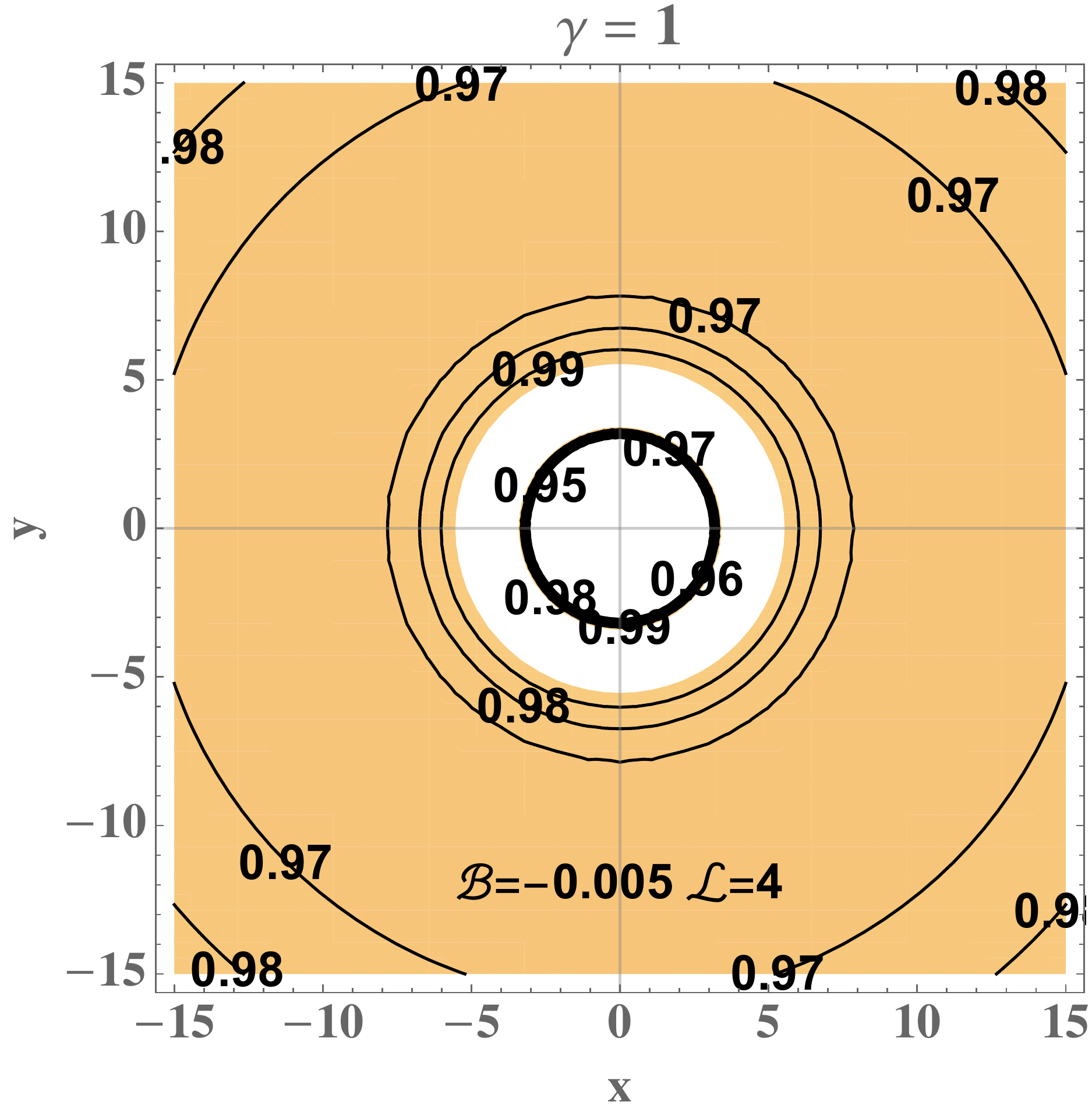}
			\caption{Equipotential slices for the effective potential in the equatorial plane ($\theta=\pi/2$), $V_{\text{eff}}(r,\pi/2;\mathcal{L},\mathcal{B})$ as a function of the Cartesian coordinates $x$ and $y$ for fixed values of $\mathcal{B}=\pm 0.005$ and $\mathcal{L}=4$. In the figure we have set $m=1$.\label{figure0}}
		\end{figure*}
	\end{center}
		
	\section{Charged particle dynamics \label{IV}}
	
	To study the dynamics of charged particles we follow Ref.~\cite{Kolos:2015iva}. The Hamiltonian for the charged particle can be written as
	\begin{equation}
	\label{IV.1}
	\mathcal{H}=\frac{1}{2}g^{\alpha\beta}(\pi_\alpha-qA_\alpha)(\pi_\beta-qA_\beta)+\frac{1}{2}m^2_0,
	\end{equation}
	where $\pi_\alpha$ is the canonical four-momentum which is related to the kinematical four-momentum $p^\mu=m_0 u^\mu$ by the relation
	\begin{equation}
	\label{IV.2}
	\pi^\mu=p^\mu+qA^\mu,
	\end{equation} 
	and satisfies Hamilton's equations 
	\begin{equation}
	\label{IV.3}
	\begin{aligned}
	\frac{dx^\mu}{d\zeta}&\equiv p^\mu=\frac{\partial \mathcal{H}}{\partial \pi_\mu},\\
	\frac{d\pi_\mu}{d\zeta}&=-\frac{\partial \mathcal{H}}{\partial x^\mu}.\\
	\end{aligned}
	\end{equation}
	The affine parameter $\zeta$ is related to the proper time of the particle by the relation $\zeta=\tau/m_0$. For the line element in Eq.~(\ref{II.1}), using Eqs.~(\ref{IV.1}) and (\ref{IV.3}), we obtain two constants of motion: the energy per unit mass $\mathcal{E}$ and the angular momentum per unit mass $\mathcal{L}$, which are given by
	\begin{equation}
	\label{IV.4}
	\begin{aligned}
	\mathcal{E}&=\frac{E}{m_0}=\left(1-\frac{2m}{r}\right)^{\gamma}\frac{dt}{d\tau}, \\
	\mathcal{L}&=\frac{L}{m_0}=r^2\sin^2\theta\left(1-\frac{2m}{r}\right)^{1-\gamma}\left[\frac{d\phi}{d\tau}+\mathcal{B}\right],
	\end{aligned}
	\end{equation}
	where we have introduced also $\mathcal{B}=qB/2m_0$. As expected, Eqs.~(\ref{IV.4}) reduce to the Schwarzschild case when $\gamma=1$.
	
	\begin{center}
		\begin{figure*}[hhh]
			\includegraphics[scale=0.4]{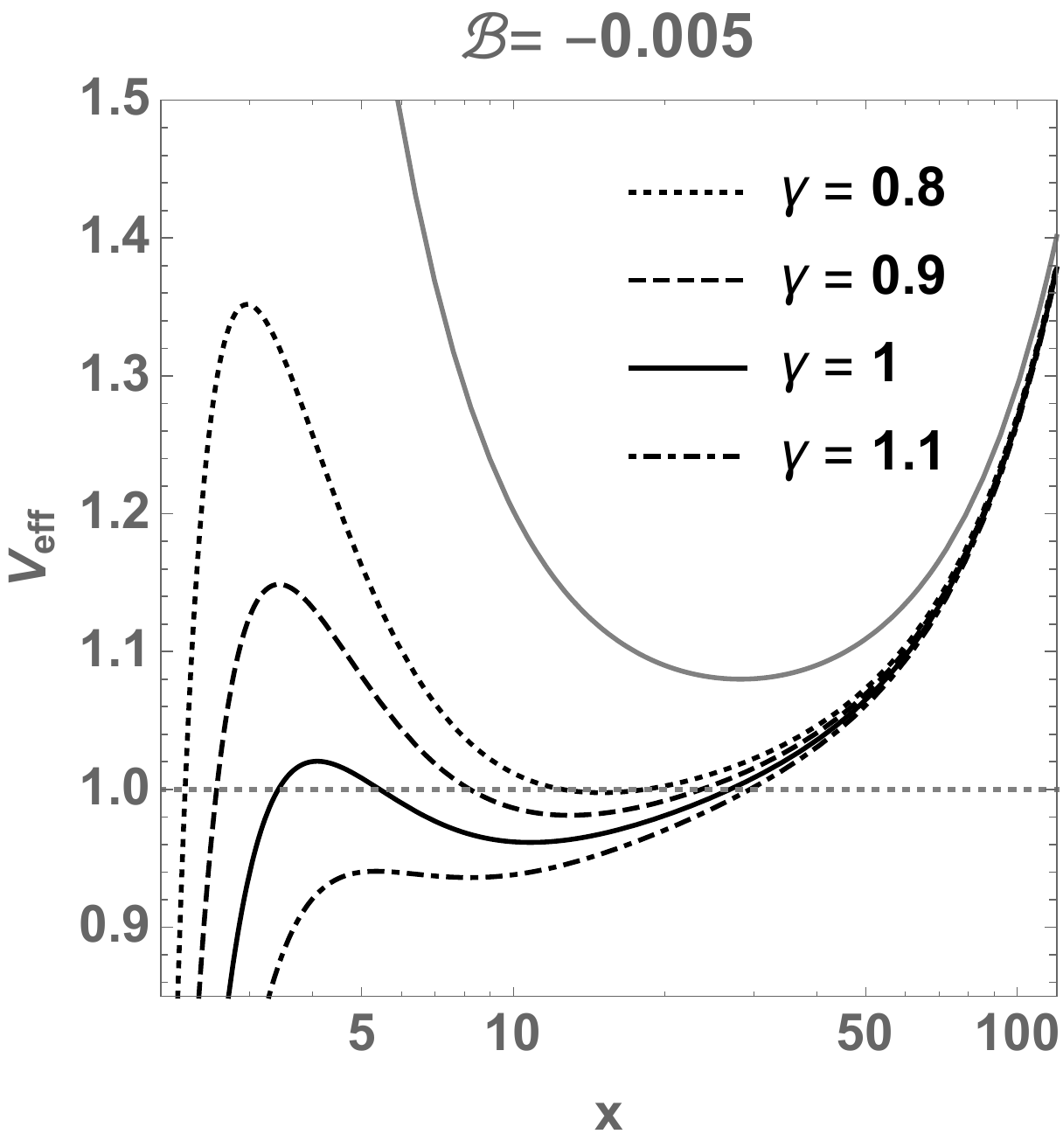}
			\hspace{0.5cm}
			\includegraphics[scale=0.4]{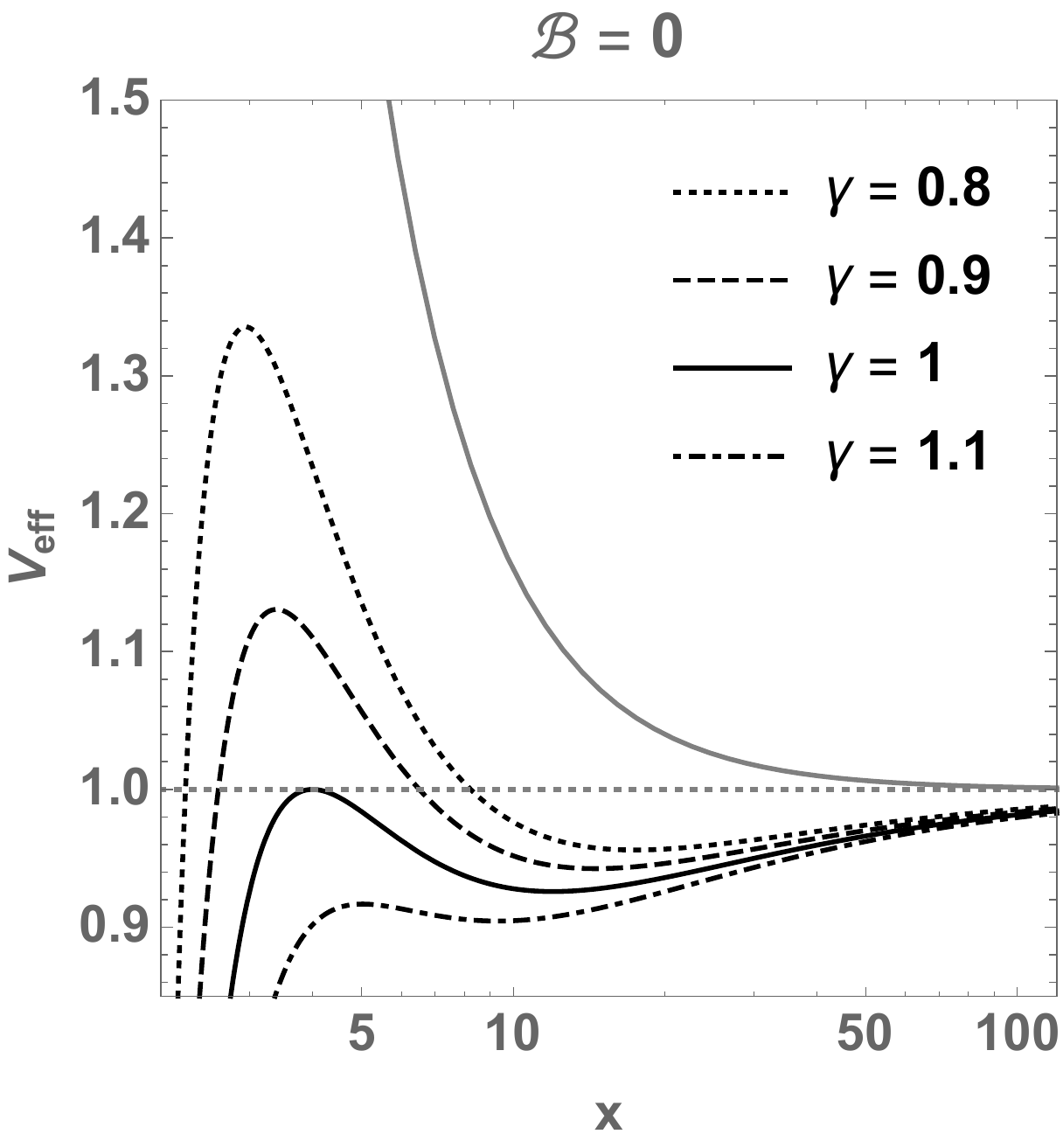}
			\hspace{0.5cm}
			\includegraphics[scale=0.4]{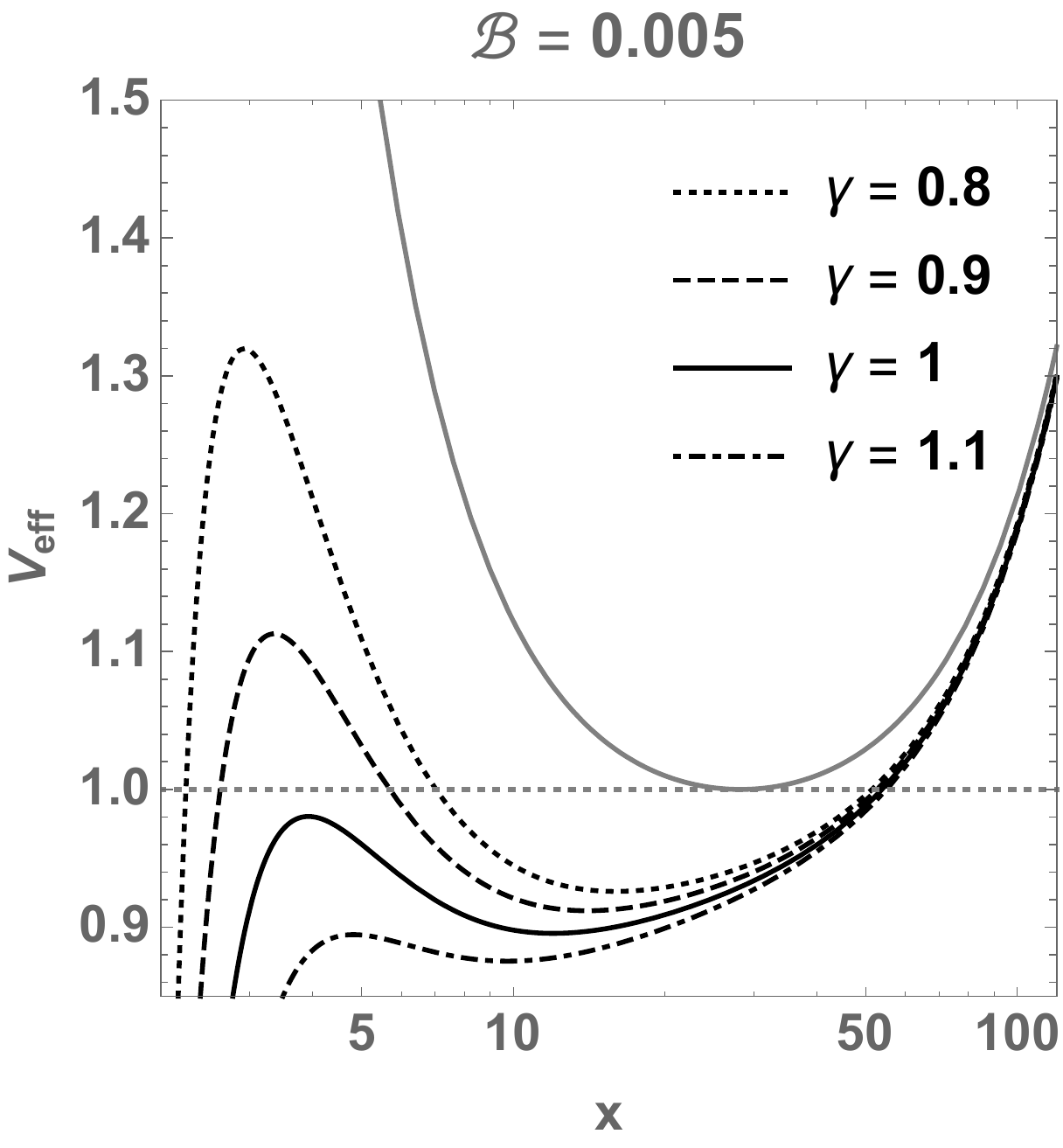}
			\caption{Sections ($y=0$) of the effective potential taken at the equatorial plane $z=0$ and at spatial infinity $z\rightarrow\infty$ (gray solid line) for different values of $\gamma$. In the figure we consider $\mathcal{B}=0$, $\mathcal{B}=\pm0.005$, $m=1$ and $\mathcal{L}=4$. \label{figureI}}
		\end{figure*}
	\end{center}
	
	By expressing the Hamiltonian in Eq.~(\ref{IV.1}) as
	
	\begin{equation}
	\label{IV.6}
	\mathcal{H}=\frac{1}{2}\frac{F(r)}{G(r)} p^2_r+\frac{1}{2}\frac{F(r)}{H(r)}p^2_\theta+\frac{m^2_0}{2F(r)}\left(V_{\text{eff}}(r,\theta;\mathcal{B},\mathcal{L})-\mathcal{E}^2\right).
	\end{equation}
	we can derive the effective potential $V_{\text{eff}}(r,\theta;\mathcal{B},\mathcal{L})$  as~\cite{Benavides-Gallego:2018htf}
	\begin{widetext}
		\begin{equation}
		\label{IV.7}
		V_{\text{eff}}(r,\theta;\mathcal{B},\mathcal{L})=\left(1-\frac{2m}{r}\right)^\gamma\left[1+\left(1-\frac{2m}{r}\right)^{\gamma-1}\left(\frac{\mathcal{L}}{r\sin\theta}-\mathcal{B}r\sin\theta \left(1-\frac{2m}{r}\right)^{1-\gamma}\right)^2\right].
		\end{equation}
	\end{widetext}
	The term in parentheses of Eq.~(\ref{IV.7}) is the central force potential for the specific
	angular momentum $\mathcal{L}$ and the electromagnetic potential energy given by the magnetic parameter $\mathcal{B}$. Once again, note that Eq.~(\ref{IV.7}) reduces to the Schwarzschild case when $\gamma=1$. 
	As usual, for a given value of $\mathcal{E}$ the particle's motion is restricted by the condition that
	\begin{equation}
	\label{IV.7a}
	\mathcal{E}^2=V_{\text{eff}}(r,\theta;\mathcal{L},\mathcal{B}).
	\end{equation}
	
	Due to the static nature of the space-time, the effective potential of the $\gamma$-metric immersed in a uniform magnetic field is symmetric with respect to the change in sign for $(\mathcal{L},\mathcal{B})$, namely replacing $(\mathcal{L},\mathcal{B})$ with $(-\mathcal{L},-\mathcal{B})$ does not change the effective potential~\cite{Kolos:2015iva,Frolov:2010mi}. This symmetry  enables us to distinguish the following two situations:
	\begin{enumerate}
		\item[a.] Minus configuration (\textbf{MC}): the magnetic field and angular momentum parameters have opposite signs and the Lorentz force is attracting the charged particle toward the axis of symmetry. This holds for $\mathcal{L}>0$, $\mathcal{B}<0$ or, equivalently, for $\mathcal{L}<0$, $\mathcal{B}>0$
		\item[b.] Plus configuration (\textbf{PC}): the magnetic field and angular momentum parameters have the same sign and the Lorentz force is repulsive, pushing the particle away from the source. In this configuration we have $\mathcal{L}>0$, $\mathcal{B}>0$ or equivalently $\mathcal{L}<0$, $\mathcal{B}<0$).
	\end{enumerate}
	Having fixed the direction of the symmetry axis $z$, a positive value of the angular momentum $\mathcal{L}>0$ means that the particle moves counter-clockwise. On the other hand, taking the particle's charge $q>0$, in the \textbf{MC}, $\mathcal{B}<0$ corresponds to the magnetic field pointing in the negative direction (downward), while in the \textbf{PC}, $\mathcal{B}>0$ corresponds to the magnetic field pointing upward the $z$-axis.

	In Fig.~\ref{figure0}, we show the equipotential slices for the effective potential at the equatorial plane for different values of $\gamma$. From the figure, it is possible to see the regions where $V_{\text{eff}}$ has a local minimum. 
	On the other hand, in Fig.~\ref{figureI}, we plot sections ($y=0$) of the effective potential $V_{\text{eff}}$ as function of the Cartesian direction $x$ for different values of $\gamma$ at a fixed value of $\mathcal{B}$. We consider two cases: $z=0$ and $z\rightarrow \infty$. To do so, we use spherical coordinates given by 
	\begin{equation}
	\begin{array}{ccc}
	x=r\sin\theta\cos\phi,&y=r\sin\theta\sin\phi,&z=r\cos\theta,
	\end{array}
	\end{equation}
	with $r=\sqrt{x^2+y^2+z^2}$. For $y=0$ (i.e. $\phi=0$), the last equation reduces to $x=r\sin\theta$ and $z=r\cos\theta$. Therefore, the effective potential takes the form,
	\begin{widetext}
	\begin{equation}
	V_{\text{eff}}(x,0,z;\mathcal{L},\mathcal{B})=\left(1-\frac{2 m}{\sqrt{x^2+z^2}}\right)^{\gamma } \left[\left(1-\frac{2 m}{\sqrt{x^2+z^2}}\right)^{\gamma -1} \left(\frac{\mathcal{L}}{x}-x \mathcal{B} \left(1-\frac{2 m}{\sqrt{x^2+z^2}}\right)^{1-\gamma }\right)^2+1\right].
	\end{equation}
	\end{widetext}
    At $z\rightarrow \infty$, the effective potential reduces to 
    \begin{equation}
   V_{\text{eff}}(x,0,z\rightarrow \infty;\mathcal{L},\mathcal{B})=1+\left(\frac{\mathcal{L}}{x}-\mathcal{B}x\right)^2,
    \end{equation}
    which is the same expression as Schwarzschild~\cite{Kolos:2015iva}.
    
	From Fig.~\ref{figureI} it can be seen that in the equatorial plane ($z=0$) notable differences in $V_{\text{eff}}$ for different values of $\gamma$ appear in the region where the role of the magnetic field is suppressed by gravity (i.e. small values of $x$). In this case, it is possible to distinguish between different values of $\gamma$. At large values of $x$, the role of the magnetic field dominates over the gravitational part of the effective potential, and differences between different values of $\gamma$ become smaller. Fig.~\ref{figureI} also shows that at $z\rightarrow\infty$, the behavior of $V_{\text{eff}}$ does not depend on $\gamma$ and is the same as in the Schwarzschild case~\cite{Kolos:2015iva} (see the solid gray lines in the figure). 
	The reason for such a behavior lies in those terms which contain the dependence on $\gamma$ in Eq.~(\ref{IV.7}). Since these terms tend to $1$ when $z\rightarrow \infty$, the effective potential reduces to the Schwarzschild case as $z\rightarrow\infty$ as shown also for example in Ref.~\cite{Kolos:2015iva}. This behavior is expected since both line elements ($\gamma$-metric and Schwarzschild) are asymptotically flat.

	According to Eq.~(\ref{IV.7}), the effective potential is a function of two variables only: $r$ and $\theta$. This is due to the axial symmetry of the system which is in turn due to the symmetry of the background magnetic field and the symmetry of the geometry. Therefore, the extrema of the effective potential, can be obtained from
	\begin{equation}
	\label{IV.8}
	\begin{array}{ccc}
	\partial_r V_{\text{eff}}(r,\theta;\mathcal{L},\mathcal{B})=0&\text{and}&\partial_\theta V_{\text{eff}}(r,\theta;\mathcal{L},\mathcal{B})=0.
	\end{array}
	\end{equation}
	From the second condition in Eq.(\ref{IV.8}), we see that all extrema of $V_{\text{eff}}(r,\theta;\mathcal{L}, \mathcal{B})$ occur in the equatorial plane. Therefore, similarly to the Schwarzschild case, there are no off-equatorial circular orbits for charged particles. 
	
	On the other hand, from the first condition in Eq.~(\ref{IV.8}) restricted to the equatorial plane, we obtain the following polynomial equation 
	\begin{equation}
	\label{IV.9}
    d\mathcal{L}^2+e\mathcal{L}+f=0\;,
	\end{equation}
	with
	\begin{equation}
	\label{IV.9a}
	\begin{aligned}
	d&=[r-m(1+2\gamma)](r-2m)^{2\gamma-2},\\\\
	e&=2m\gamma\mathcal{B}r^{1+\gamma}(r-2m)^{\gamma-1},\\\\
	f&=-(r-m)r^{2\gamma+2}\mathcal{B}^2-m\gamma r^{1+\gamma}(r-2m)^{\gamma}.
	\end{aligned}
	\end{equation}
	
	Equation (\ref{IV.9}) is quadratic for the specific angular momentum. Therefore, if we solve for $\mathcal{L}$, we find that circular orbits are given by 
    \begin{widetext}
	\begin{equation}
	\label{IV.10}
	\mathcal{L}_{E\pm}=(r-2m)^{1-\gamma}\left[\frac{-m\mathcal{B}\gamma r^{\gamma+1}\pm(r-2m)^{1-\gamma}r^{\frac{1+\gamma}{2}}\mathcal{F}(r;\mathcal{B})}{r-m(1+2\gamma)}\right]\;,
	\end{equation}
	with
	\begin{equation}
	\mathcal{F}(r;\mathcal{B})=\sqrt{(r-2m)^{2\gamma-3}\left[m(r-2m)^\gamma(r-m(1+2\gamma))\gamma+\mathcal{B}^2(r-2m)r^{\gamma+1}(m(1+\gamma)-r)^2\right]}.
	\end{equation}
	\end{widetext}
	
	Equation~(\ref{IV.9}) can be used to obtain the innermost stable circular orbit (ISCO). There are two ways to compute the ISCO: either by solving  the equation $\partial^2_r V_{\text{eff}}=0$, or by means of the local extrema $\mathcal{L}_{E(\text{ex})}$ of Eq.~(\ref{IV.10}). Hence, after using the implicit function theorem on Eq.~(\ref{IV.9}), we find the local extremum of $\mathcal{L}_{E\pm}(r;\mathcal{B})$ is given by
	\footnote{We considered the function
		\begin{equation*}
		\begin{aligned}		 
		\mathcal{W}&=\mathcal{L}^2 [r-m(2 \gamma +1)] (r-2m)^{2 \gamma -2}+2m\gamma  \mathcal{B} r^{\gamma +1} (r-2m)^{\gamma -1}\mathcal{L}\\
		& -m\gamma  r^{\gamma +1} (r-2m)^{\gamma -1} -(r-m) \mathcal{B}^2 r^{2 \gamma +2},
		\end{aligned}
		\end{equation*}
		from which $d\mathcal{L}/dr=-(\partial \mathcal{W}/\partial r)/(\partial \mathcal{W}/\partial \mathcal{L})$, where $\partial\mathcal{W}/\partial \mathcal{L}\neq 0$. Thus, using the condition $d\mathcal{L}/dr=0$ and solving for $\mathcal{L}$ we obtain Eq.~(\ref{IV.11}).}. 
	\begin{widetext}
		\begin{equation}
		\label{IV.11}
		\mathcal{L}_{E(\text{ex})}=\frac{-2m\gamma\mathcal{B}r^\gamma(r-2m)^{1-\gamma}[(r-m)\gamma-m] +(r-2m)^{1-\gamma}\mathcal{G}(r;\mathcal{B})}{(r-2m\gamma)(2\gamma-1)} \, ,
		\end{equation} 
		where the function $\mathcal{G}(r;\mathcal{B})$ is given by
		\begin{equation}
		\label{IV.12}
		\mathcal{G}(r;\mathcal{B})=\sqrt{4m^2\mathcal{B}^2r^{2\gamma}\gamma^2[m+\gamma(m-r)]^2-r^\gamma(r-2m)(r-2m\gamma)(2\gamma-1)\delta} \, ,
		\end{equation}
		with
		\begin{equation}
		\label{IV.11a}
		\delta=-\mathcal{B}^2r^{1+\gamma}(3r-2m+2(r-m)\gamma)+2m(r-2m)^{\gamma-2}\gamma(m+\gamma(m-r)) \, .
		\end{equation}
	\end{widetext}
	
	\begin{table}[hhh]
		\label{tableII}
		\centering
		\begin{tabular}{|c|c|c|c|c|c|}
			\hline
			$\gamma$& $\mathcal{B}=-0.2$ & $\mathcal{B}=-0.1$& $\mathcal{B}=0$ & $\mathcal{B}=0.1$&$\mathcal{B}=0.2$\\
			\hline
			& $r_{\text{ISCO}}$ & $r_{\text{ISCO}}$& $r_{\text{ISCO}}$ & $r_{\text{ISCO}}$ &$r_{\text{ISCO}}$\\
			\hline
			0.6&3.028&3.235&3.6944&2.987&2.618\\
			\hline
			0.7&3.411&3.609&4.3045&3.284&2.817\\
			\hline
			0.8&3.759&3.922&4.8832&3.549&3.033\\
			\hline
			0.9&4.066&4.291&5.4464&3.802&3.227\\
			\hline
			1&4.4139&4.6319&6.0000&3.9827&3.3552\\
			\hline
			1.1&4.763&4.981&6.54722&4.205&3.487\\
			\hline
		\end{tabular}
		\caption{\label{tabI} Numerical values of the innermost stable circular orbits ($r_{\text{ISCO}}$) as a function of $\gamma$ and $\mathcal{B}$. We set $m=1$, which is equivalent to using the rescaling $r\rightarrow r/m$ in the line element.}
	\end{table}
	\begin{center}
		\begin{figure}[b!]
			\includegraphics[scale=0.5]{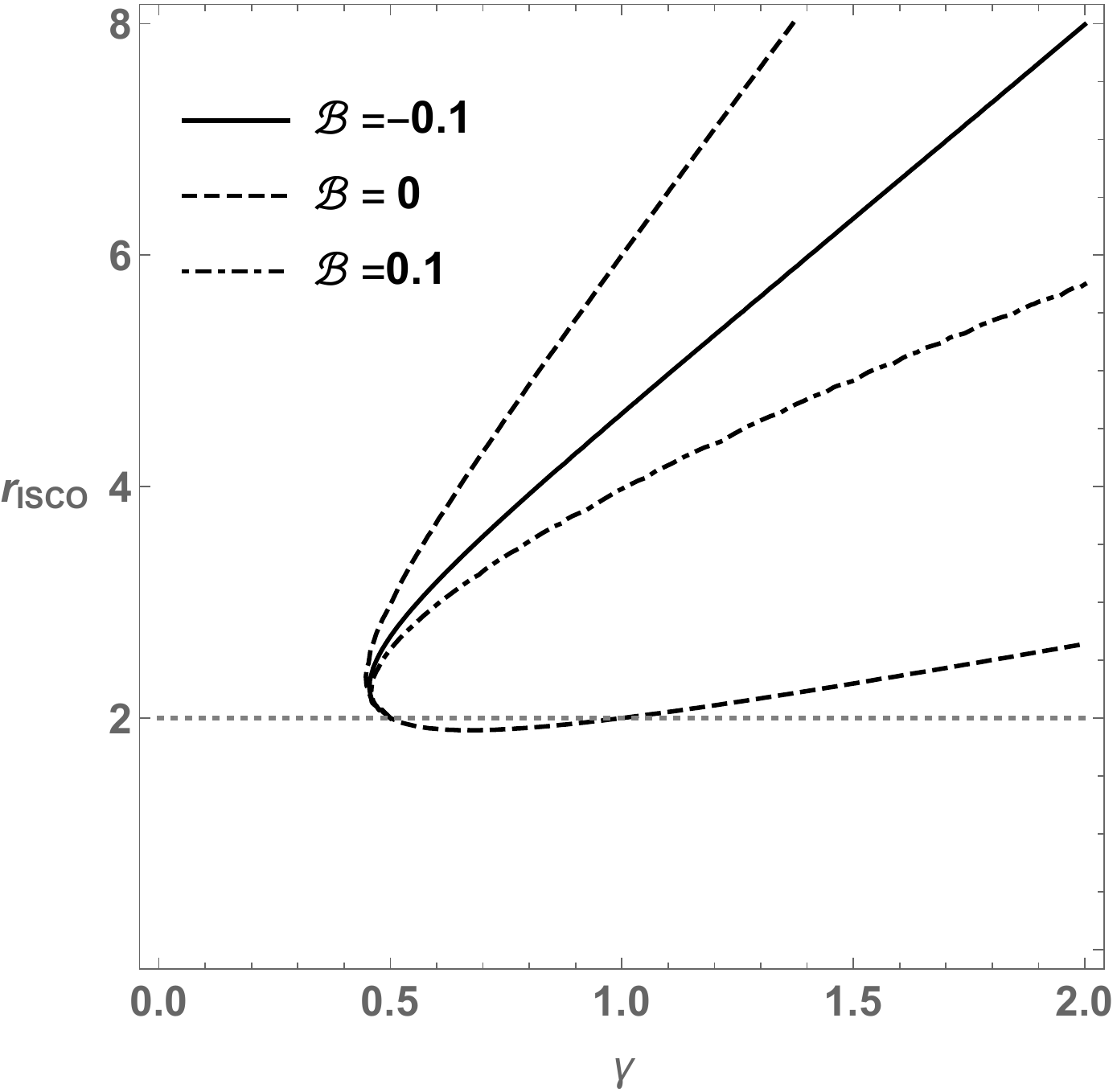}
			\caption{Location of $r_{\text{ISCO}}$ as a function of $\gamma$ for different values of $\mathcal{B}$. Again, we set $m=1$. \label{figureIA}}
		\end{figure}
	\end{center}
	The ISCO is located at the point of intersection of the functions $\mathcal{L}_{\text{E(ex)}}$ and $\mathcal{L}_{\text{E+}}$. 
	In Table~\ref{tabI} some ISCO radii are evaluated for different values of $\gamma$ and $\mathcal{B}$. In Fig.~\ref{figureIA}, we show the behavior of $r_{\text{ISCO}}$ as a function of $\gamma$ for different values of $\mathcal{B}$. The case without an external magnetic field (i.e. the dashed line in Fig.~\ref{figureIA}) was studied in Ref.~\cite{Chowdhury:2011aa}. There it was shown that the dependence of the ISCO radius upon the $\gamma$ parameter for neutral particles could be divided into three different regions: (i) For $\gamma < 1/\sqrt{5}$ there is no ISCO, (ii) for $ 1/\sqrt{5}<\gamma <1/2$, there are two disjoint regions for stable circular orbits (i.e. two separate allowed values of the ISCO), (iii) for $\gamma>1/2$, there is only one ISCO, similarly to the spherically symmetric case. 
	A similar situation was described in Ref.~\cite{Benavides-Gallego:2018htf}. 
	
	\begin{center}
		\begin{figure*}[t]
			\includegraphics[scale=0.37]{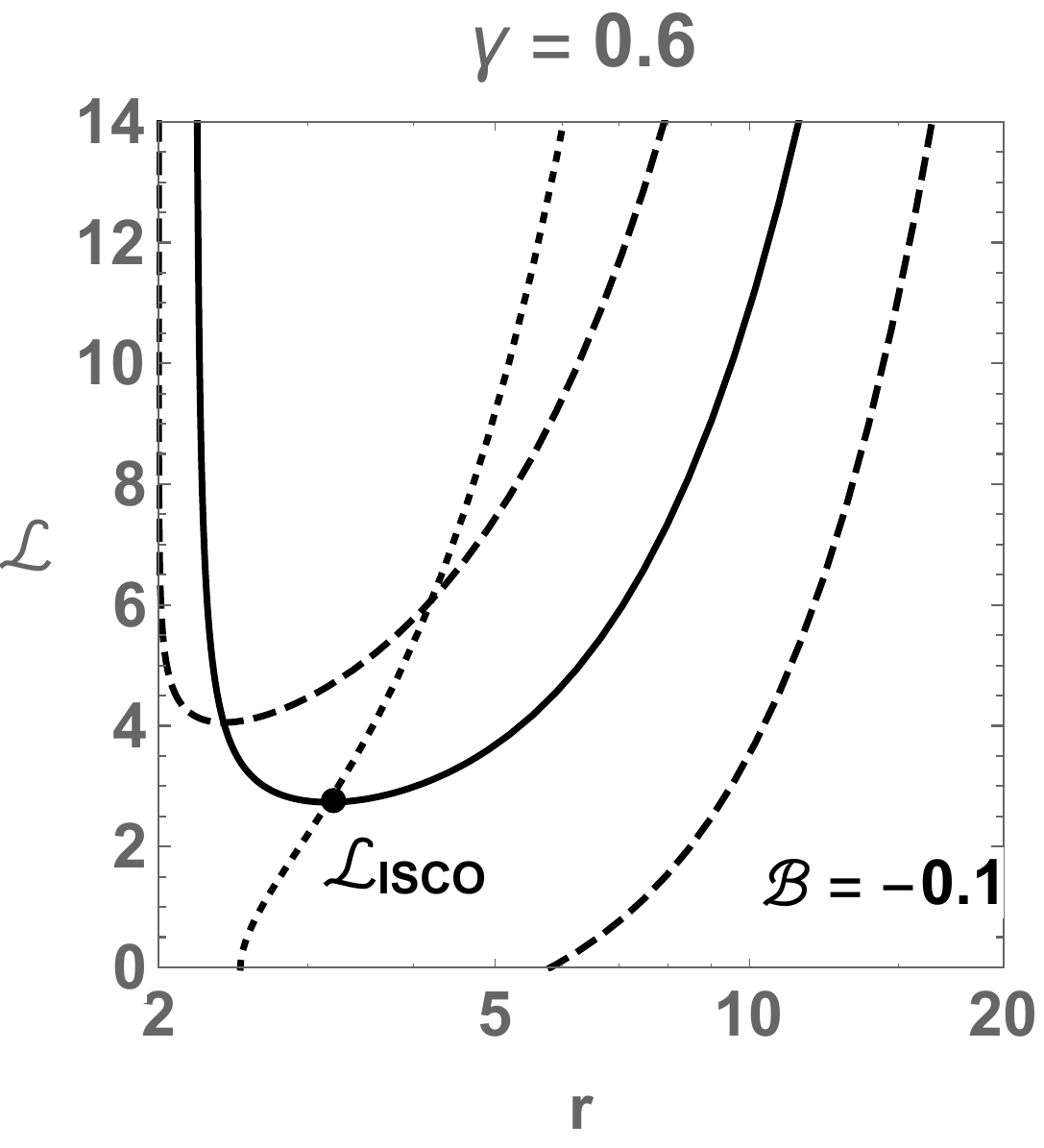}
			\hspace{0.7cm}
			\includegraphics[scale=0.37]{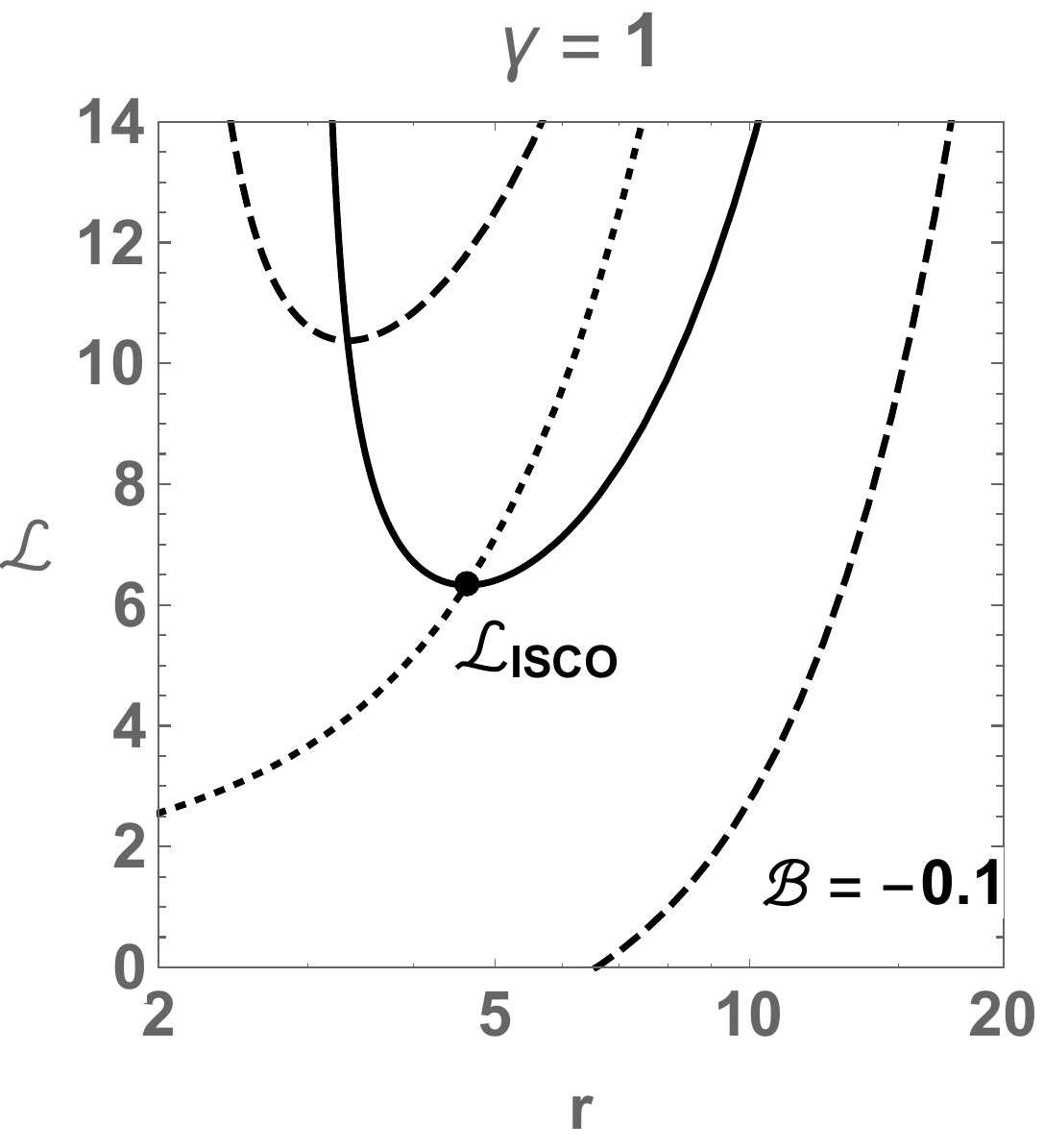}
			\hspace{0.7cm}
			\includegraphics[scale=0.37]{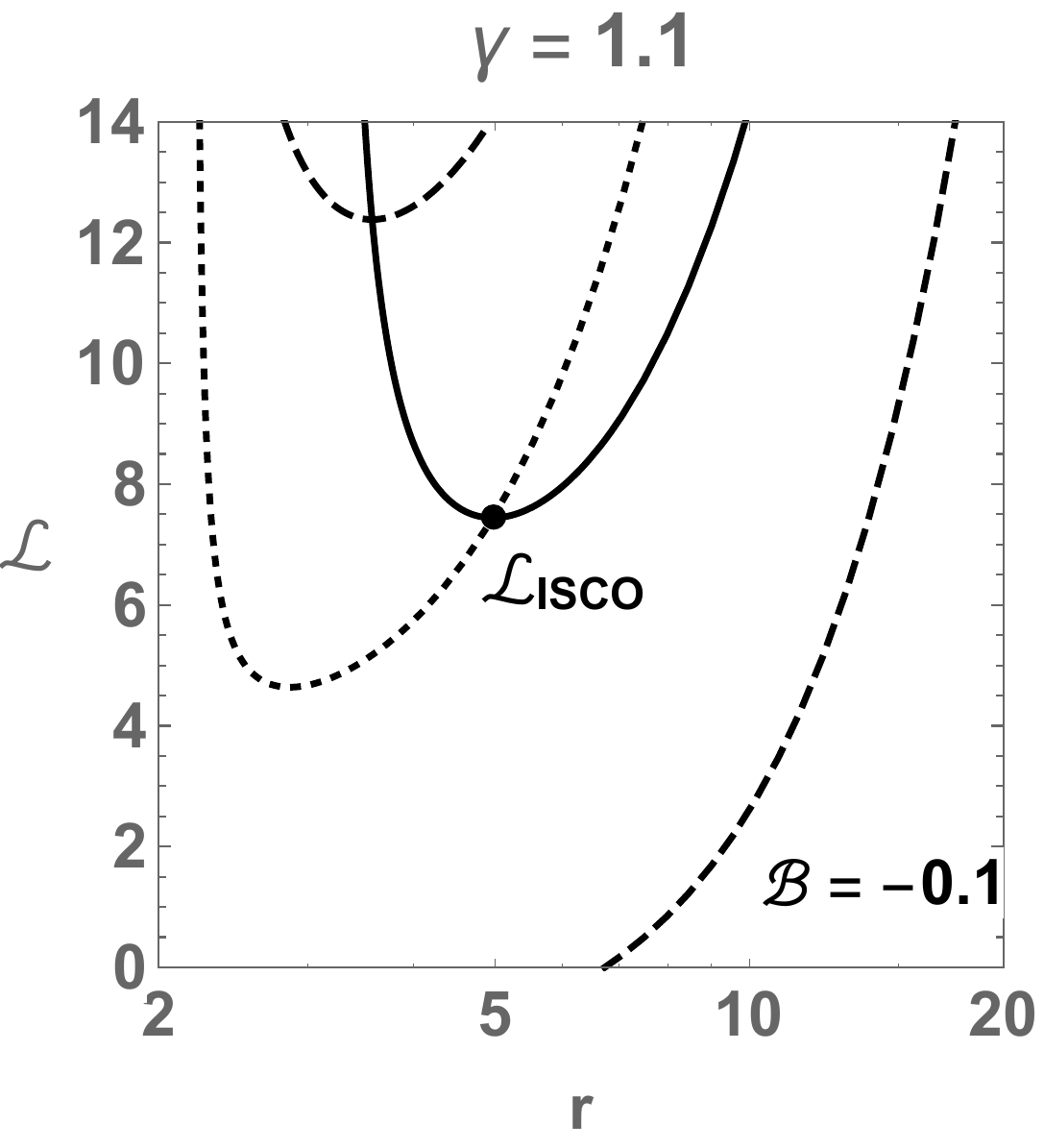}\\
			\includegraphics[scale=0.37]{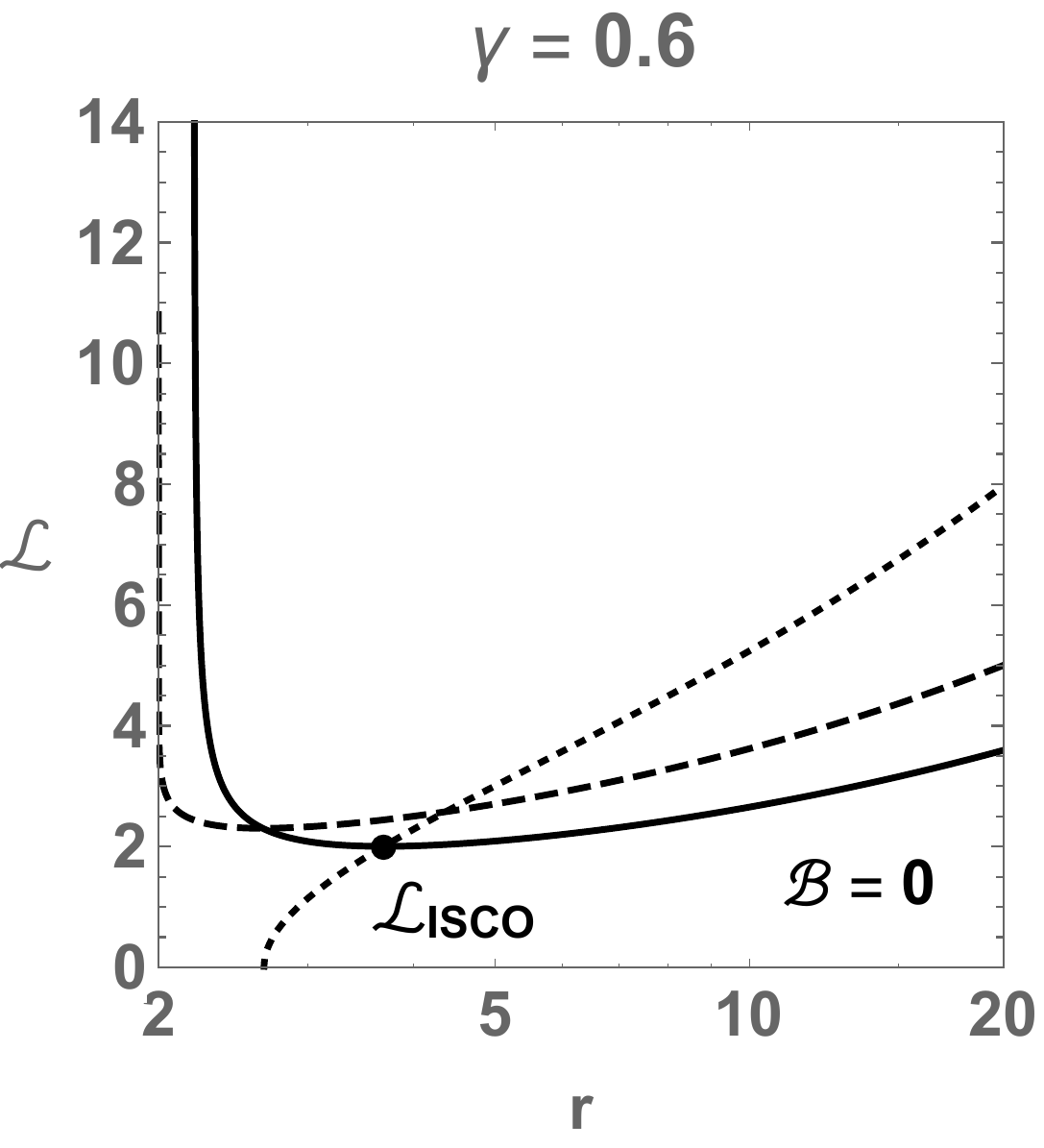}
			\hspace{0.7cm}
			\includegraphics[scale=0.37]{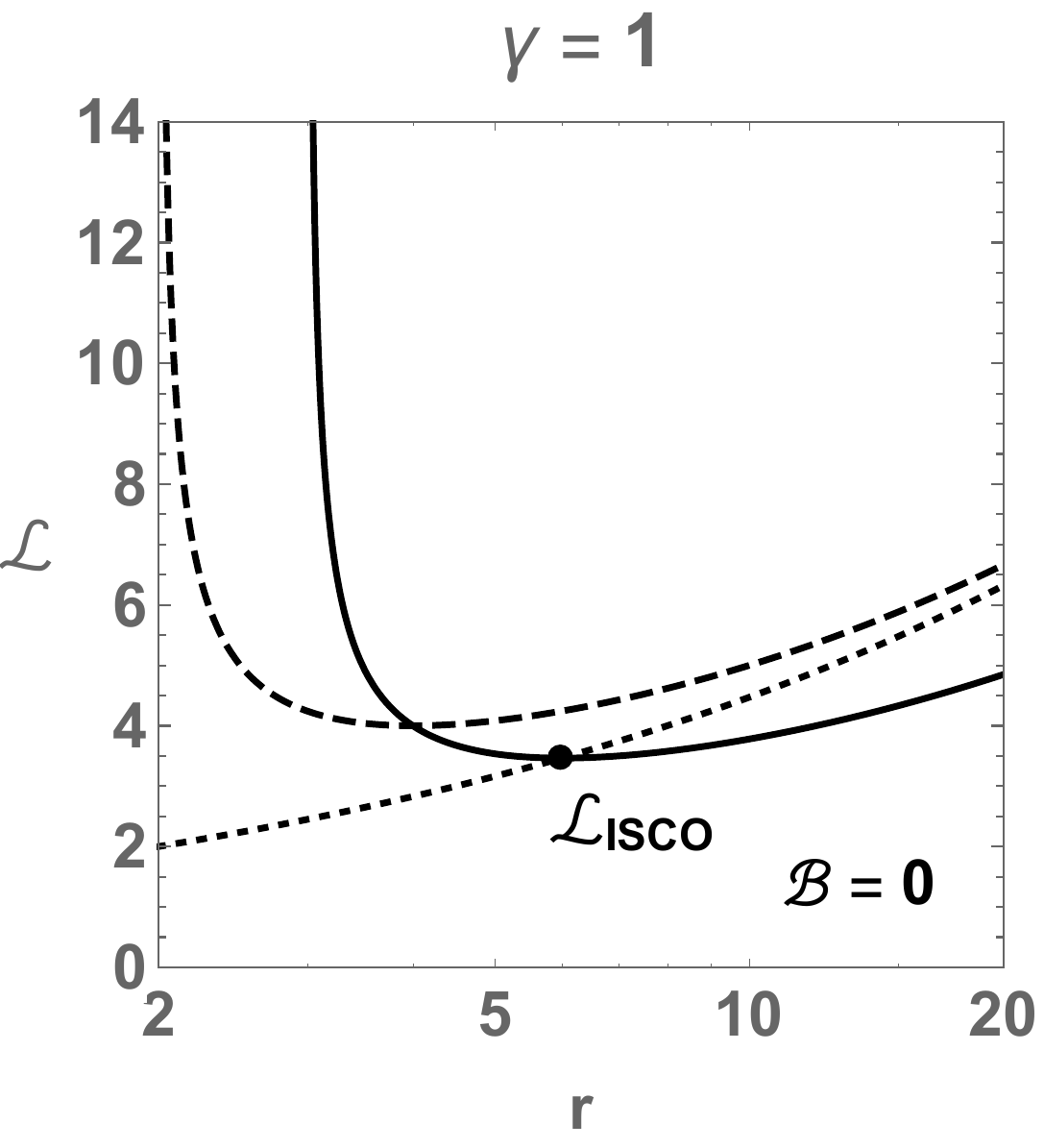}
			\hspace{0.7cm}
			\includegraphics[scale=0.37]{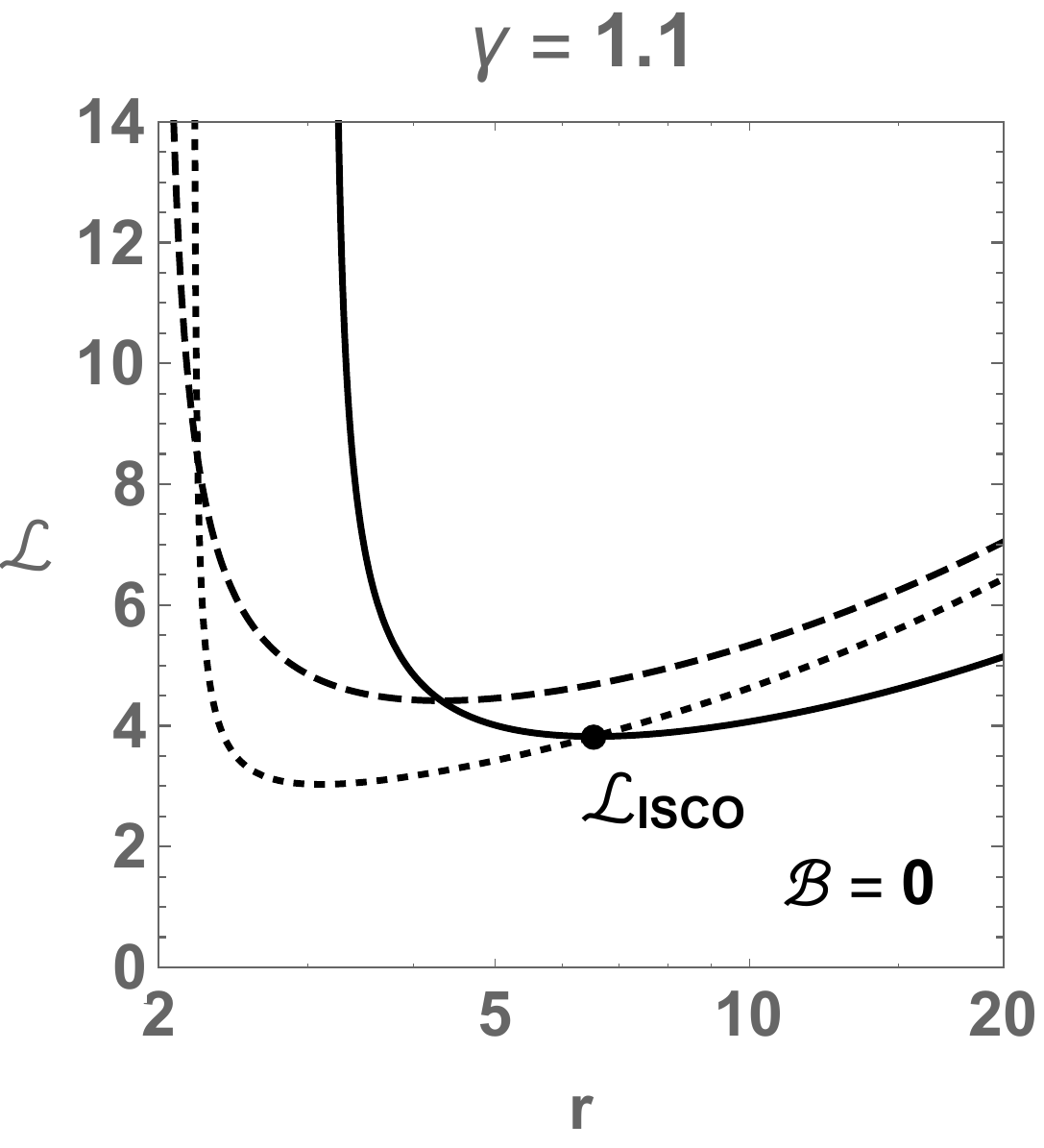}\\
			\includegraphics[scale=0.37]{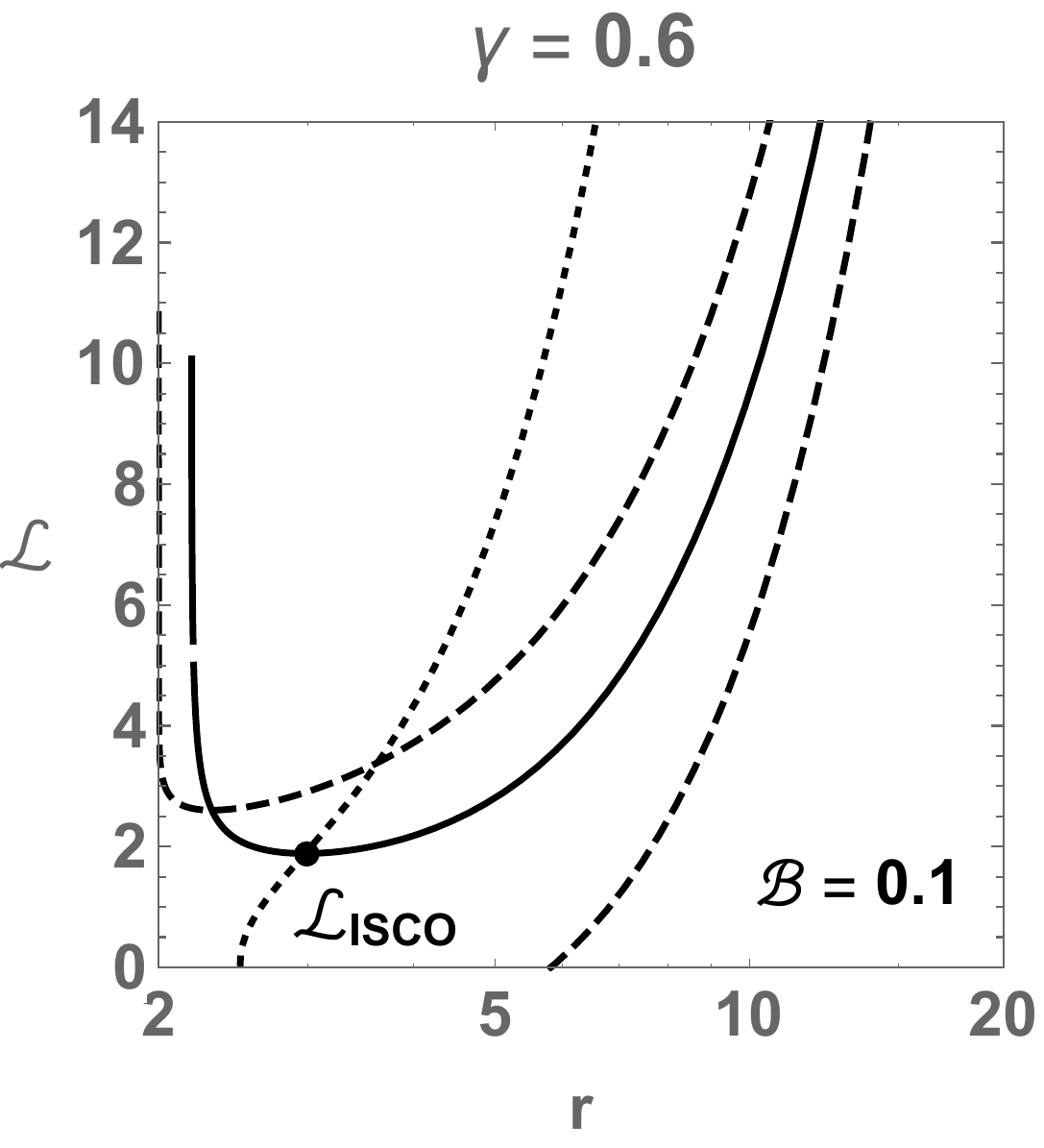}
			\hspace{0.7cm}
			\includegraphics[scale=0.37]{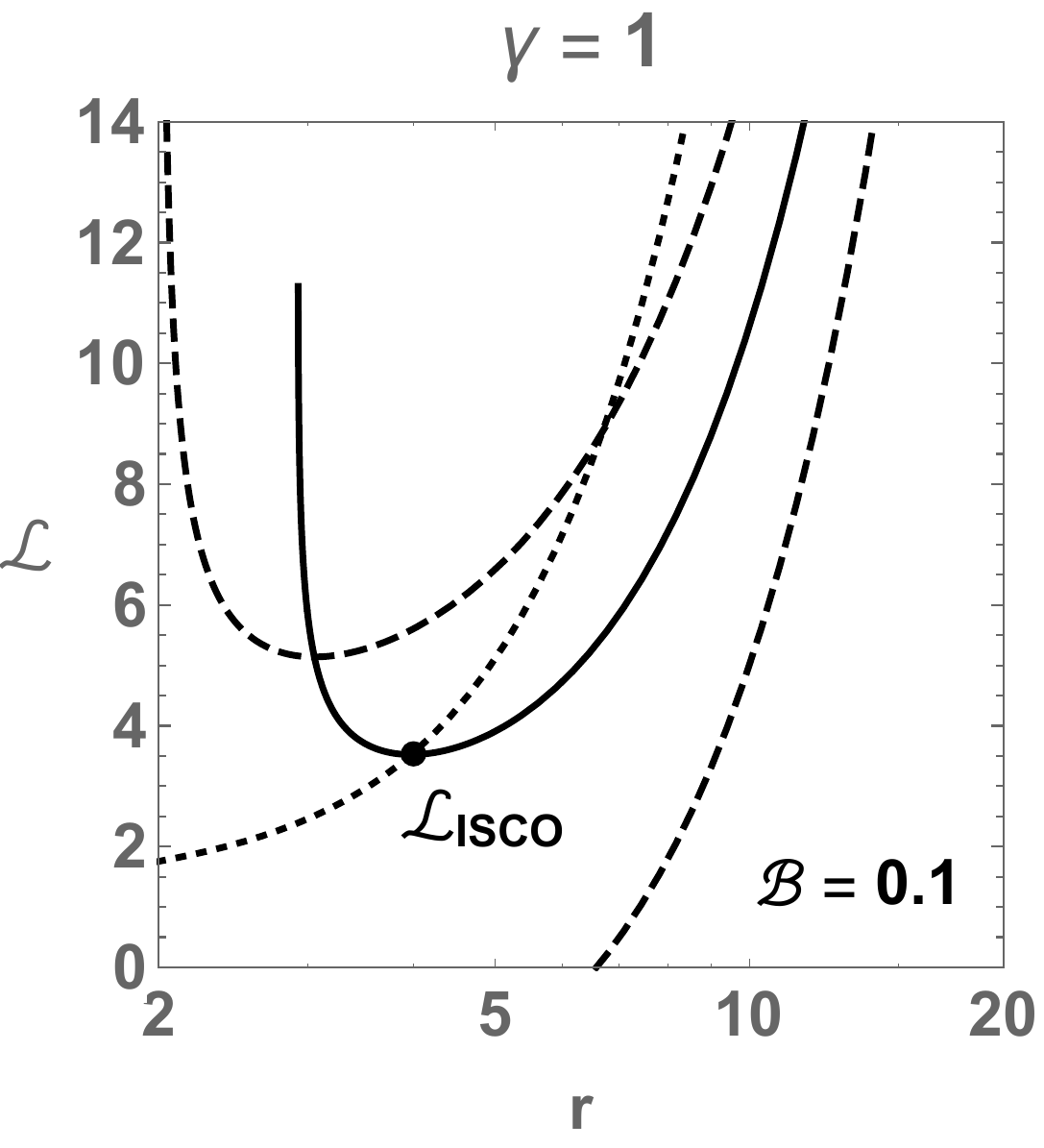}
			\hspace{0.7cm}
			\includegraphics[scale=0.37]{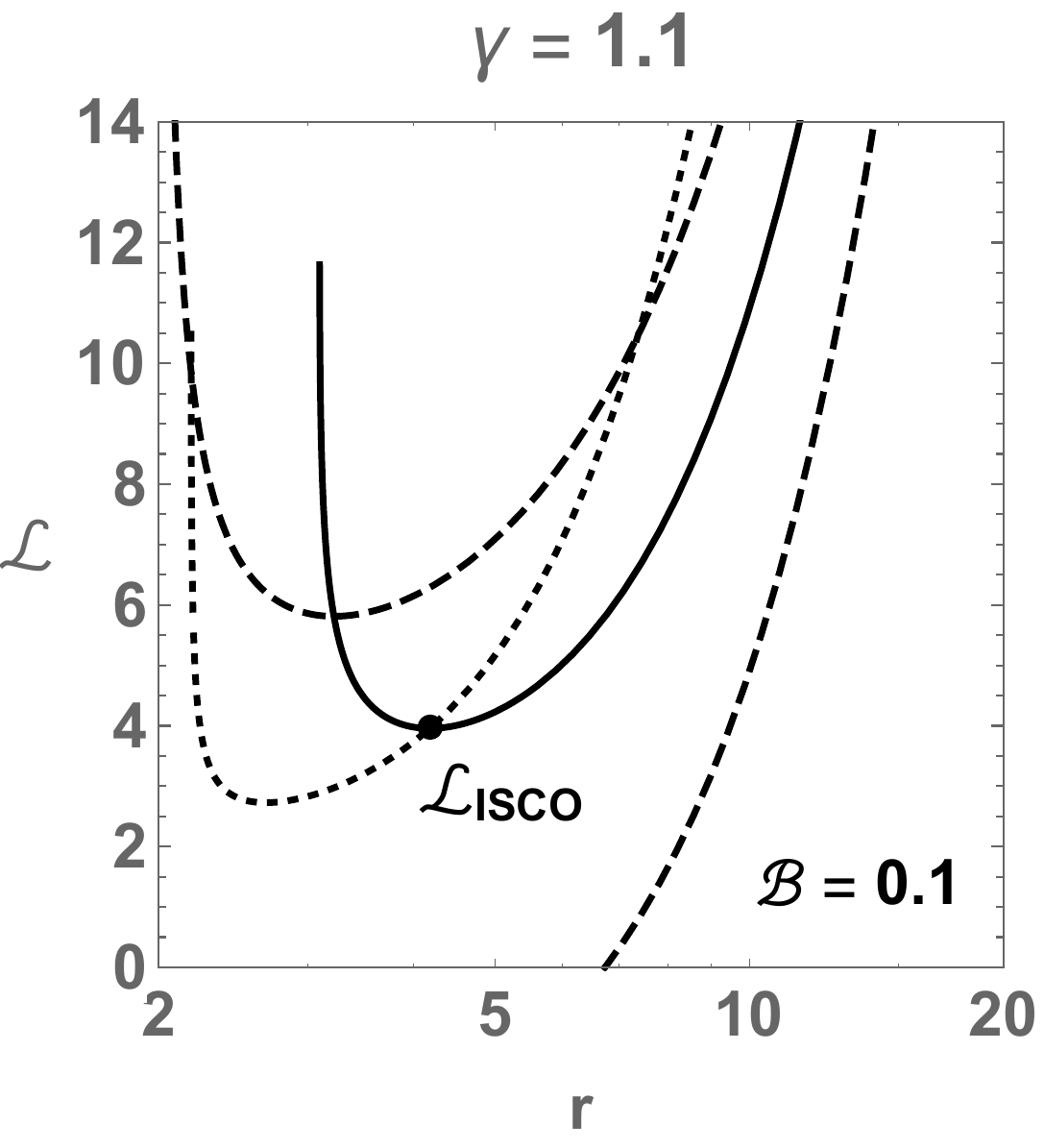}
			\caption{Plots of $\mathcal{L}_{\text{E}\pm}$ (solid line), $\mathcal{L}_{\text{E(ex)}}$ (dotted line), $\mathcal{L}_{\text{L}_\pm}$ (dashed line) as a function of $r$ for different values of $\gamma$. The intersection of $\mathcal{L}_{\text{E}\pm}$ with $\mathcal{L}_{\text{E(ex)}}$ provides location and angular momentum for a particle at the ISCO. In the figure we consider $\mathcal{B}=0$, $\mathcal{B}=\pm0.1$ and set $m=1$.\label{figureII}}
		\end{figure*}
	\end{center}
	
	In Fig.~\ref{figureIA}, we show the value of $r_\text{ISCO}$ as function of $\gamma$ for $\mathcal{B}=\pm0.1$ and compare it with the case without magnetic field. 
	In both cases, the effect of the magnetic field on the value of the ISCO is to reduce it with respect to the case $\mathcal{B}=0$. In addition, the ISCO for $\mathcal{B}=-0.1$ (i.e. the {\bf MC} configuration) is larger than the ISCO radius for $\mathcal{B}=0.1$ (i.e. the {\bf PC} configuration). Hence, the presence of a uniform magnetic field allows particles to move in stable circular orbits closer to the infinitely redshifted surface $r=2m$.
	This behavior can be observed in Fig.~\ref{figureII}, where we plot the behavior of $\mathcal{L}_{\text{E}_+}$ and $\mathcal{L}_{\text{E(ex)}}$.   
	
	Similarly to the case of Schwarzschild immersed in a uniform magnetic field, the motion of charged particles in the $\gamma$-space-time is also bounded in the radial direction near the equatorial plane due to the term proportional to $\mathcal{B}$ in the effective potential. Nevertheless, this term vanishes for $\theta=0$ and so particles can escape along the polar direction, i.e. the $z$ direction, toward infinity. 
	
	In order to find the condition on the particle's energy for the particle to escape, we consider the effective potential at $z\rightarrow\infty$ along a given direction $x$. Then $V_{\text{eff}}(x;z\rightarrow\infty)$ is given by (see gray solid line in Fig.~\ref{figureI})
	\begin{equation}
	\label{IV.15}
	V_{\text{eff}}(x;z\rightarrow\infty)=1+\frac{\mathcal{L}^2}{x^2}-2\mathcal{L}\mathcal{B}+\mathcal{B}^2x^2. 
	\end{equation} 
	The minimum of Eq.~(\ref{IV.15}) is located at $x=|\mathcal{L}/\mathcal{B}|$ and at this value of $x$ the energy is 
	\begin{equation}
	\label{IV.16}
	\mathcal{E}_{\text{escape}}=\sqrt{1+2|\mathcal{L}\mathcal{B}|-2\mathcal{L}\mathcal{B}}.
	\end{equation}  
	Therefore, the condition for particles to escape is given by 
	\begin{equation}
	\label{IV.17}
	\mathcal{E}\geq\mathcal{E}_{\text{scape}}=
	\begin{cases}
	\begin{array}{ccc}
	1 \hspace{1cm}&&\text{if}\hspace{0.5cm}\mathcal{B}\geq 0\\\\
	\sqrt{1-4\mathcal{BL}}&& \text{if}\hspace{0.5cm}\mathcal{B}<0\\
	\end{array}
	\end{cases}
	\end{equation}
	The exact same condition was found by M.~Kološ et.~al in \cite{Kolos:2015iva}. This is due to the fact that Schwarzschild and $\gamma$-metric have the same behavior at $z\rightarrow\infty$ (see Eq.~(\ref{IV.15}) and Fig.~\ref{figureI}). 
	
	On the other hand, for charged particles on bound orbits 
	the energy condition in Eq.~(\ref{IV.7a}) describes what is called a ``\textit{lake-like}'' region where the particles are trapped between two radii. The condition for such a region is given by 
	\begin{equation}
	\label{IV.18}
	\mathcal{E}<\mathcal{E}_{\text{scape}}.
	\end{equation}
	Since the specific energy $\mathcal{E}$ is related by the specific angular momentum $\mathcal{L}$, the trapped region can be expressed by the condition
	\begin{center}
		\begin{figure*}[h!]	
			\includegraphics[scale=0.4]{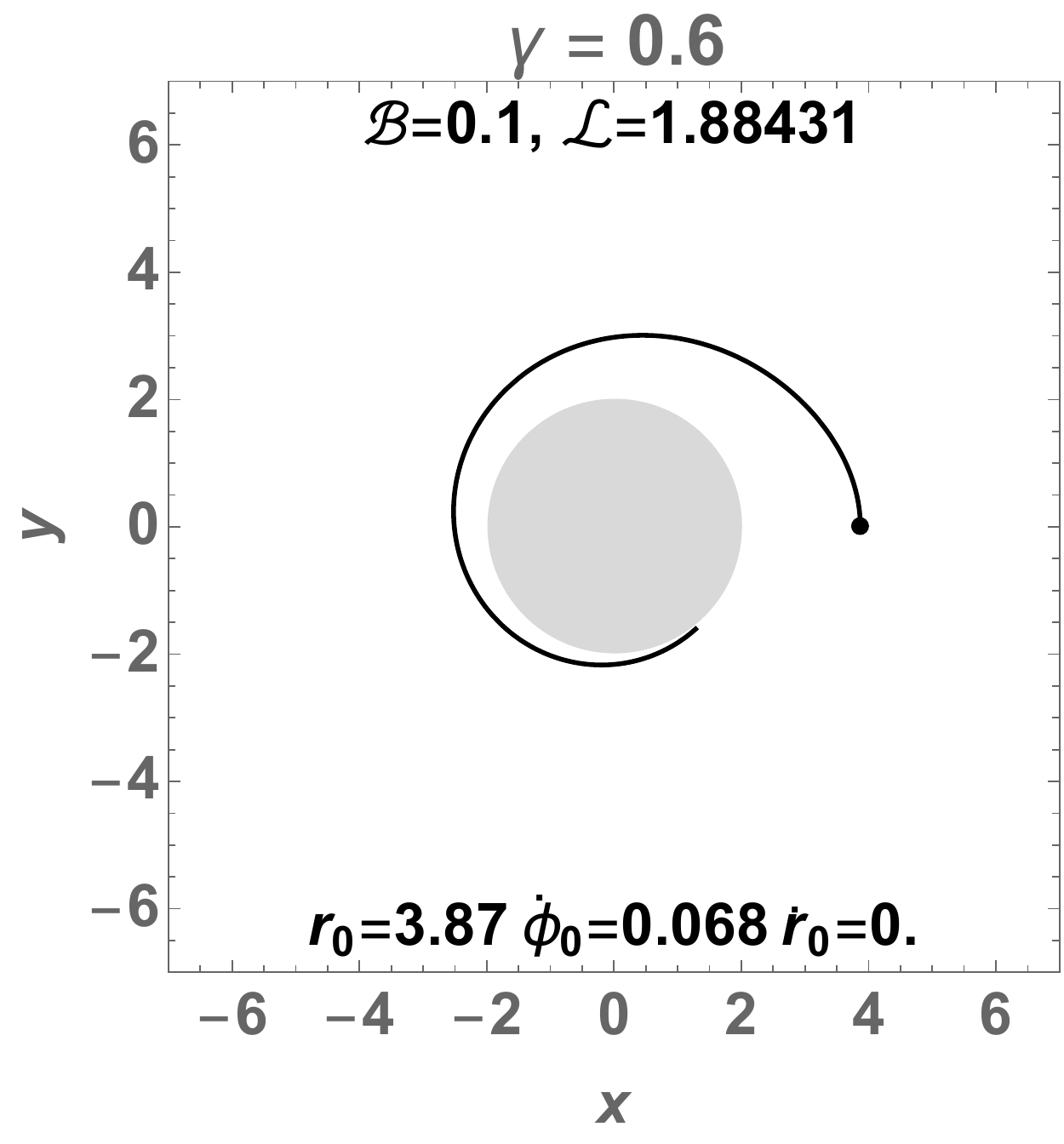}
			\hspace{0.7cm}
			\includegraphics[scale=0.4]{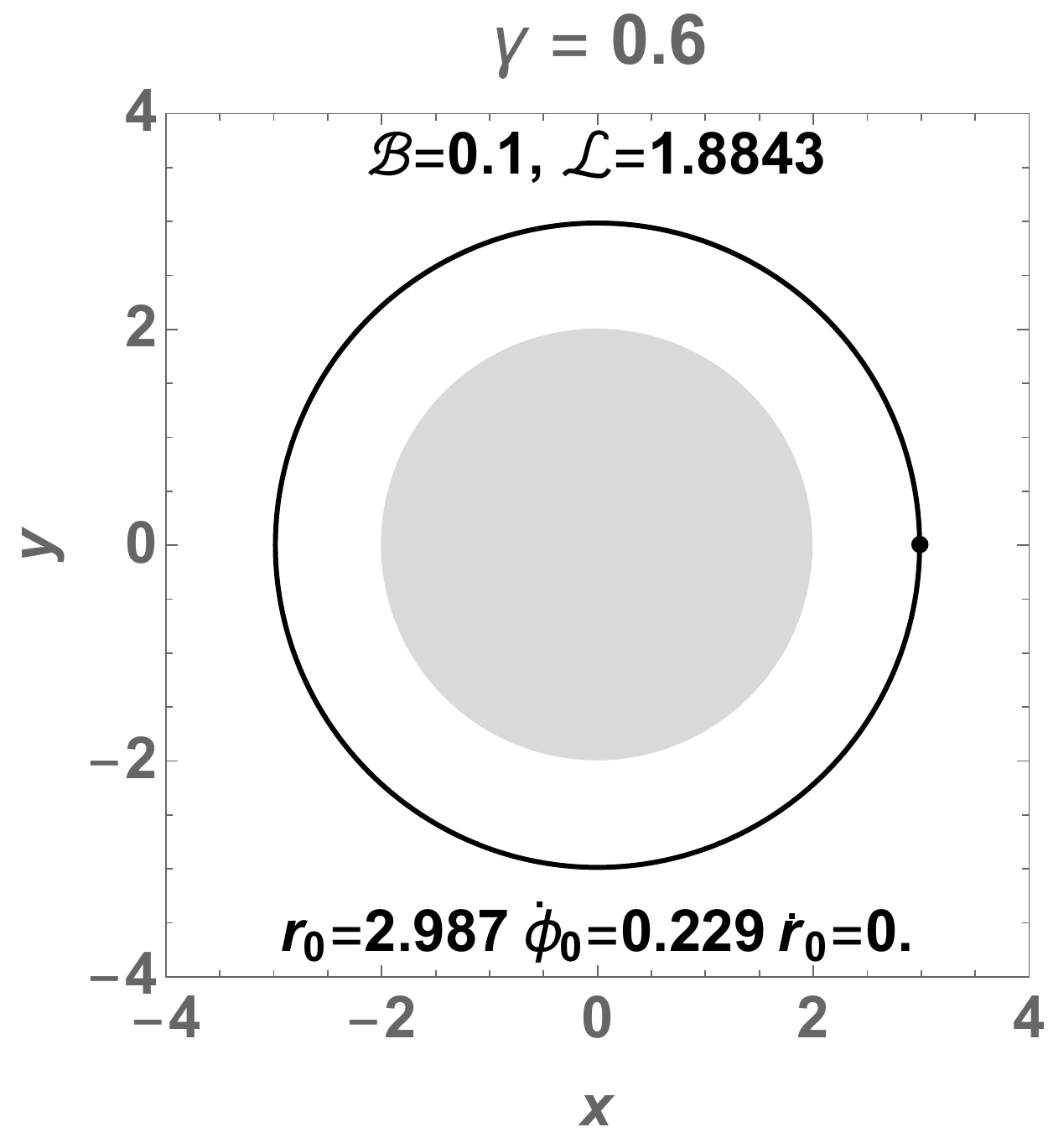}
			\hspace{0.7cm}
			\includegraphics[scale=0.4]{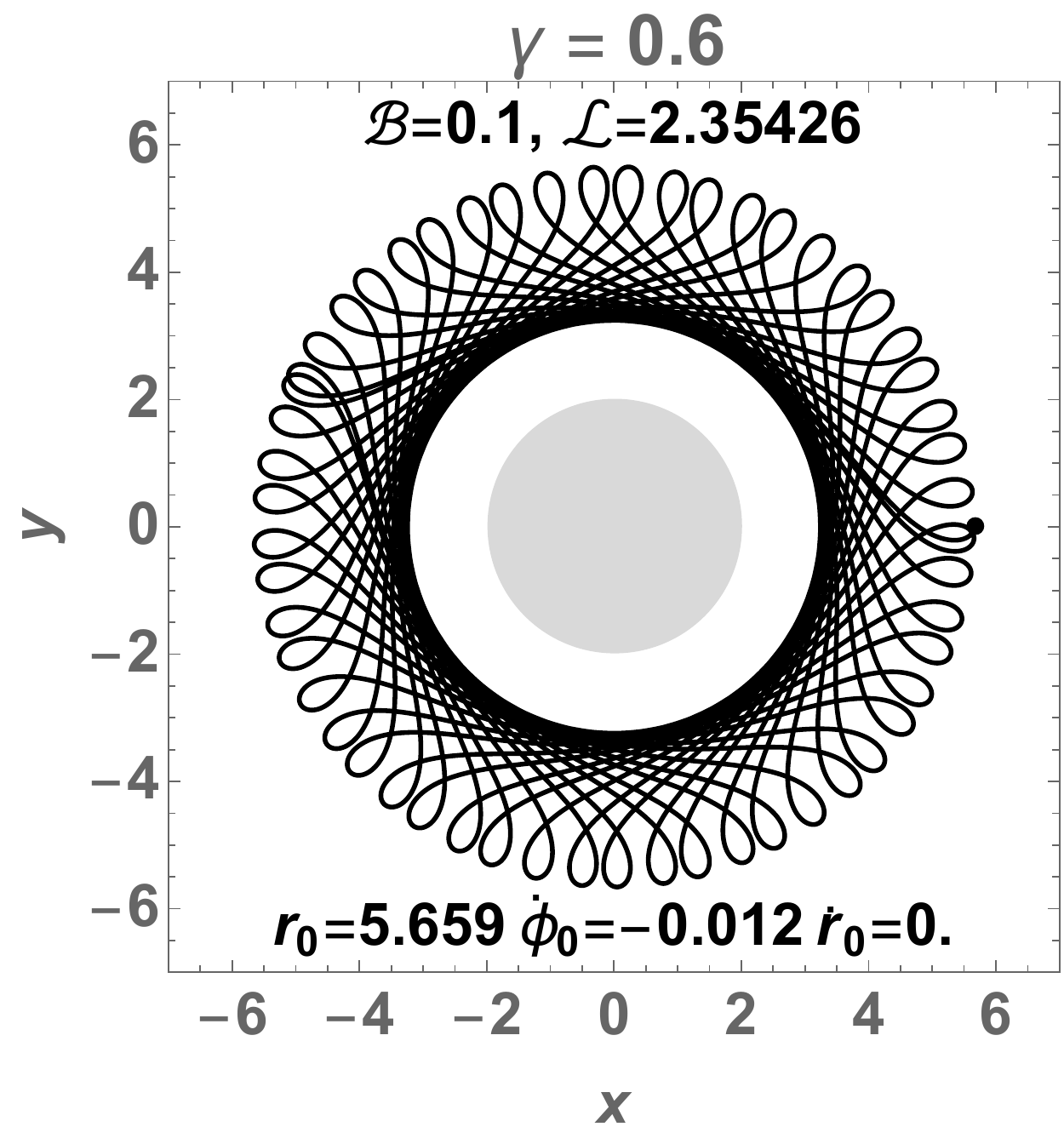}\\
			\includegraphics[scale=0.4]{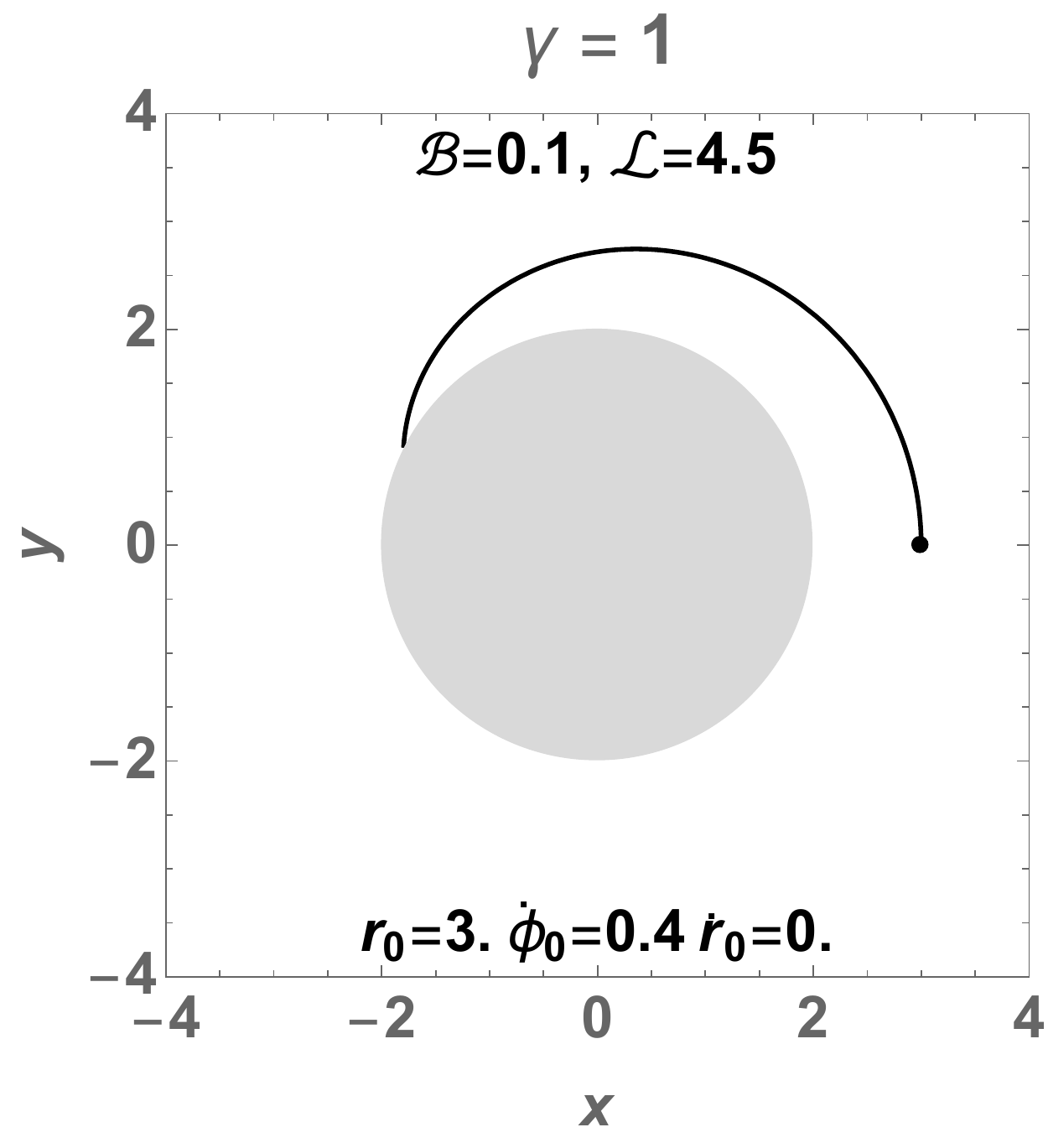}
			\hspace{0.7cm}
			\includegraphics[scale=0.4]{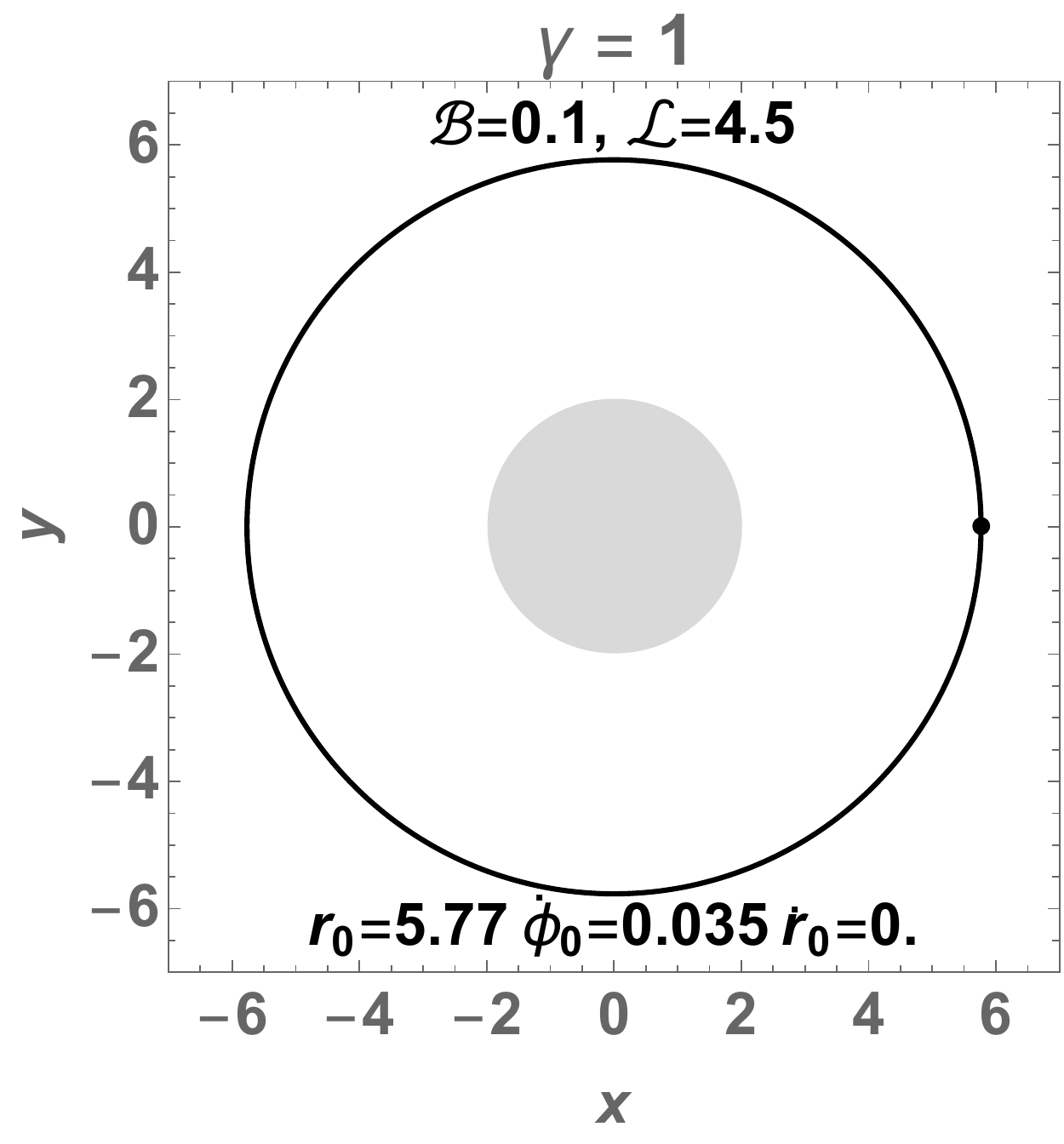}
			\hspace{0.7cm}
			\includegraphics[scale=0.43]{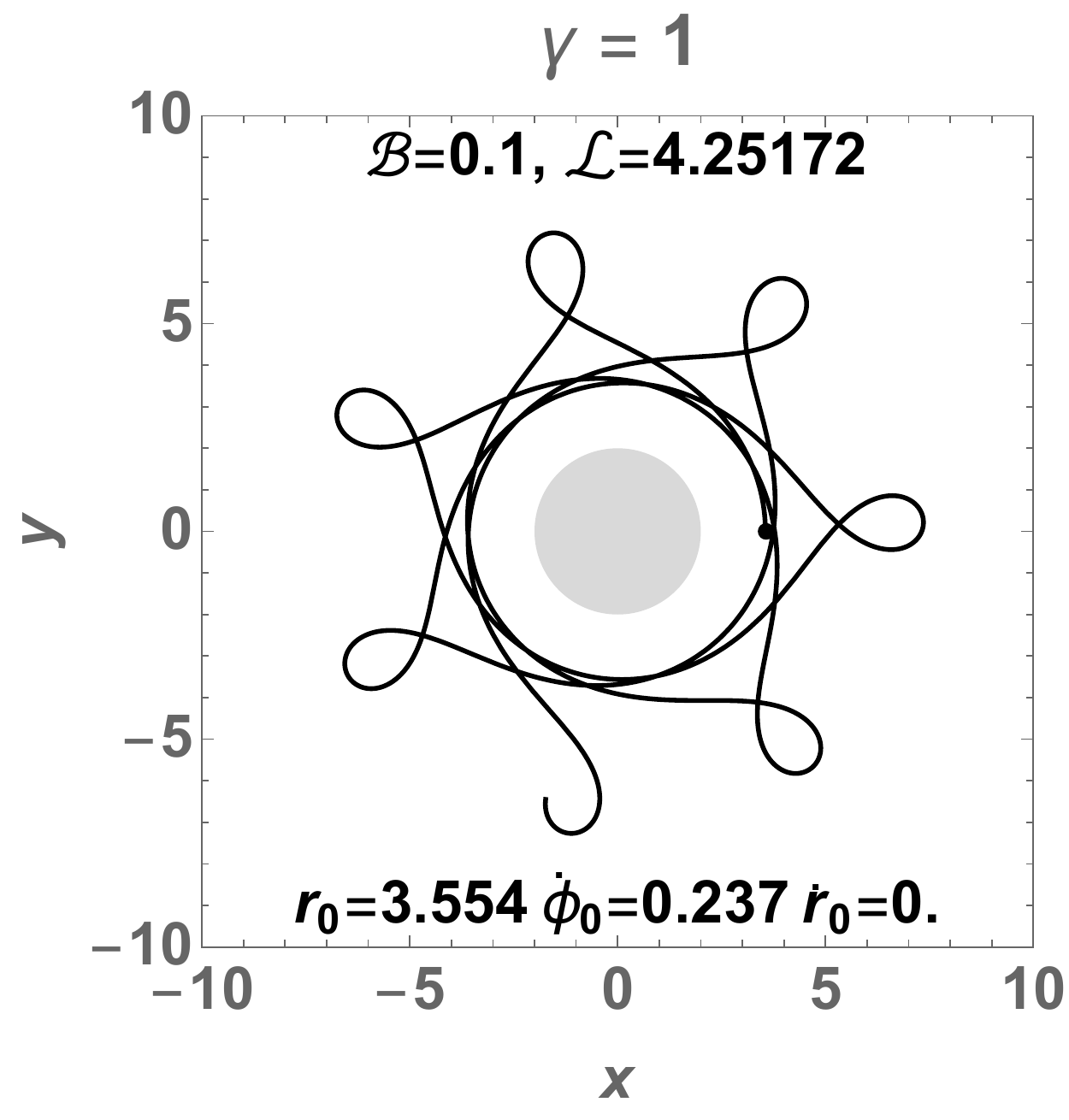}\\
			\includegraphics[scale=0.4]{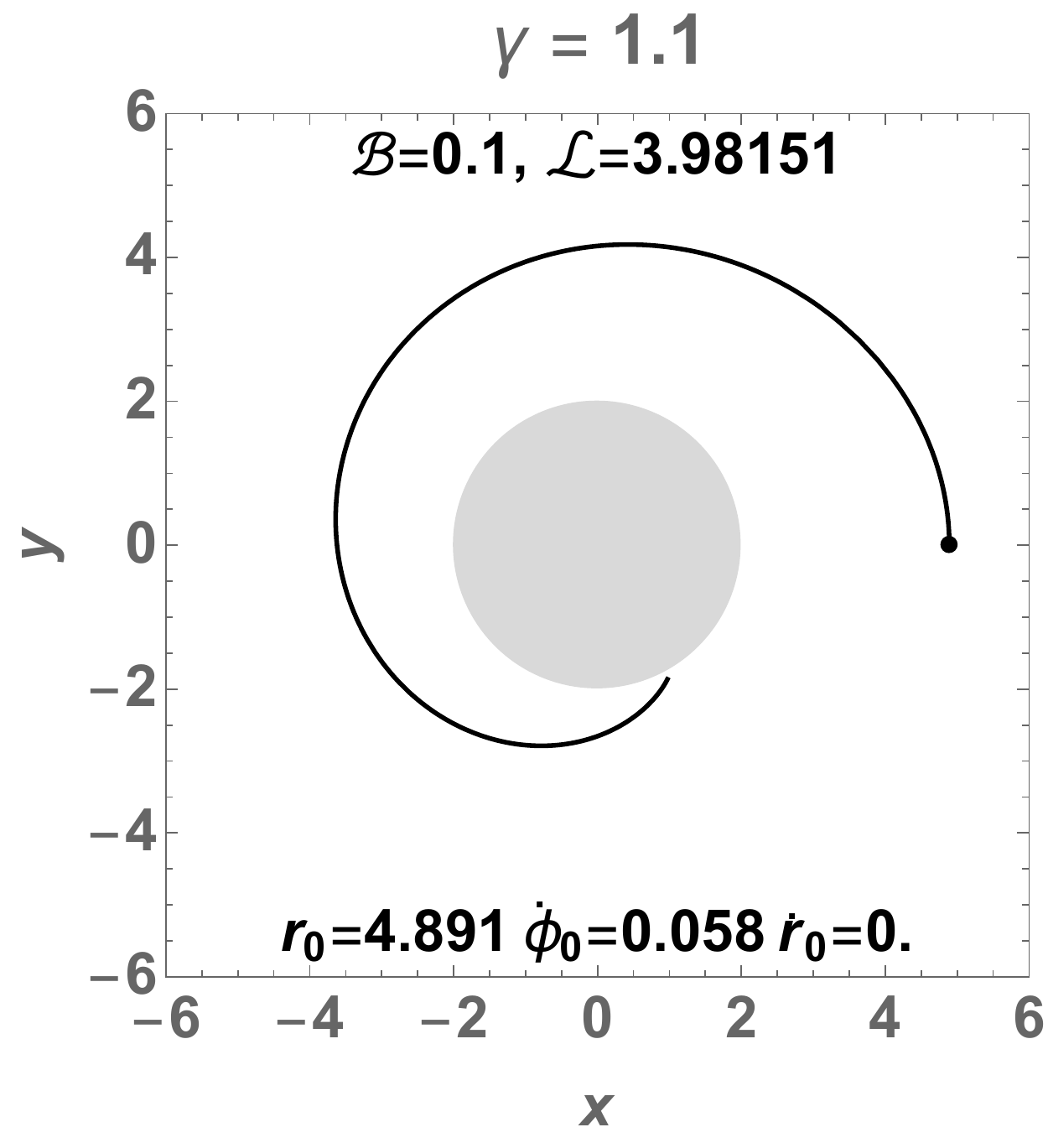}
			\hspace{0.7cm}
			\includegraphics[scale=0.4]{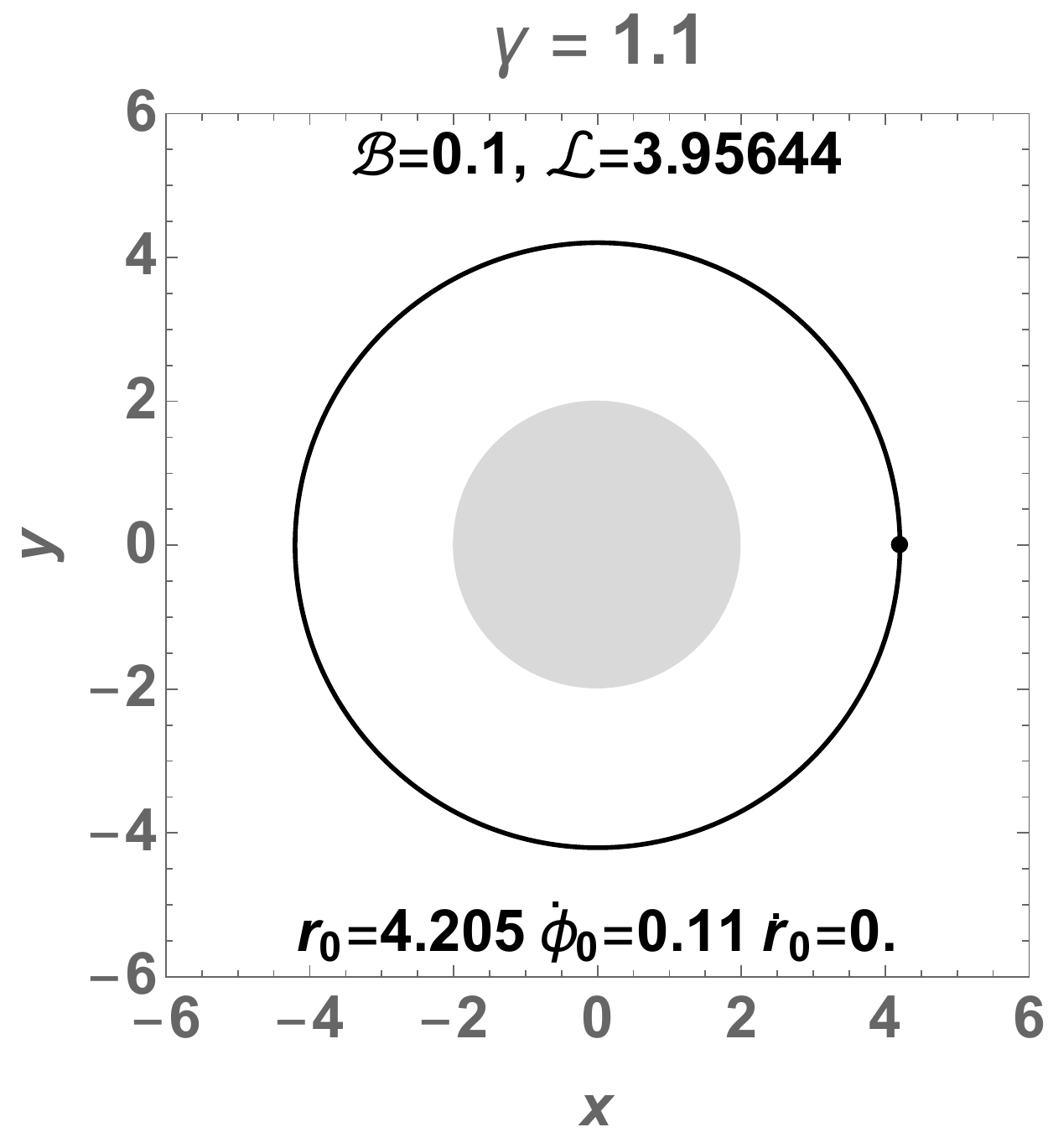}
			\hspace{0.7cm}
			\includegraphics[scale=0.4]{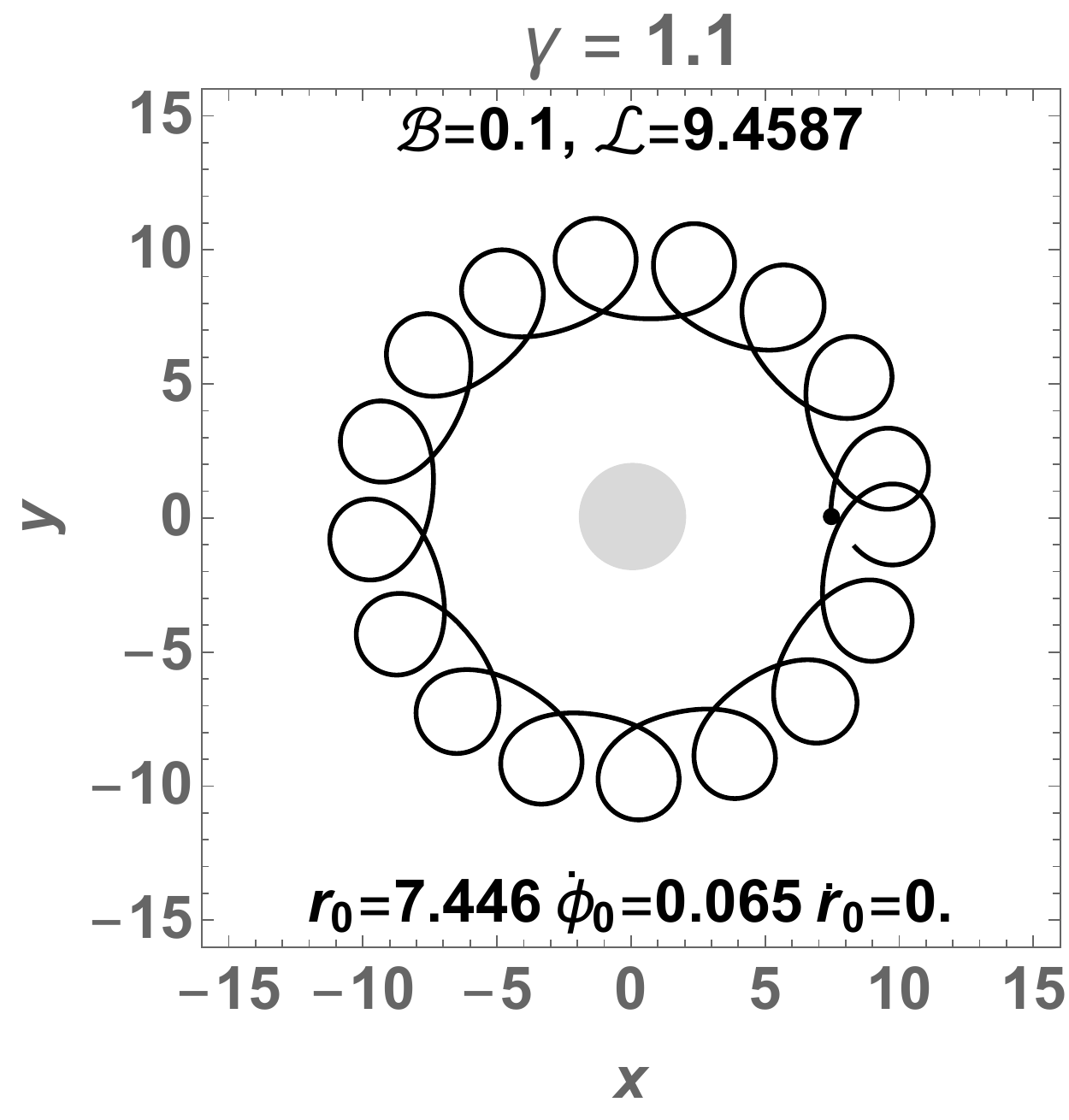}
			\caption{Charged particle motion in the equatorial plane ($\theta=\pi/2$, $\dot{\theta}=0$) in the \textbf{PC} case (i.e. $\mathcal{L}\mathcal{B}>0$) for different values of $\gamma$, $\mathcal{L}$ and $\mathcal{B}$. The motion of the particle is plotted from a given initial radius $r_0$ (the initial angle $\phi$ is irrelevant due to symmetry), radial velocity $\dot{r}_0$ and angular velocity $\dot{\phi}_0$. We set $m=1$. Notice that curly motion is allowed in this case, as opposed to the {\bf MC} case. \label{figure3}}
		\end{figure*}
	\end{center}
	\begin{center}
		\begin{figure*}[h!]	
			\includegraphics[scale=0.4]{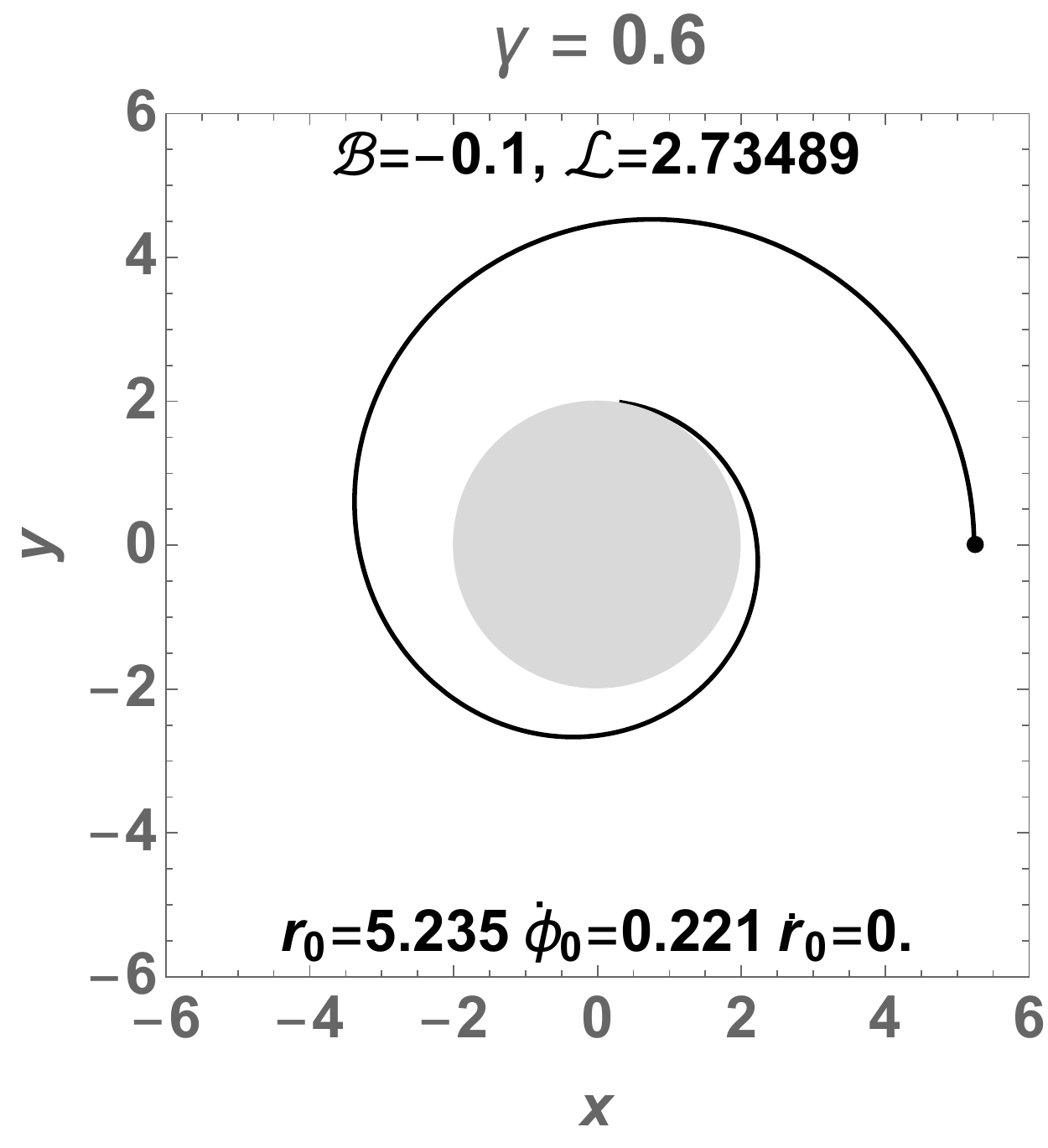}
			\hspace{0.7cm}
			\includegraphics[scale=0.4]{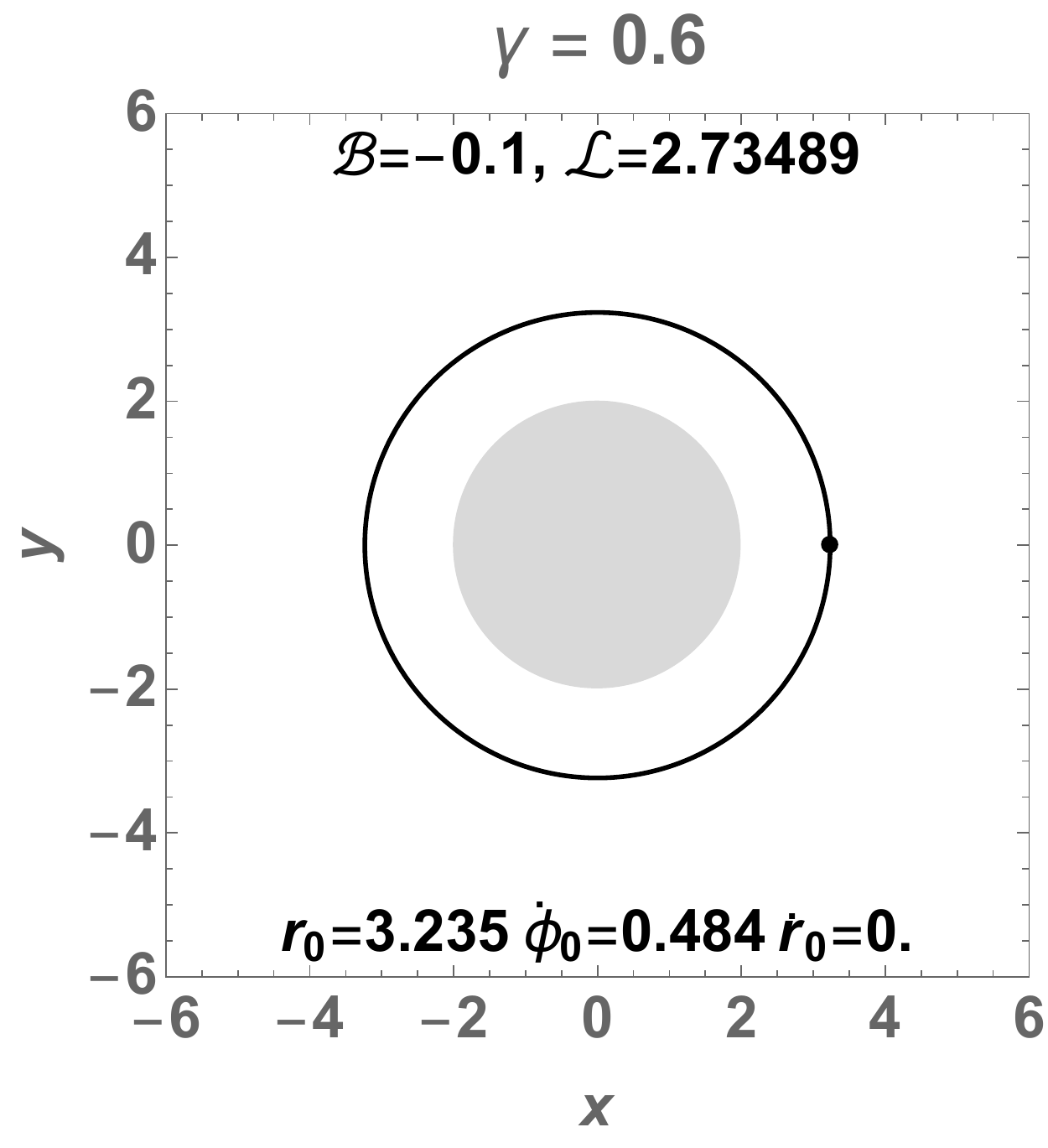}
			\hspace{0.7cm}
			\includegraphics[scale=0.4]{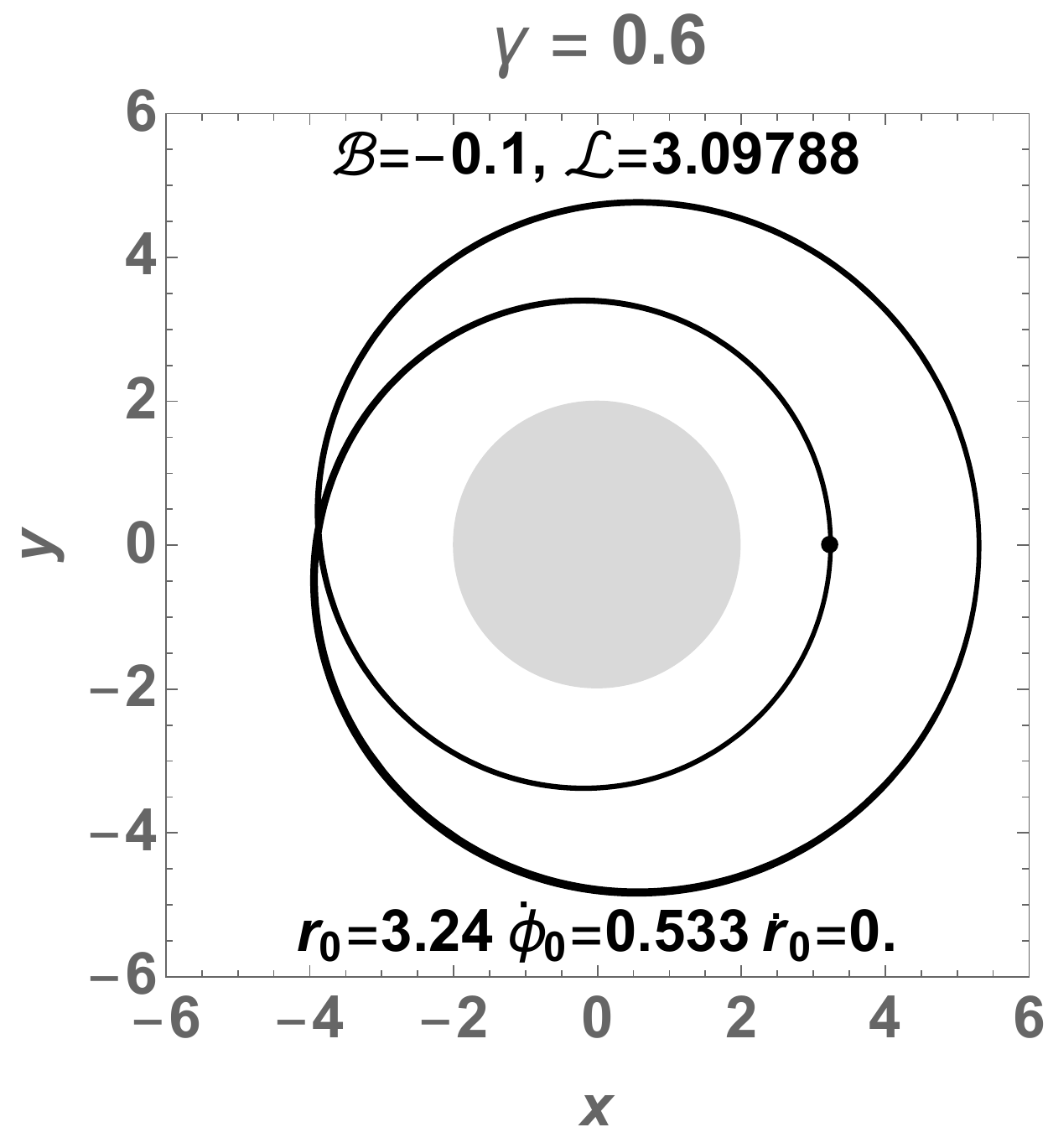}\\
			\includegraphics[scale=0.4]{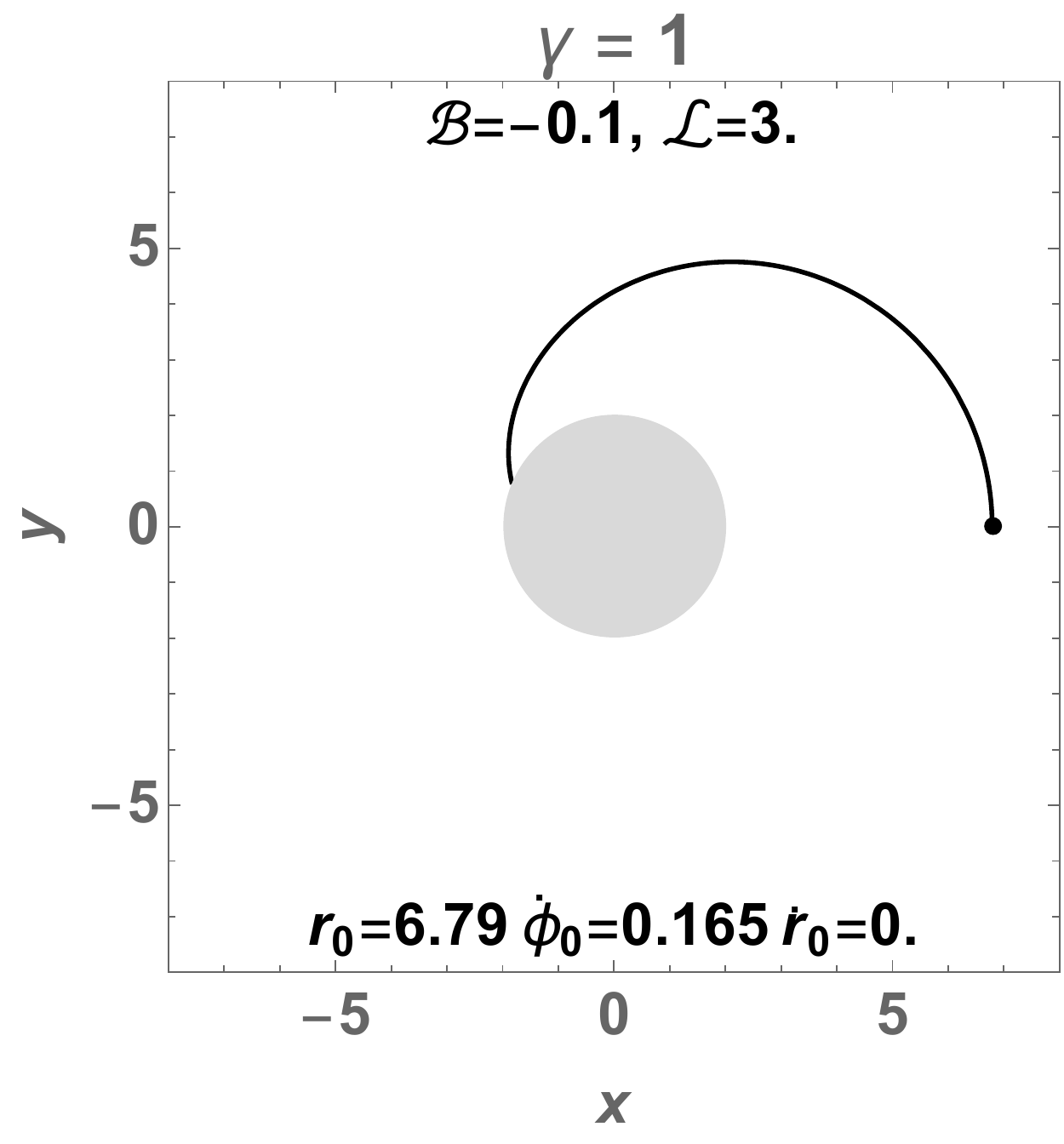}
			\hspace{0.7cm}
			\includegraphics[scale=0.4]{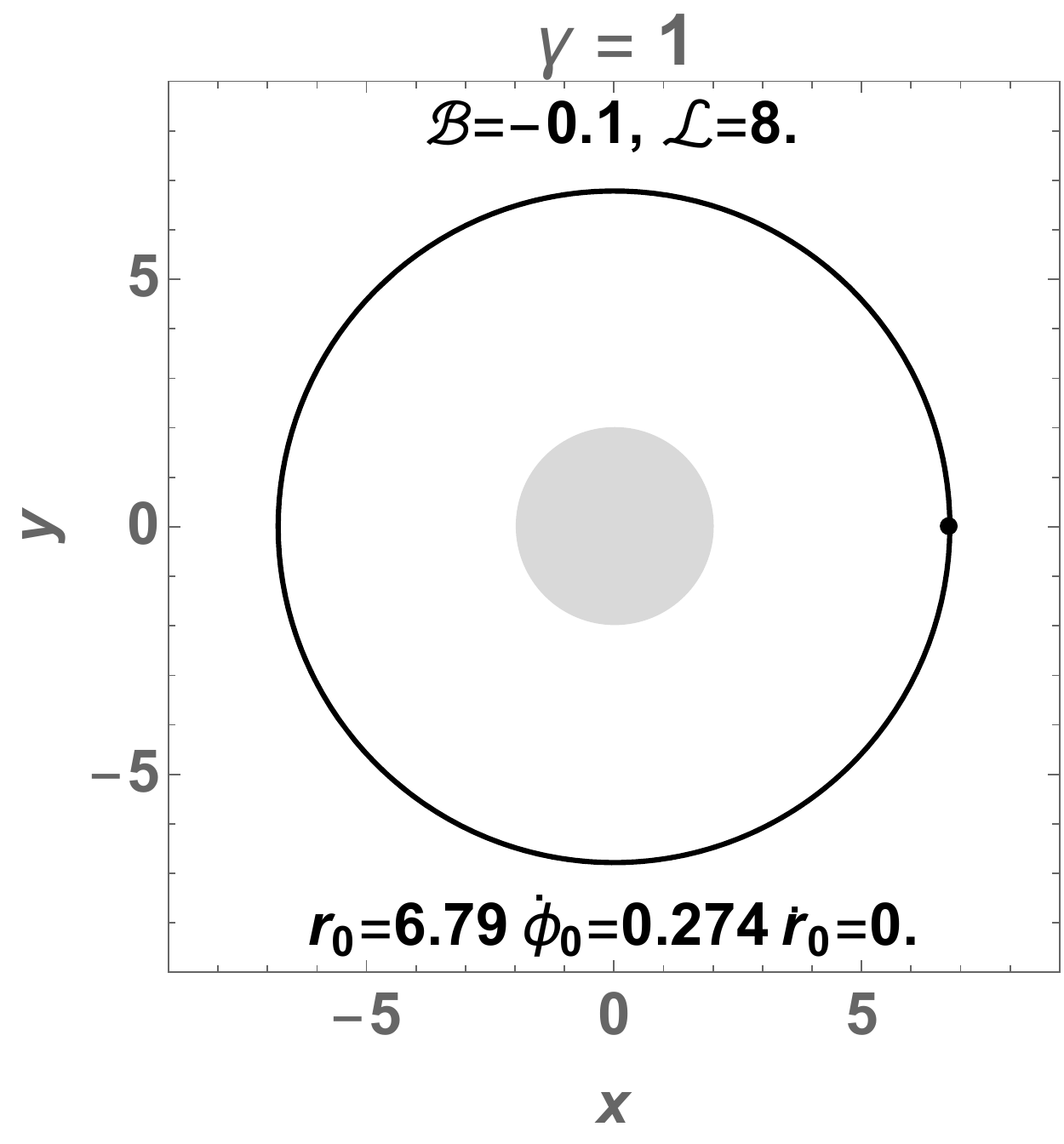}
			\hspace{0.7cm}
			\includegraphics[scale=0.4]{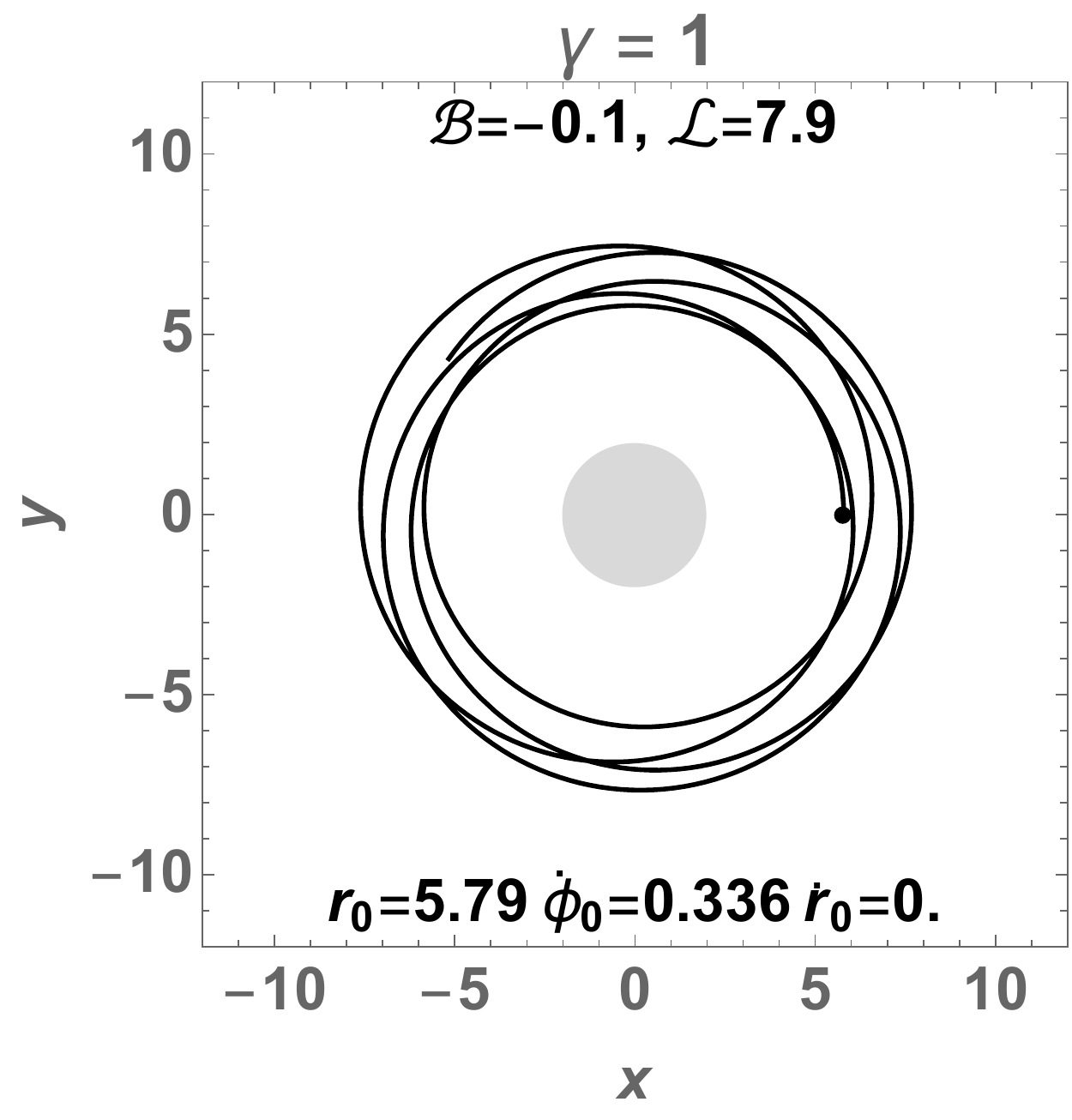}\\
			\includegraphics[scale=0.4]{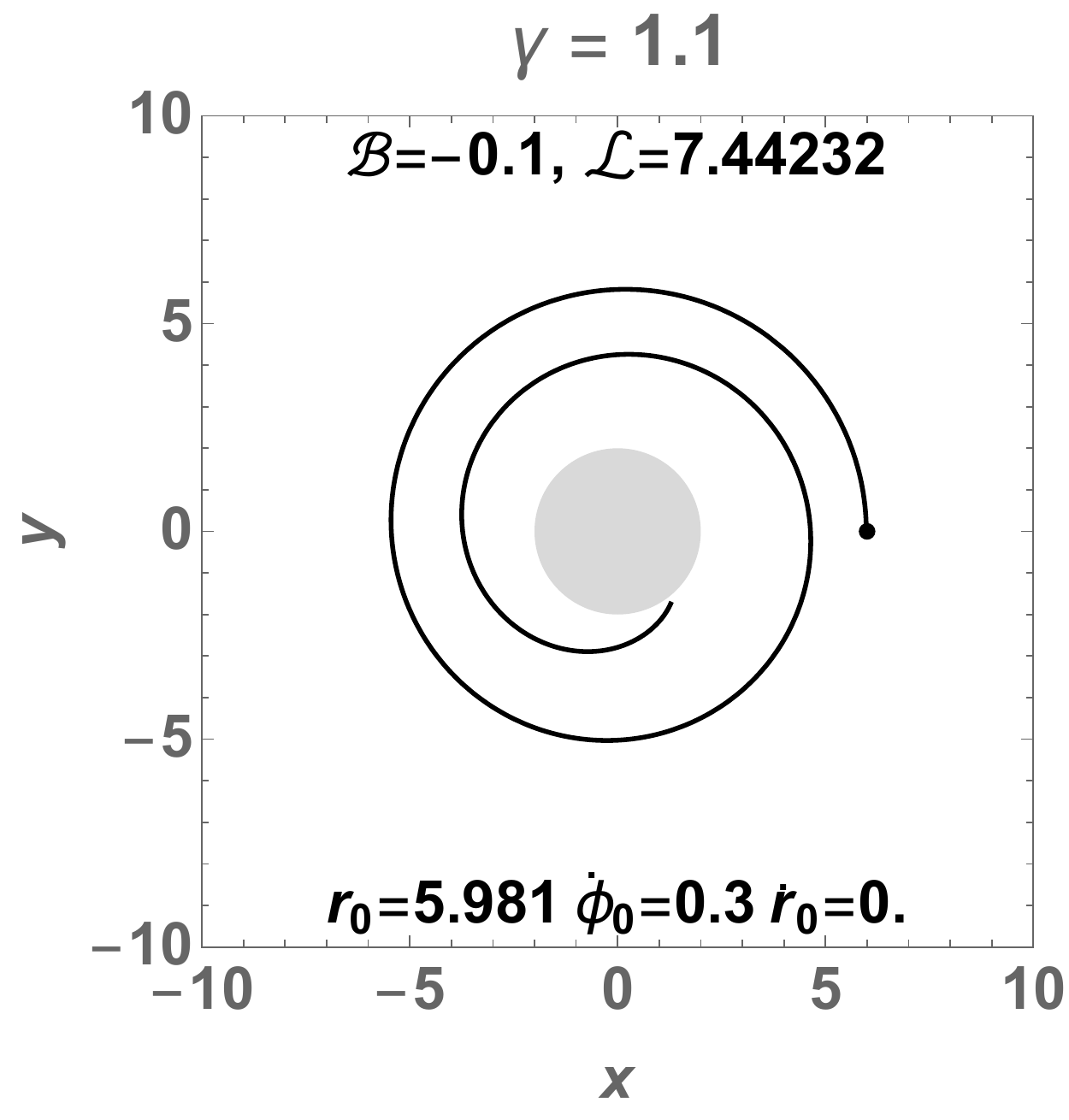}
			\hspace{0.7cm}
			\includegraphics[scale=0.4]{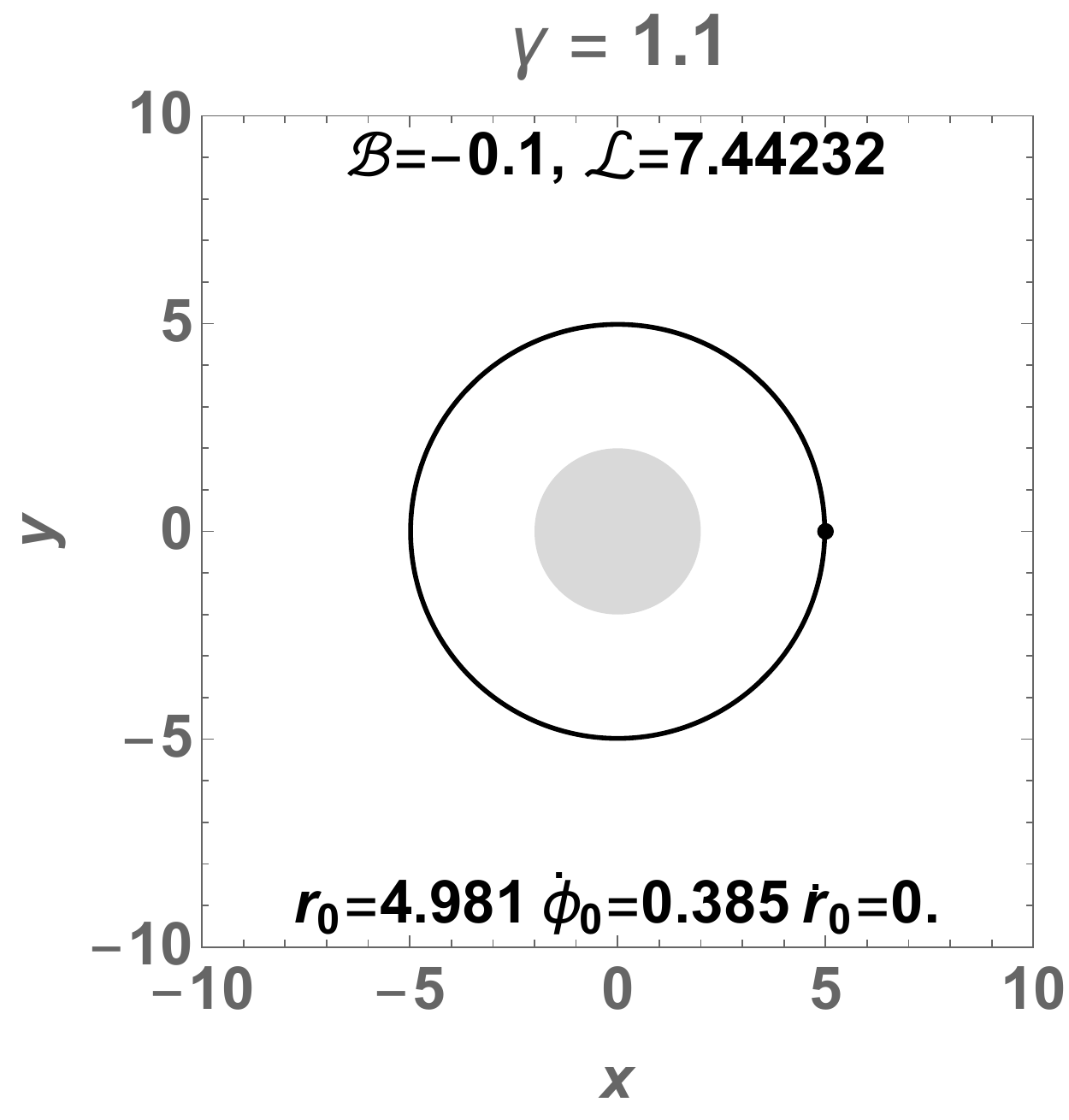}
			\hspace{0.7cm}
			\includegraphics[scale=0.4]{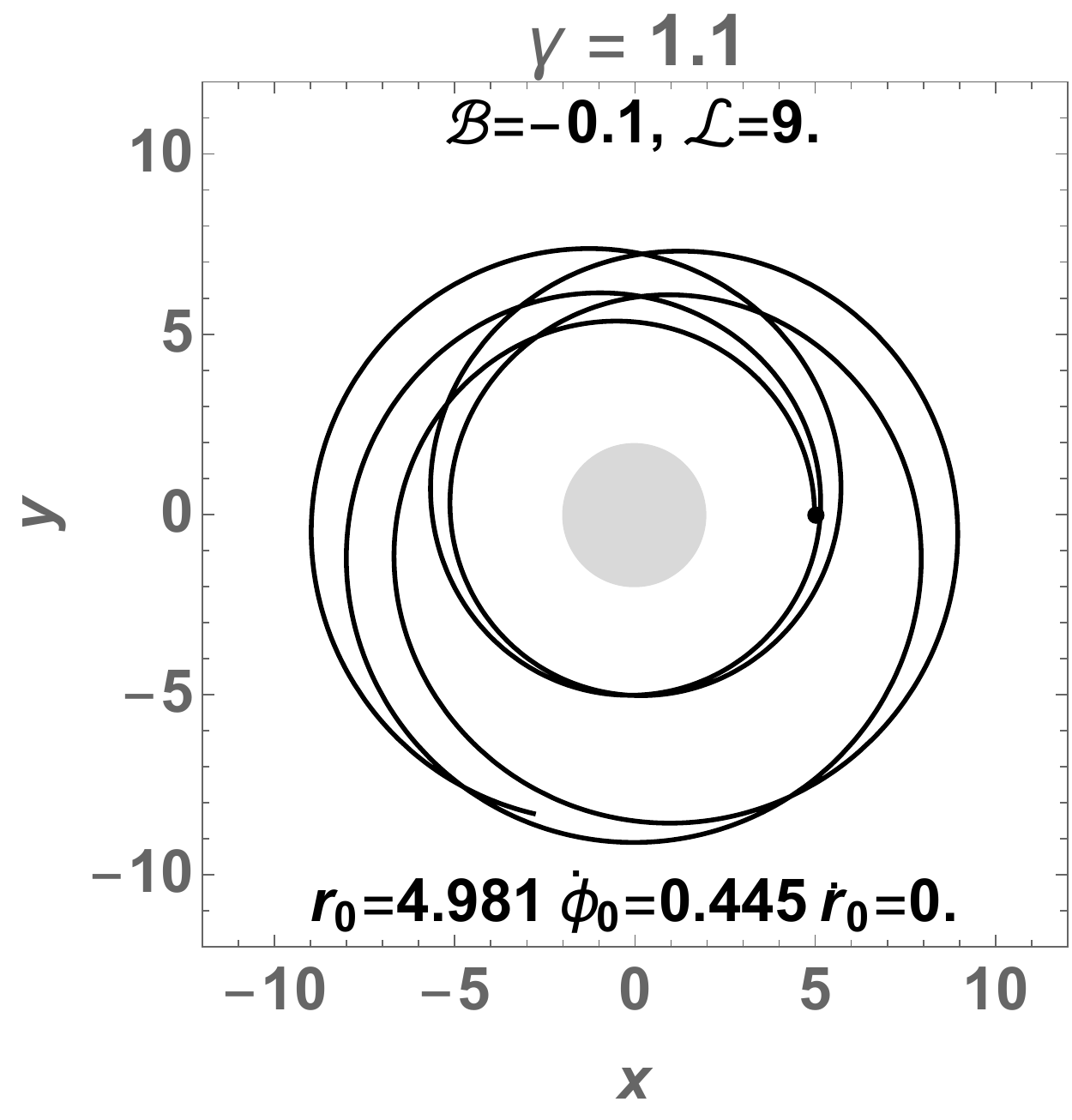}
			\caption{Charged particle motion in the equatorial plane ($\theta=\pi/2$, $\dot{\theta}=0$) in the \textbf{MC} for different values of $\gamma$, $\mathcal{L}$ and $\mathcal{B}$. The motion of the particle is plotted from a given initial radius $r_0$ (the initial angle $\phi$ is irrelevant due to symmetry), radial velocity $\dot{r}_0$ and angular velocity $\dot{\phi}_0$. We set $m=1$,
				\label{figure3a}}
		\end{figure*}
	\end{center}
	\begin{equation}
	\label{IV.19}
	\mathcal{L}_{\text{L}_-}<\mathcal{L}<\mathcal{L}_{\text{L}_+},
	\end{equation}
	where the values of $\mathcal{L}_{\text{L}_-}$ and $\mathcal{L}_{\text{L}_+}$ are given by the condition $\mathcal{E}=\mathcal{E}_{\text{escape}}$. Therefore, for $\mathcal{B}\geq 0$ we have
	\begin{widetext}
	\begin{equation}
	\label{IV.20}
	\mathcal{L}_{\text{L}\pm}=\frac{(r-2m) \mathcal{B} r^{\gamma +1}\pm r^{\gamma }\sqrt{(r-2m) r^{1-\gamma} \left(r^{\gamma }-(r-2m)^{\gamma }\right)} }{(r-2m)^{\gamma }}\;,
	\end{equation}	
	and  for $\mathcal{B}<0$, $\mathcal{L}_{\text{L}_-}$ and $\mathcal{L}_{\text{L}_+}$ become 
		\begin{equation}
		\label{IV.21}
		\mathcal{L}_{\text{L}_\pm}=\frac{-\mathcal{B} r^{\gamma +1} \left(2 r^{\gamma } (r-2m)^{1-\gamma }-(r-2m)\right)\pm r^{2 \gamma -1} \mathcal{K}(r,\mathcal{B})}{(r-2m)^{\gamma }}\;,
		\end{equation}	
		with
		\begin{equation}
		\label{IV.21A}
		\mathcal{K}(r,\mathcal{B})=\sqrt{r^{3 (1-\gamma )} (r-2m)^{1-2 \gamma } \left(r^{\gamma }-(r-2m)^{\gamma }\right) \left(4 (r-2m) \mathcal{B}^2 r^{2 \gamma +1}+(r-2m)^{2 \gamma }\right)}\;.
		\end{equation}
	\end{widetext}
	\begin{center}
		\begin{figure*}[t]	
			\includegraphics[scale=0.4]{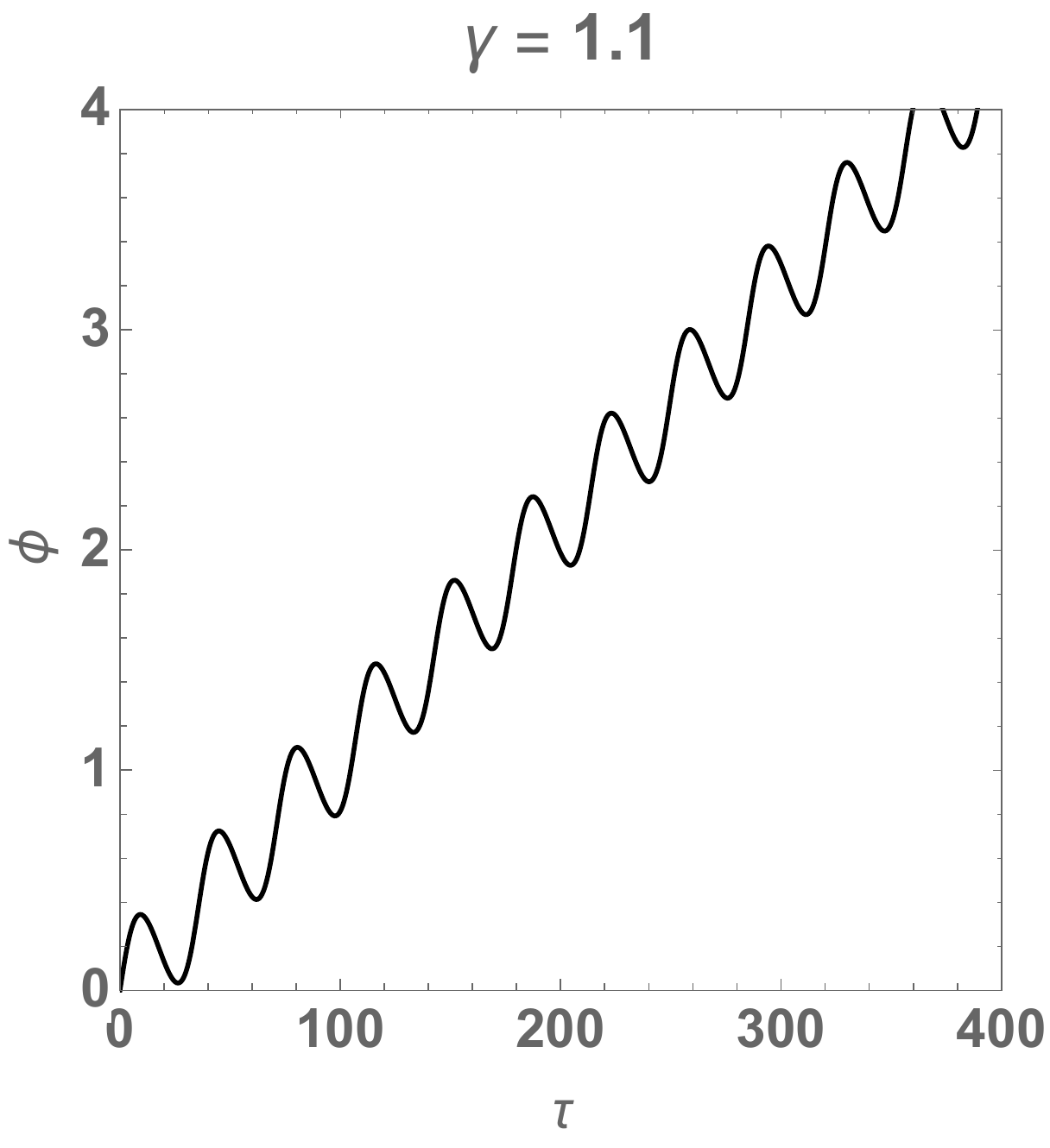}
			\hspace{0.7cm}
			\includegraphics[scale=0.42]{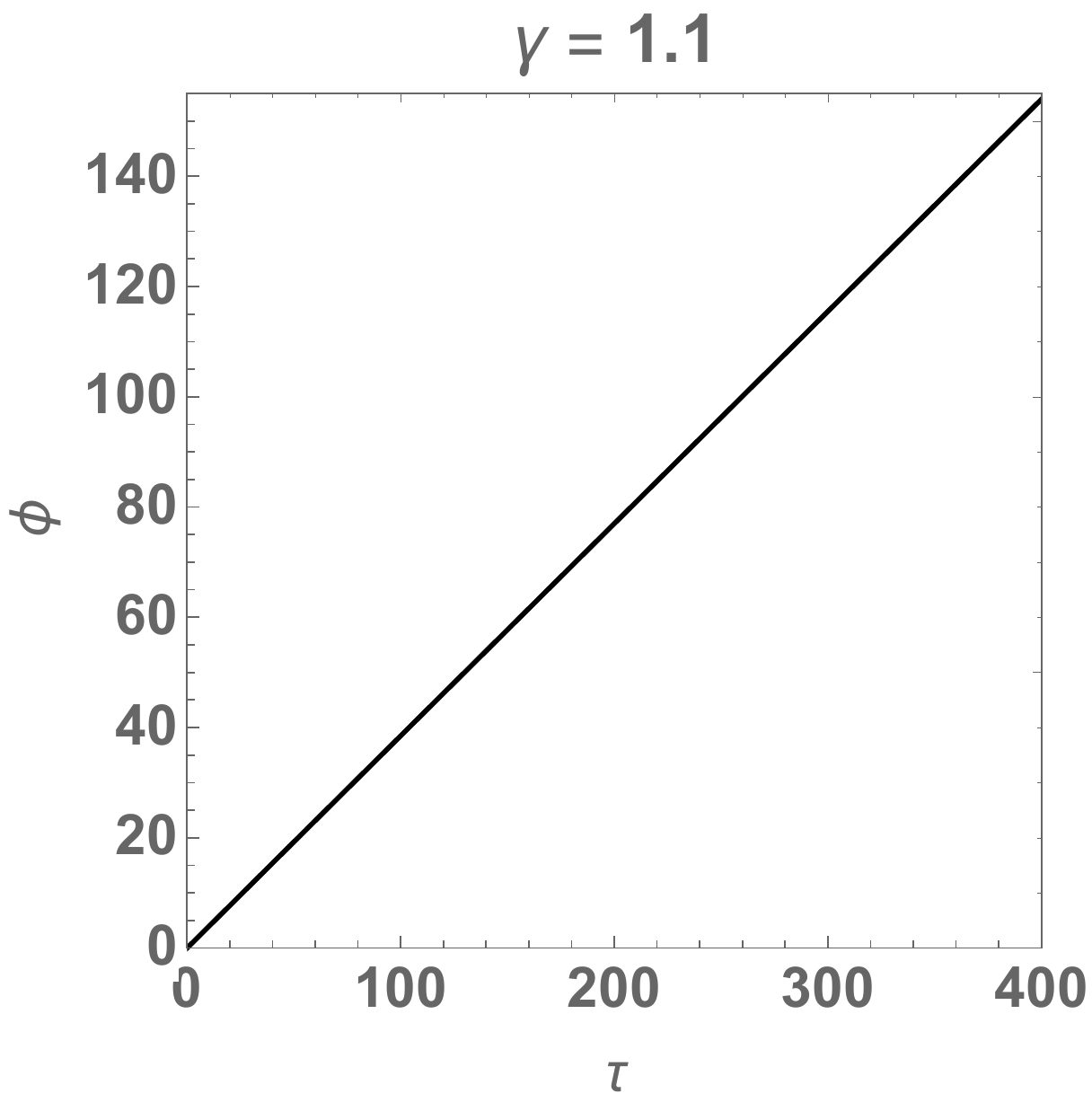}
			\hspace{0.7cm}
			\includegraphics[scale=0.42]{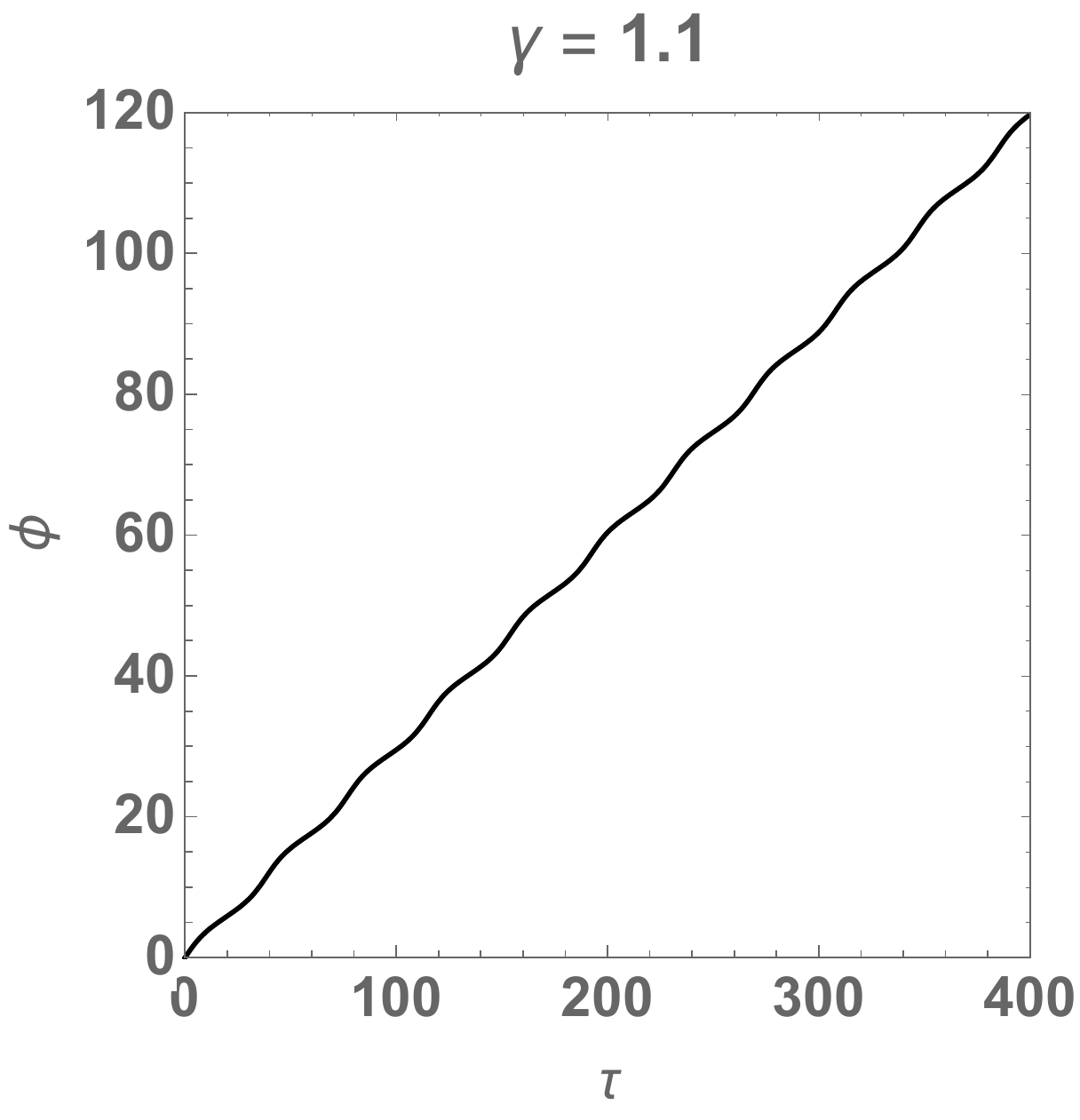}
				\caption{Plots of $\phi$ coordinate for curled (left panel), circular (center) and non-curled orbits (right panel). The plots are obtained from the data of the curled trajectory shown in the third column, last row of Fig.~\ref{figure3} (left panel), and the orbits shown in the second and third columns in the last row of Fig.~\ref{figure3a} (middle and right panels), respectively. We set $m=1$.\label{figure3b}}
			\end{figure*}
		\end{center}
		\begin{center}
		\begin{figure*}[t]
			\includegraphics[scale=0.35]{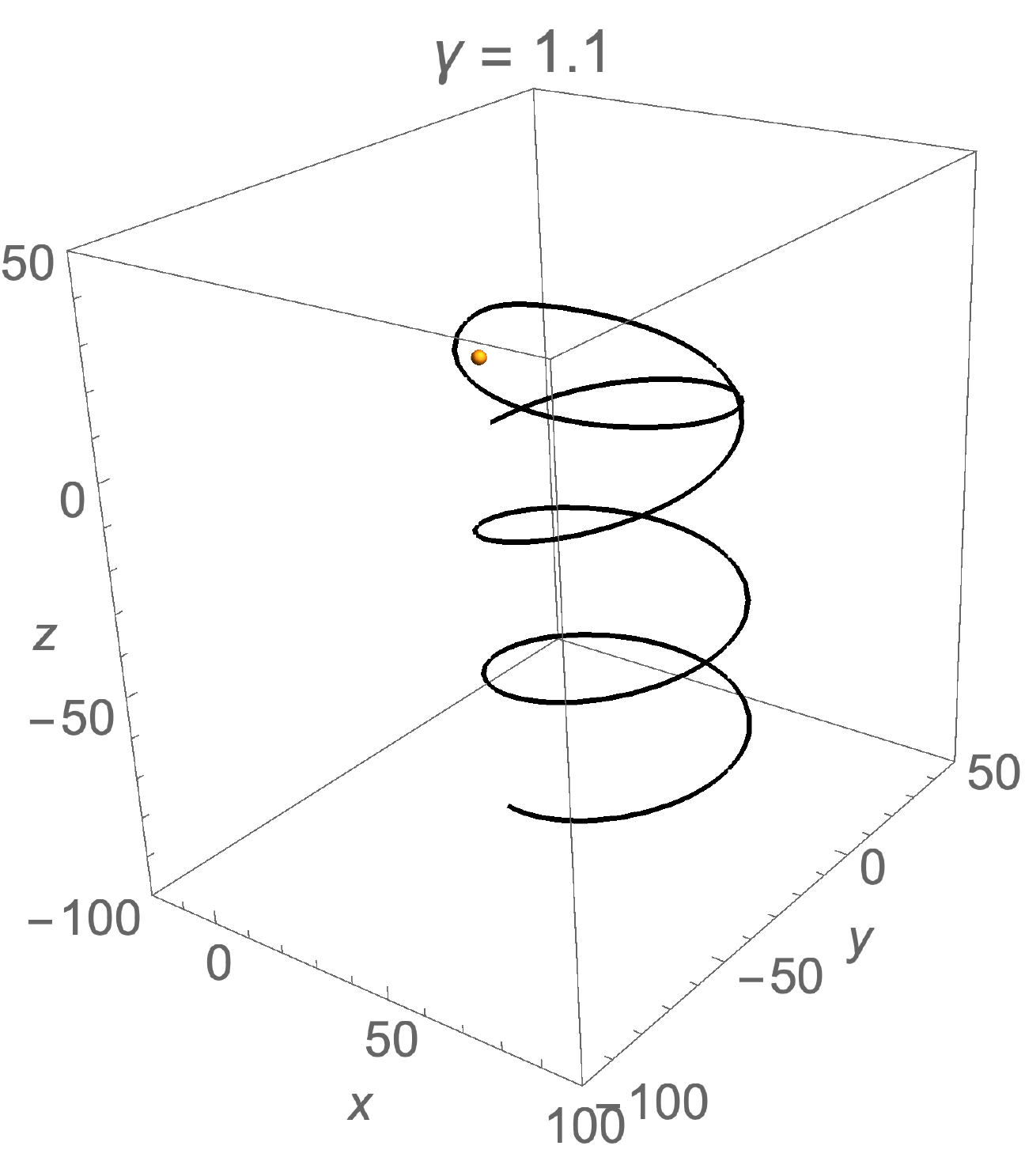}
			\hspace{0.7cm}
			\includegraphics[scale=0.35]{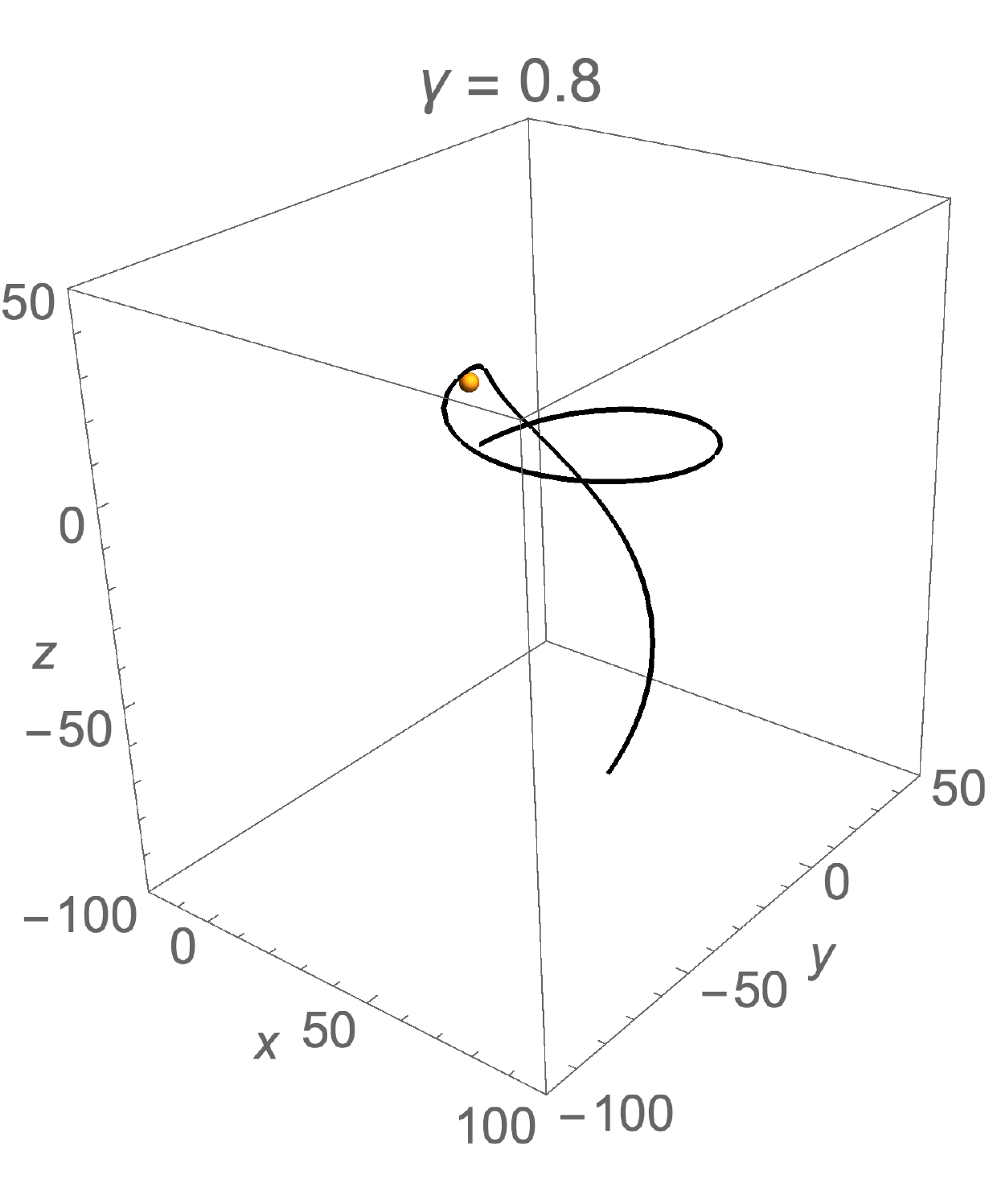}
			\hspace{0.7cm}
			\includegraphics[scale=0.45]{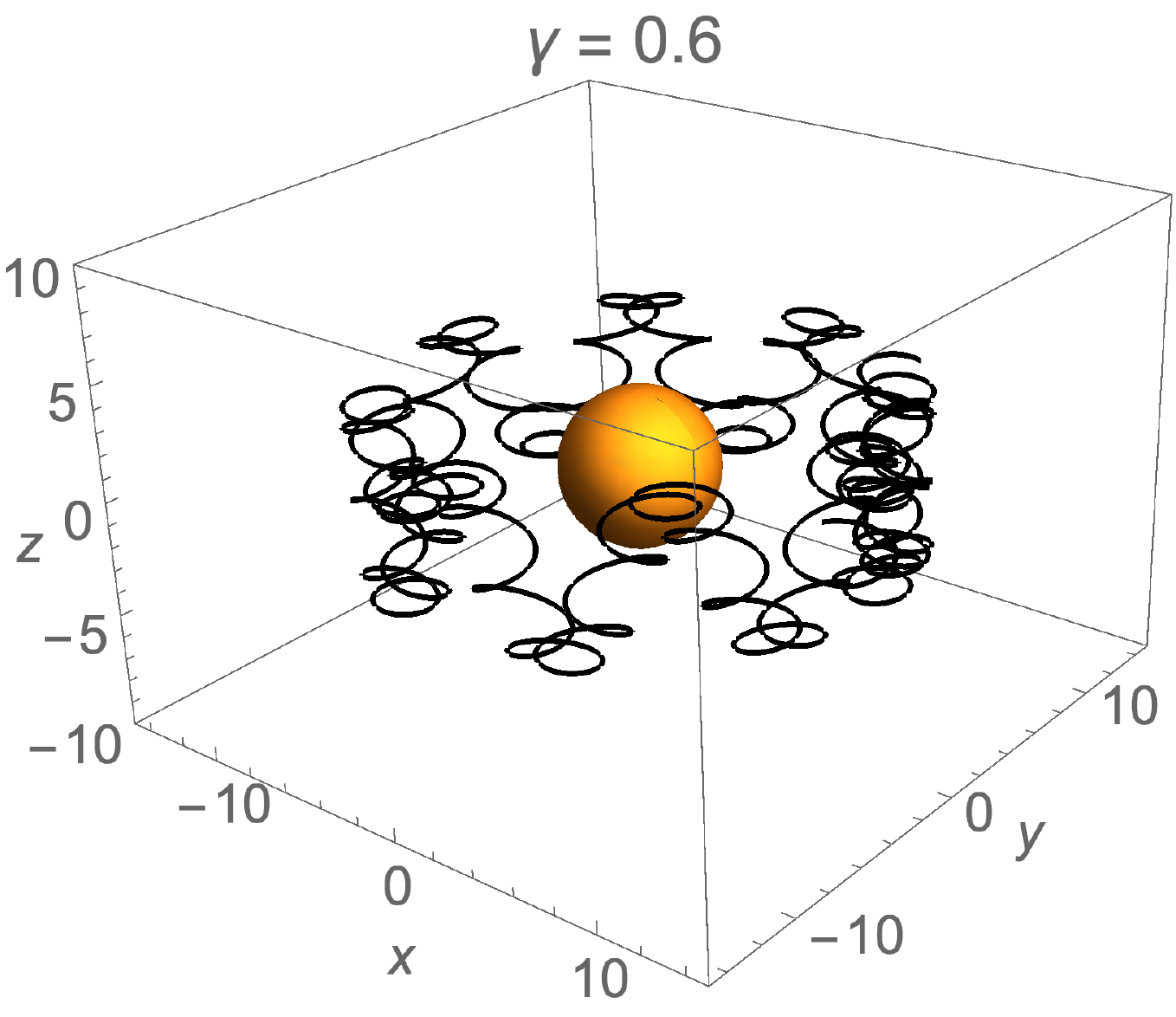}
			\caption{Left panel: example of a escape trajectory. The initial conditions are $r_0=16$, $\dot{r}_0=0$, $\theta_0=3$, $\phi_0=0$, and $\dot{\phi}_0=0$ with $\mathcal{B}=0.1$ and $\mathcal{L}=7.1$.  Center: another example of a escaped trajectory  with the same initial conditions as in the left panel but for $\gamma<1$. Right panel: example of a trajectory with a small perturbation ($\dot{\theta}_0\neq 0$). The initial conditions are $\dot{\theta}_0=0.01$, $\dot{r}_0=0$, $\mathcal{B}=0.1$, $\mathcal{L}=10$, $r_0=9$, and $\theta_0=\pi/2$. We set $m=1$.\label{figure3c}}
		\end{figure*}
	\end{center}
	
	In Fig.~\ref{figureII}, we show the behavior of $\mathcal{L}_{\text{E}\pm}$, $\mathcal{L}_{\text{E(ex)}}$, and the behavior of the so-called ``{\em lake-like}'' angular momentum functions $\mathcal{L}_{\text{L}_\pm}$ as functions of $r$ for different values of $\gamma$ and $\mathcal{B}$. In the figure, we see that the extremum function $\mathcal{L}_{\text{E}\pm}$, which determines the circular orbits in the equatorial plane, diverges when the charged particles move close to the photon sphere. In the figure, we can also see the value of $\mathcal{L}_{\text{ISCO}}$ from the intersection of $\mathcal{L}_{\text{E}\pm}$ with $\mathcal{L}_{\text{E(ex)}}$. Note that the position of $r_\text{ISCO}$ is affected by the presence of the magnetic field. Moreover, stable and unstable circular orbits are located at $r>r_{\text{ISCO}}$ and $r<r_{\text{ISCO}}$, respectively. On the other hand, regarding the lake angular momentum functions, we see that the region between $\mathcal{L}_{\text{L-}}$ and $\mathcal{L}_{\text{L+}}$ is smaller for $\mathcal{B}>0$ than the region in the case $\mathcal{B}<0$ (for a given value of $\gamma$). 
	Finally, for fixed values of $\mathcal{B}\neq0$, the region between $\mathcal{L}_{\text{L-}}$ and $\mathcal{L}_{\text{L+}}$ becomes broader as $\gamma$ increases.

	\subsection{Charged particle trajectories}

	The trajectories of charged particles immersed in a uniform magnetic field can be obtained by solving the equations of motion. Typically this can be done following two possible procedures, namely i) from the Lorentz equation or ii) from the Hamiltonian formalism. In the following, we use the former method. Then, a particle with (rest) mass $m_0$ and charge $q$ immersed in an electromagnetic field $\mathrm{F}_{\rho\sigma}$ satisfies the Lorentz force law~\cite{MTW,Wald:1984rg}
	\begin{equation}
	\label{IV.22}
	\frac{d^2x^\mu}{d\tau^2}+\Gamma^{\mu}_{\;\;\alpha\beta}\frac{dx^\alpha}{d\tau}\frac{dx^\beta}{d\tau}=\frac{q}{m_0}g^{\mu\rho}\mathrm{F}_{\rho\sigma}\frac{dx^\sigma}{d\tau}.
	\end{equation}
	where $\tau$ is the affine parameter of the particle's trajectory and $\mathrm{F}_{\rho\sigma}$ is the electromagnetic Faraday tensor. In general, the tensor $\mathrm{F}_{\rho\sigma}$ is defined, in terms of the vector potential $A_\alpha$, as
	\begin{equation}
	\label{IV.23}
	\mathrm{F}_{\rho\sigma}=\partial_\rho A_\sigma-\partial_\sigma A_\rho,
	\end{equation}
	and in the case of the $\gamma$-metric immersed in a uniform magnetic field the only non vanishing components of vector potential are given by Eq.~\eqref{III.15} so that the non-vanishing components of the electromagnetic tensor, expressed in $(t,r,\theta,\phi)$ coordinates, take the form 
	\begin{equation}
	\label{IV.25}
	\begin{aligned}
	\tilde{\mathrm{F}}_{r\phi}&=-\tilde{\mathrm{F}}_{\phi r}=2\mathcal{B} \sin ^2\theta(r-2m)^{-\gamma } r^{\gamma }[r-(\gamma+1)m],\\\\
	\tilde{\mathrm{F}}_{\theta\phi}&=-\tilde{\mathrm{F}}_{\phi\theta}=2\mathcal{B} \sin\theta \cos\theta (r-2m)^{1-\gamma } r^{\gamma +1},
	\end{aligned}
	\end{equation}
	where we have defined $\tilde{\mathrm{F}}_{\rho\sigma}=q\mathrm{F}_{\rho\sigma}/m_0$. Note again that for $\gamma=1$, Eq.~(\ref{IV.25}) reduce to the case of Schwarzschild immersed in a uniform electromagnetic field~\cite{Kolos:2015iva}. 
	
	Using the conservation of angular momentum from Eq.~(\ref{IV.4}), we can obtain the  equation of motion for the axial coordinate $\phi$. This equation has the form
	\begin{equation}
	\label{IVC1}
	\frac{d\phi}{d\tau}=\frac{\mathcal{L}}{r^2}\left(1-\frac{2m}{r}\right)^{\gamma-1}-\mathcal{B}\;,
	\end{equation}
	while the equation of motion for the radial component takes the form

	\begin{widetext}
		\begin{equation}
		\left(\frac{dr}{d\tau}\right)^2=\left(1- \frac{m}{r}\right)^{\gamma ^2-1} \left[\mathcal{E}^2-\left(1-\frac{2 m}{r}\right)^{\gamma }-\frac{\left(r \mathcal{B} (2 m-r)+\mathcal{L} \left(1-\frac{2 m}{r}\right)^{\gamma }\right)^2}{r (2 m+r)}\right]\;.
		\end{equation}
	\end{widetext}

	Given the difficulty of solving analytically the equations in general one needs to resort to numerical analysis.
	In Figs.~\ref{figure3} and \ref{figure3a}, we show some examples of trajectories for charged particles on the equatorial plane. We considered both \textbf{PC} and \textbf{MC} configurations, and different values of $\gamma$. From the figures, it is possible to identify the infalling trajectories, and the circular and bounded orbits. 
	In the case of bound orbits, the radial coordinate is constrained in the interval $r_a\leq r\leq r_p$, where $r_a$ and $r_p$  are the apoapsis and periapsis, respectively~\cite{Kolos:2015iva}. From the figures, we can see two types of bounded trajectories: curled and non-curled. The former is obtained when the coordinate $\phi$ decreases during the particle motion, in contrast to non-curled trajectories, where $\phi$ always increases (see also Fig.~\ref{figure3b}). Furthermore, it is important to point out that these trajectories are specific to the charged particle motion in the asymptotically uniform magnetic fields, and cannot occur in the case of uncharged particles, and more precisely they are possible only in the \textbf{PC} configuration. Similar curly trajectories were found in the Schwarzschild space-time immersed in a magnetic field~\cite{Kolos:2015iva, Frolov:2010mi,Narzilloev:2019hep}.  
	
	It is known that the motion of an electrically neutral test particle is always restricted to the equatorial plane and that a freely moving particle can escape to infinity only in this plane, even in the presence of an external magnetic field.
	Nevertheless, when we consider charged particles in the presence of an asymptotically uniform magnetic field, it is possible to have escape trajectories, which end at spatial infinity as $z \rightarrow \infty$ evolving along the magnetic field lines, as can be seen in the left and central panels of Fig.~\ref{figure3c}.
	
    Out of the equatorial plane, particle motion tends to be chaotic~\cite{LukesGerakopoulos:2012pq, Kopacek:2014moa, Kopacek:2010yr,Sota:1995ms}. The chaotic motion is related to the variations in the coordinate $\theta$. Nevertheless, as we will discuss in the next section, harmonic oscillations may occur when bounded orbits are located close to the equatorial plane. In Fig.~\ref{figure3c}, left panel, we show one example.
	

	\section{Harmonic oscillations of charged test particles}\label{V}

	It is known that if a particle is displaced slightly from the radius of a stable circular orbit, the particle will start to oscillate around its equilibrium value. For sufficiently small displacement, it will execute simple harmonic motion\cite{Wald:1984rg}. This oscillating motion is governed by the epicyclic frequencies. 
	
	\subsection{Epicyclic frequencies}
	To obtain the epicyclic frequencies for the $\gamma$ space-time immersed in a uniform magnetic field, we follow the analysis in Ref.~\cite{Toshmatov:2019qih}. In this sense, we also restrict our analysis to the linear regime and consider radial and vertical oscillations separately.  
	
	The Hamiltonian given in Eq.~(\ref{IV.1}) can be expressed in the following form
	\begin{equation}
	\label{V1}
	\mathcal{H}=\frac{1}{2}g^{rr}p^2_r+\frac{1}{2}g^{\theta\theta}p^2_\theta+\frac{m^2_0}{2}
	\left(g^{tt}\mathcal{E}^2+g^{\phi\phi}(\mathcal{L}-\mathcal{B}g_{\phi\phi})^2+1\right).
	\end{equation}
	The normalization condition $u_\beta u^\beta=-1$ (where $u^\beta$ is the $4-$velocity of the charged particle), leads to $\mathcal{H}=0$. Hence, Eq.~(\ref{V1}) reduces to
	\begin{equation}
	\label{V2}
	g^{rr}p^2_r+g^{\theta\theta}p^2_\theta=-m^2_0
	\left(g^{tt}\mathcal{E}^2+g^{\phi\phi}(\mathcal{L}-\mathcal{B}g_{\phi\phi})^2+1\right),
	\end{equation}
	from which we can express the left-hand side as
	\begin{equation}
	\label{V3}
	g_{rr}\left(\frac{dr}{d\zeta}\right)^2+g_{\theta\theta}\left(\frac{d\theta}{d\zeta}\right)^2=
	-\Pi(r,\theta),
	\end{equation}
	where we have used the relation $p^\mu=m_0u^\mu$ and defined the function $\Pi(r,\theta)$ as 
	\begin{equation}
	\label{V4}
	\Pi(r,\theta)=g^{tt}\mathcal{E}^2+g^{\phi\phi}(\mathcal{L}-\mathcal{B}g_{\phi\phi})^2+1.
	\end{equation}
	In the case of charged particles moving on a plane with $\theta_0= \text{const}$ (as is the case for motion in the equatorial plane), Eq.~(\ref{V3}) takes the form,
	\begin{equation}
	\label{V5}
	g_{rr}\left(\frac{dr}{d\zeta}\right)^2=\mathcal{R}(r)=-\Pi(r,\theta_0).
	\end{equation}
	On the other hand, if we consider charged particles moving at a fixed radial distance $r=r_0$ with $\theta\neq0$, then Eq.~(\ref{V3}) reduces to 
	\begin{equation}
	\label{V6}
	g_{\theta\theta}\left(\frac{d\theta}{d\zeta}\right)^2=\Theta(\theta)=-\Pi(r_0,\theta).
	\end{equation}
	Therefore, test charged particles will move on a circular orbit $r=r_0$ in the equatorial $\theta=\theta_0=\pi/2$ if~\cite{Toshmatov:2019qih}
	\begin{equation}
	\label{V7a}
	\begin{array}{cc}
	\mathcal{R}(r_0)=0,&\partial_r\mathcal{R}(r)|_{r_0}=0,
	\end{array}
	\end{equation}
	and
	\begin{equation}
	\label{V7b}
	\begin{array}{cc}
	\Theta(\theta_0)=0,&\partial_\theta\Theta(\theta)|_{\theta_0}=0.
	\end{array}
	\end{equation}
	\begin{center}
		\begin{figure*}[t]
			\includegraphics[scale=0.32]{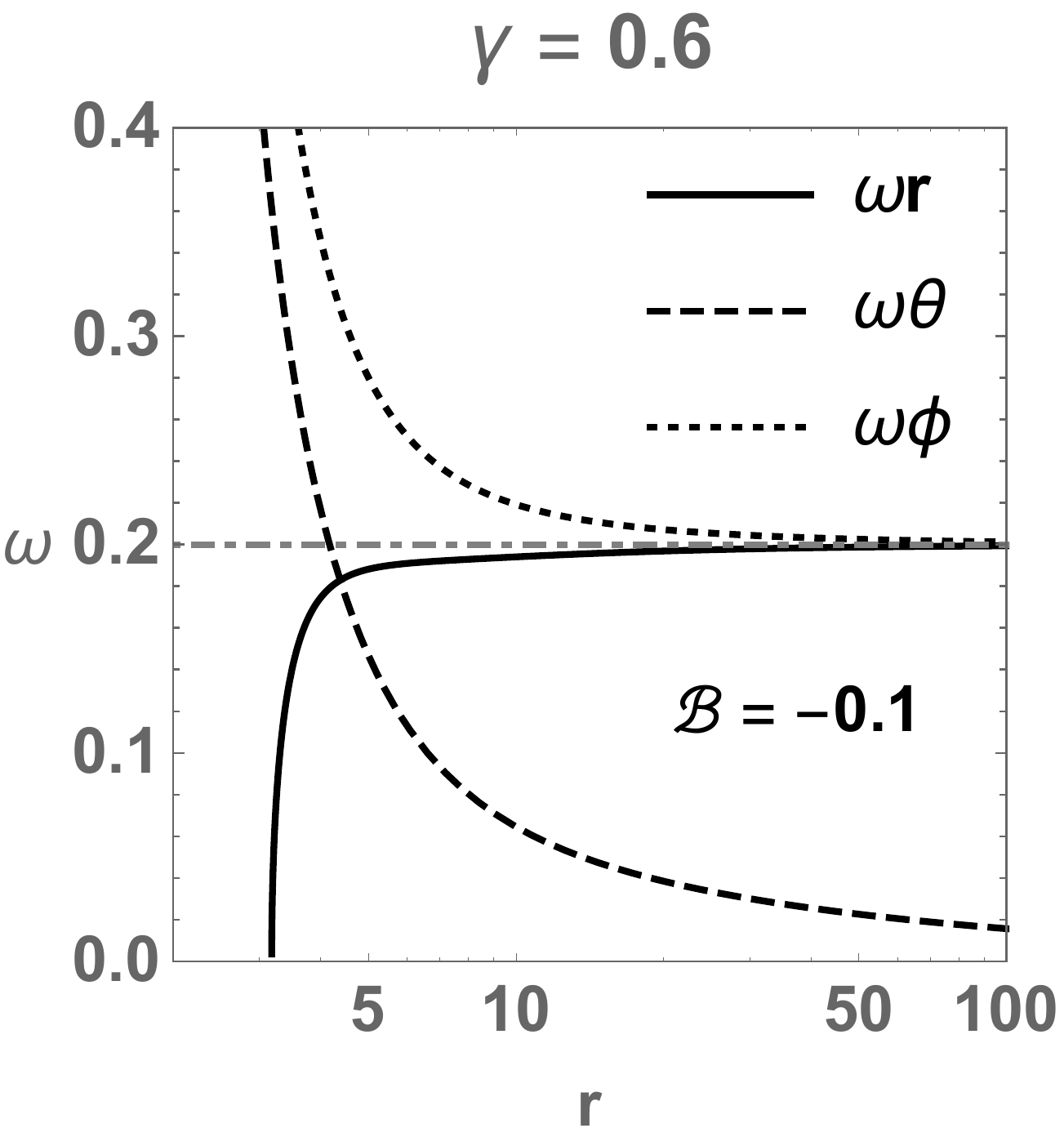}
			\hspace{0.8cm}
			\includegraphics[scale=0.32]{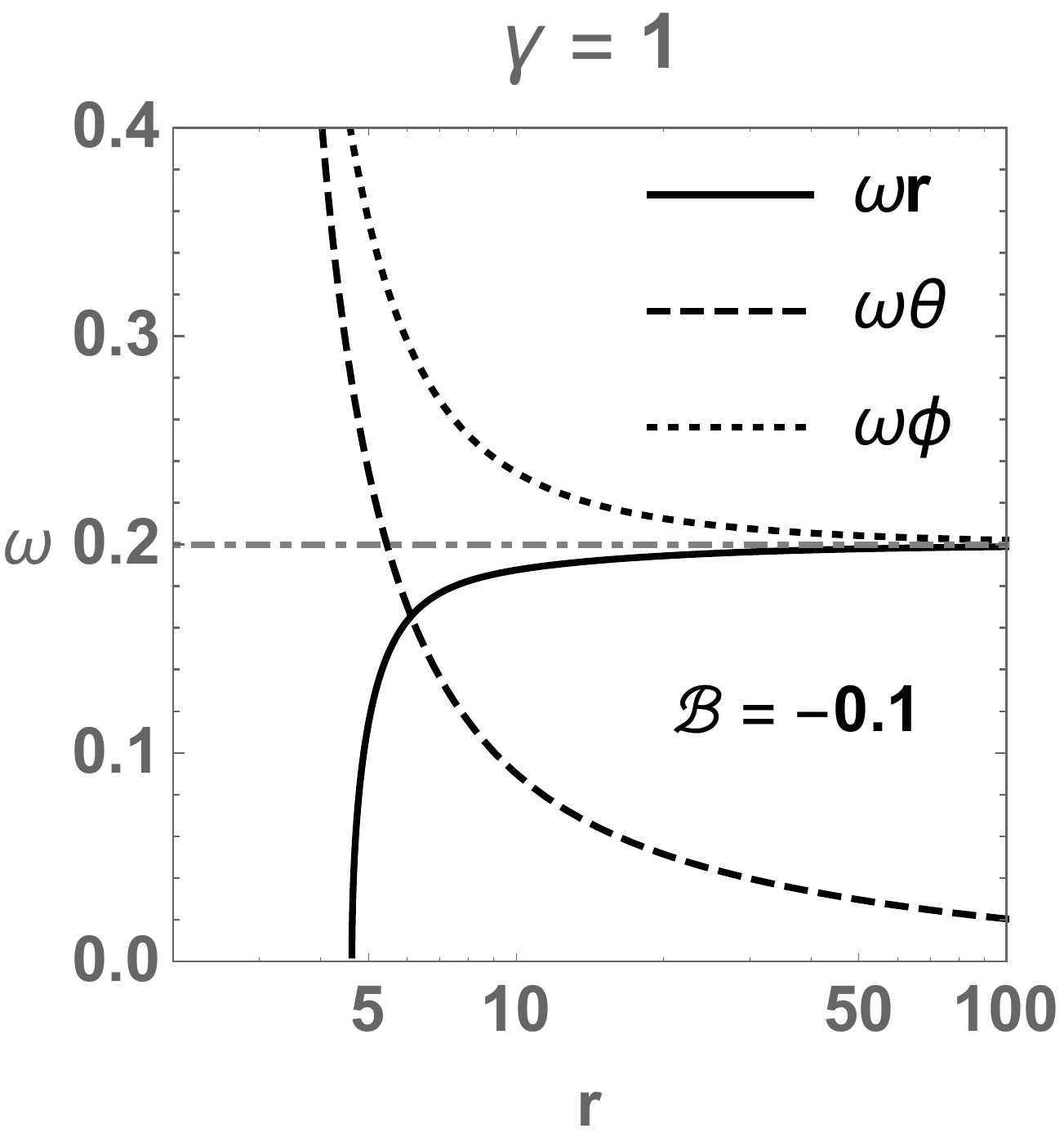}
			\hspace{0.8cm}
			\includegraphics[scale=0.32]{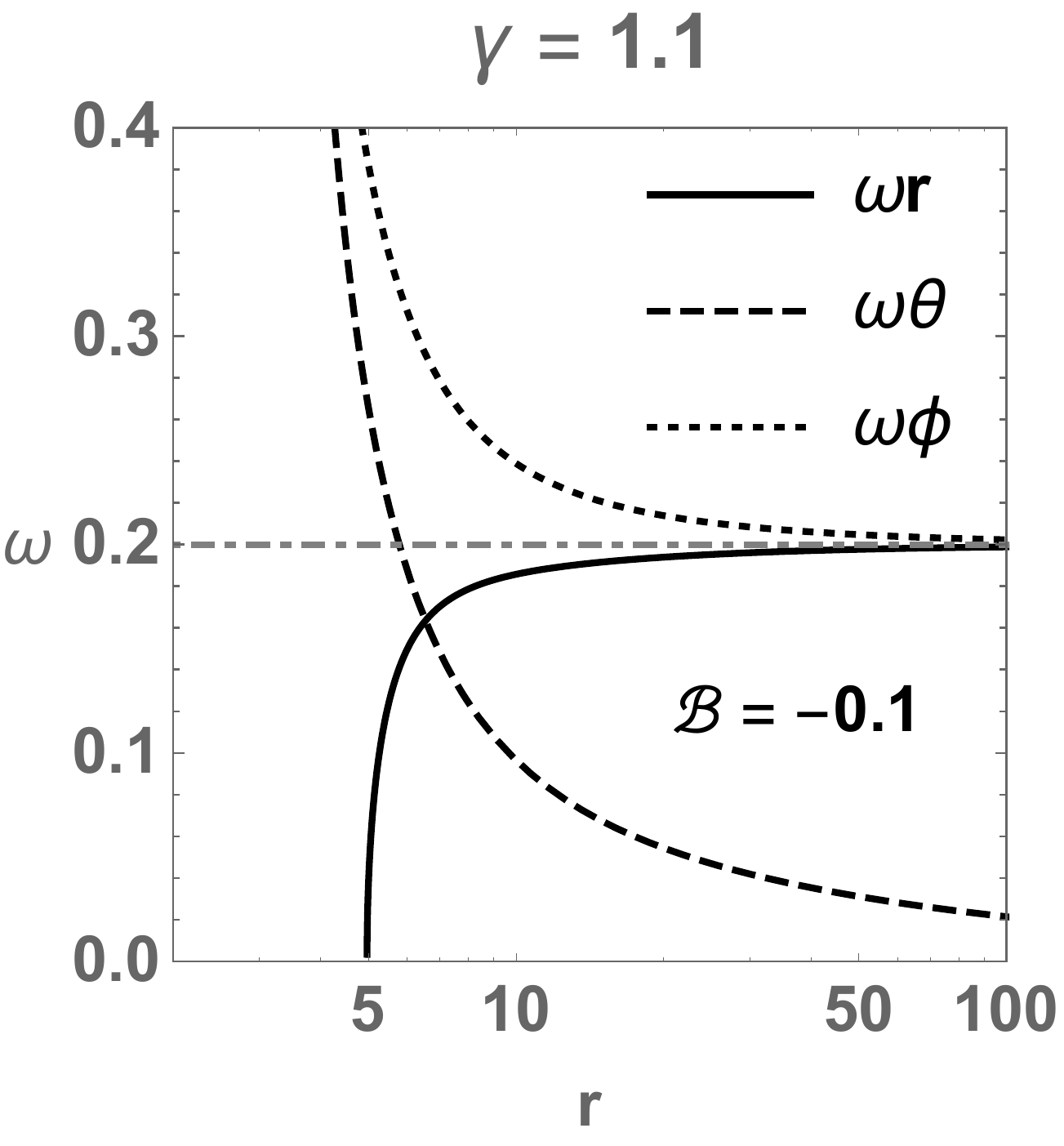}	\\
			\includegraphics[scale=0.32]{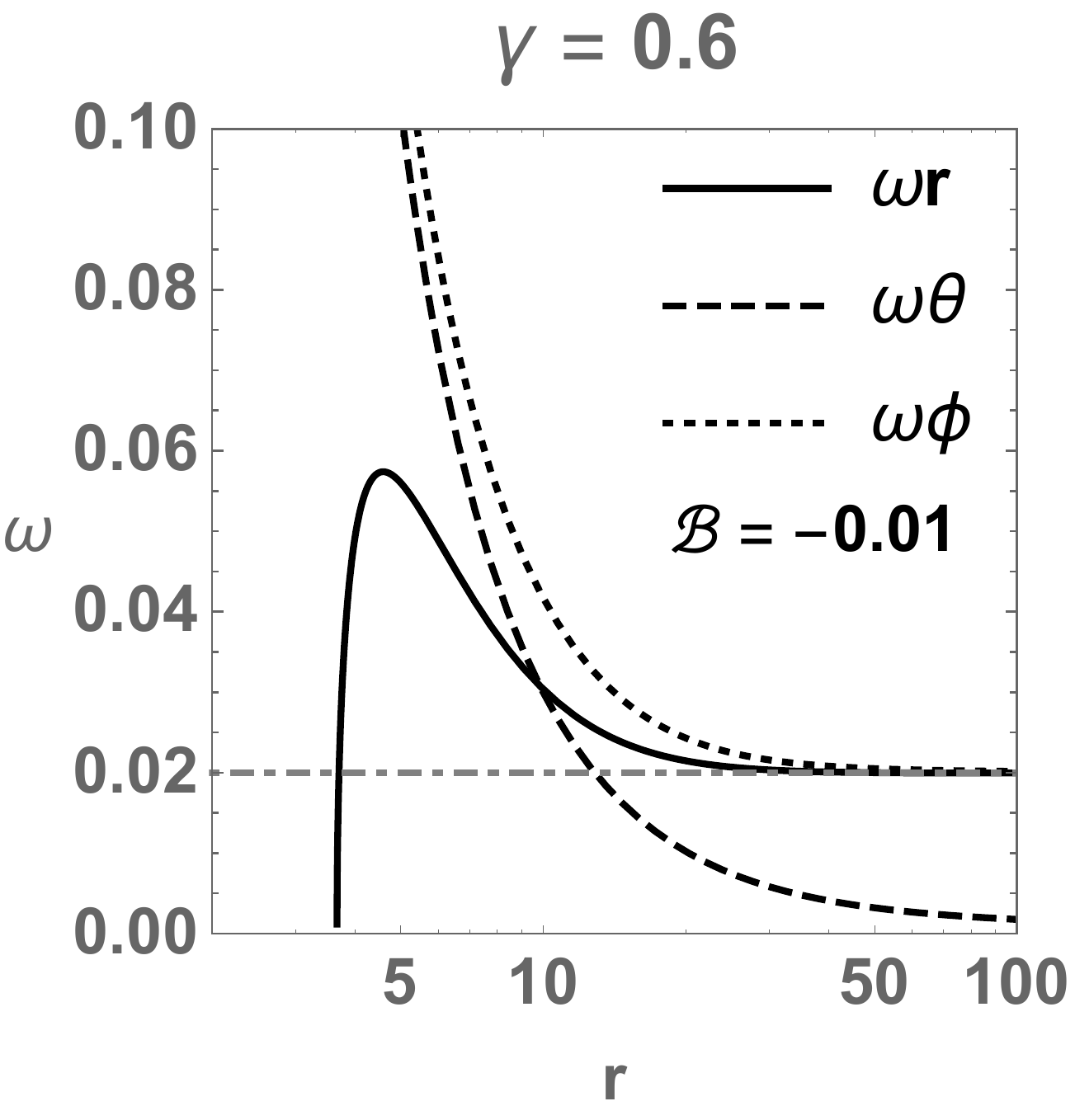}
			\hspace{0.8cm}
			\includegraphics[scale=0.32]{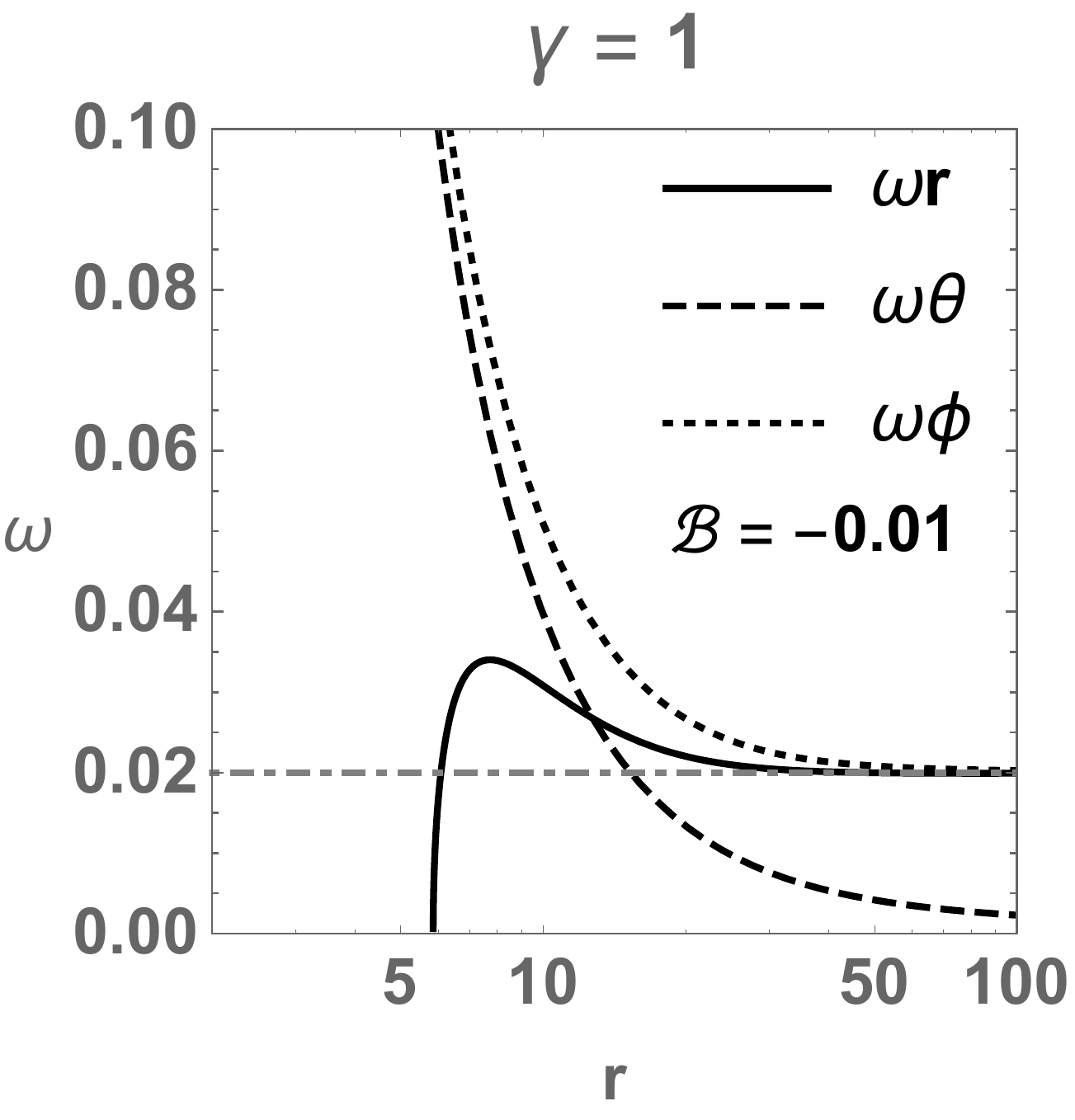}
			\hspace{0.8cm}
			\includegraphics[scale=0.32]{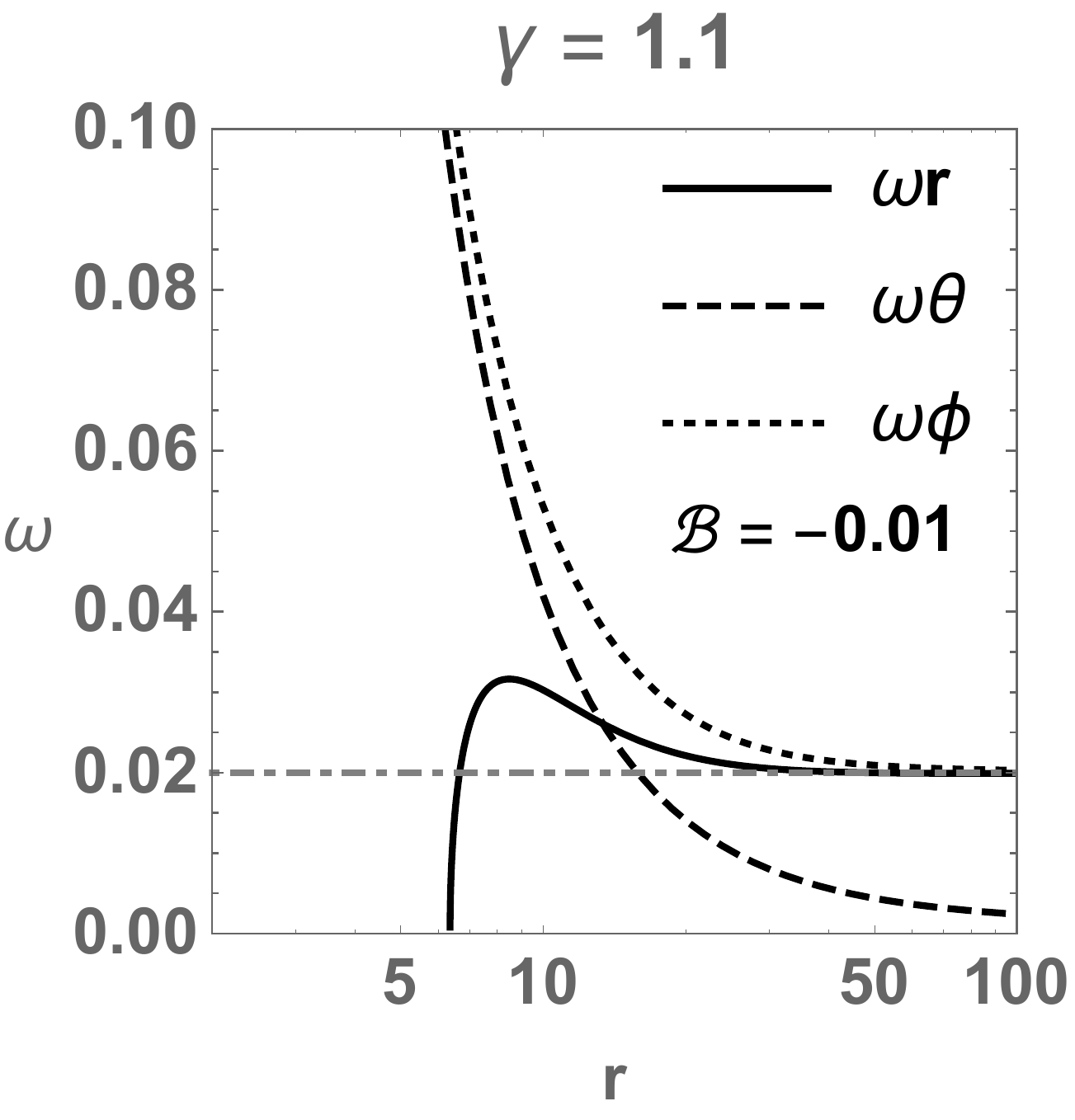}\\
			\includegraphics[scale=0.32]{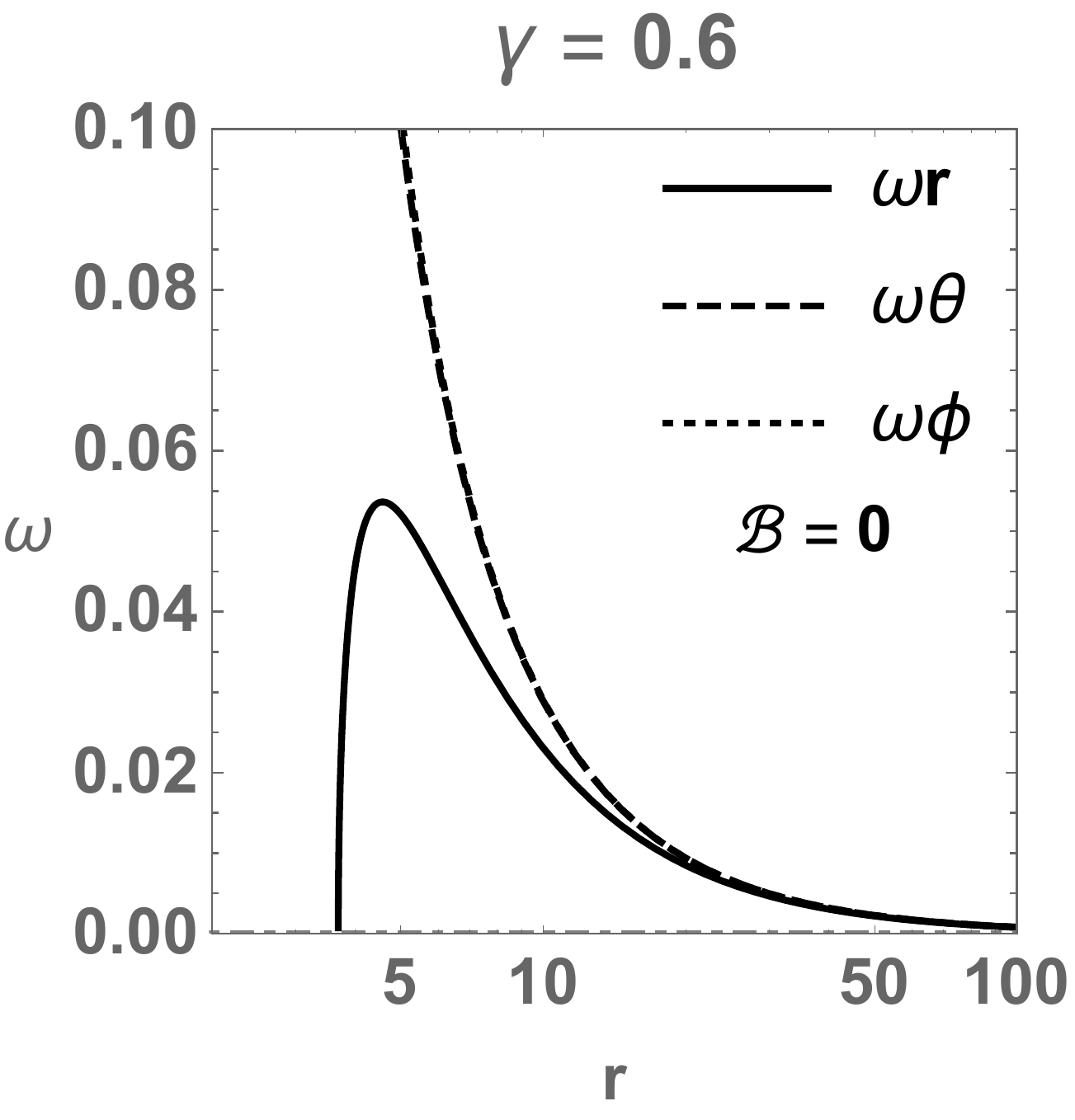}
			\hspace{0.8cm}
			\includegraphics[scale=0.32]{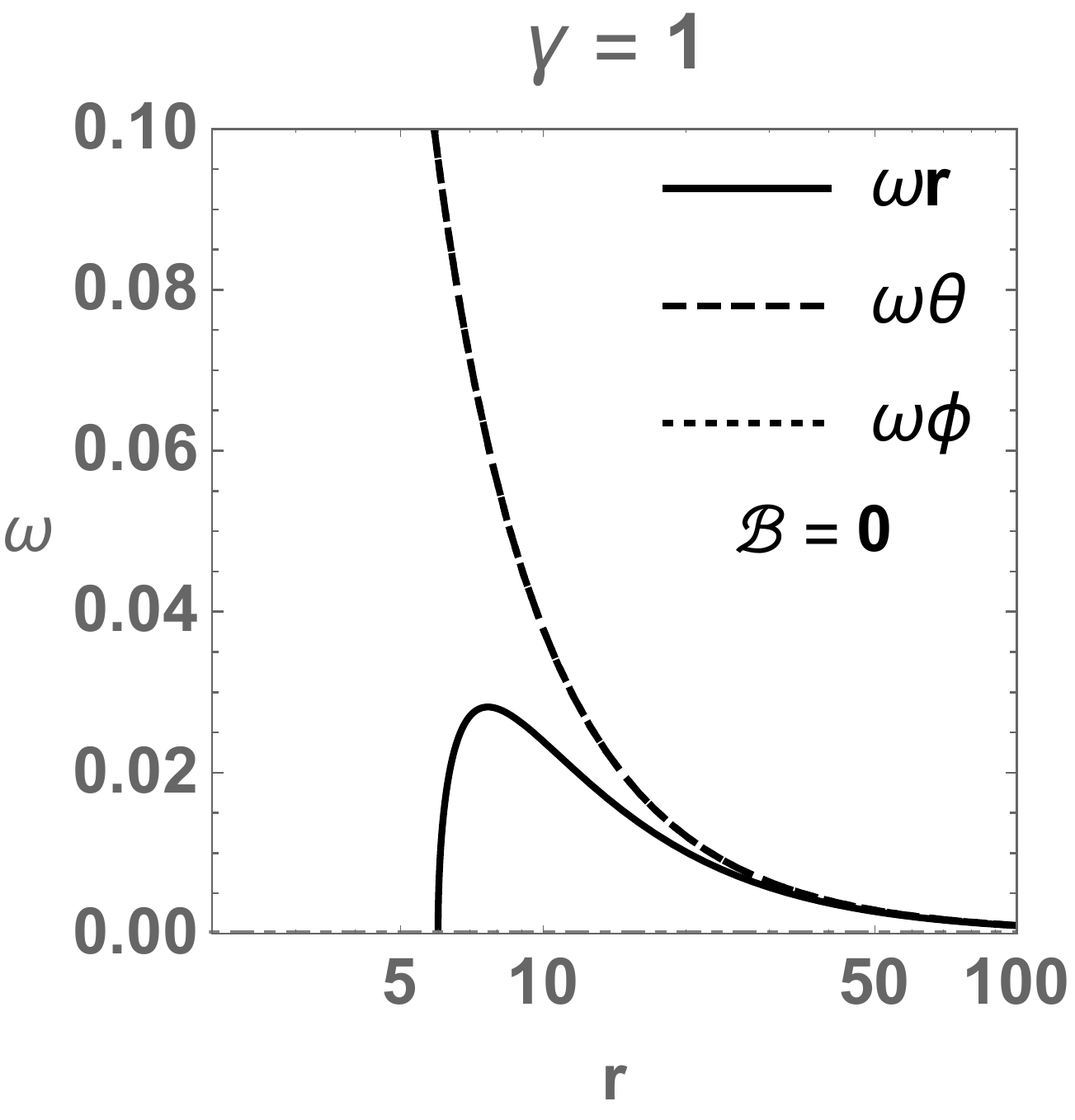}
			\hspace{0.8cm}
			\includegraphics[scale=0.32]{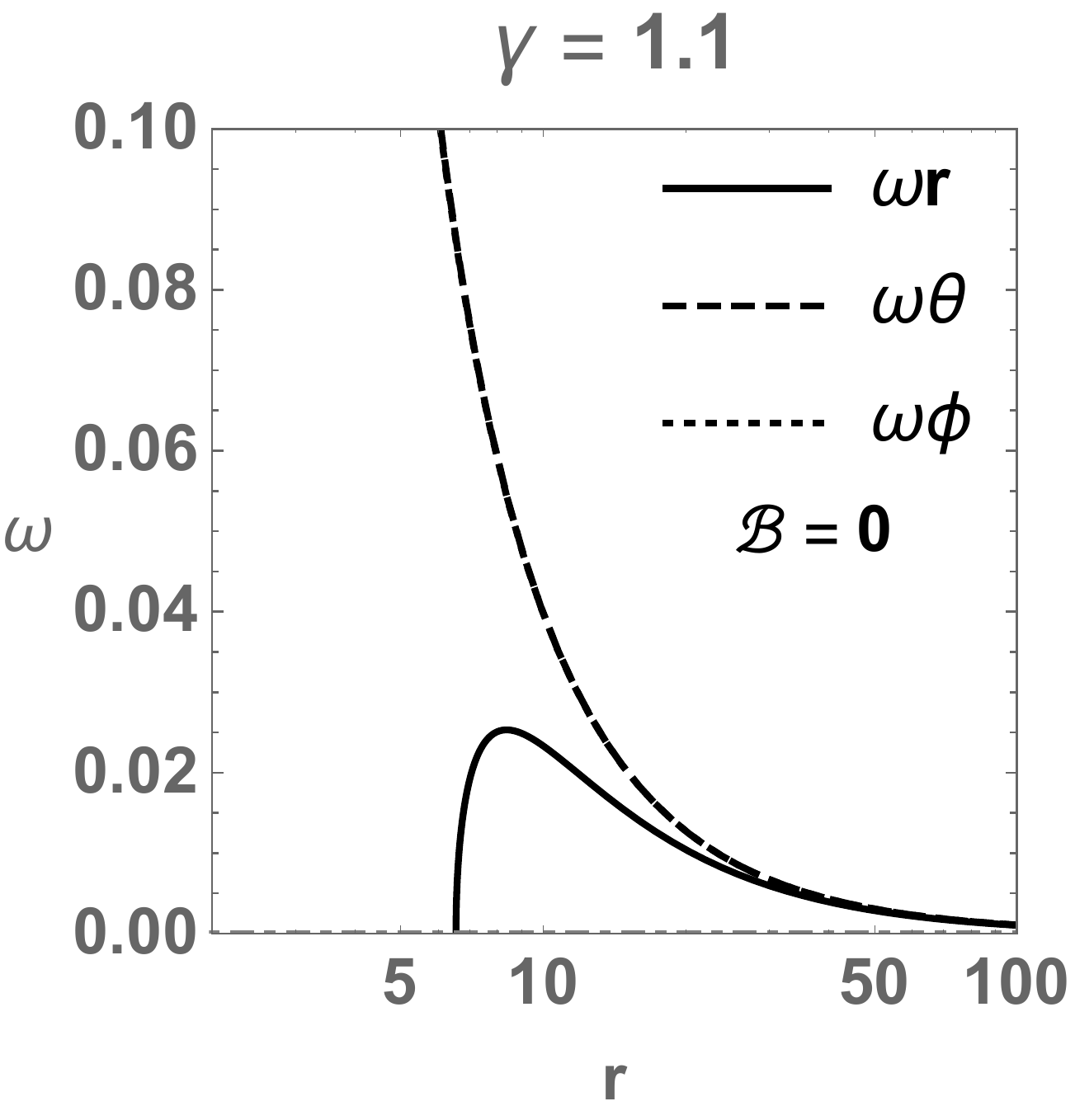}\\
			\includegraphics[scale=0.32]{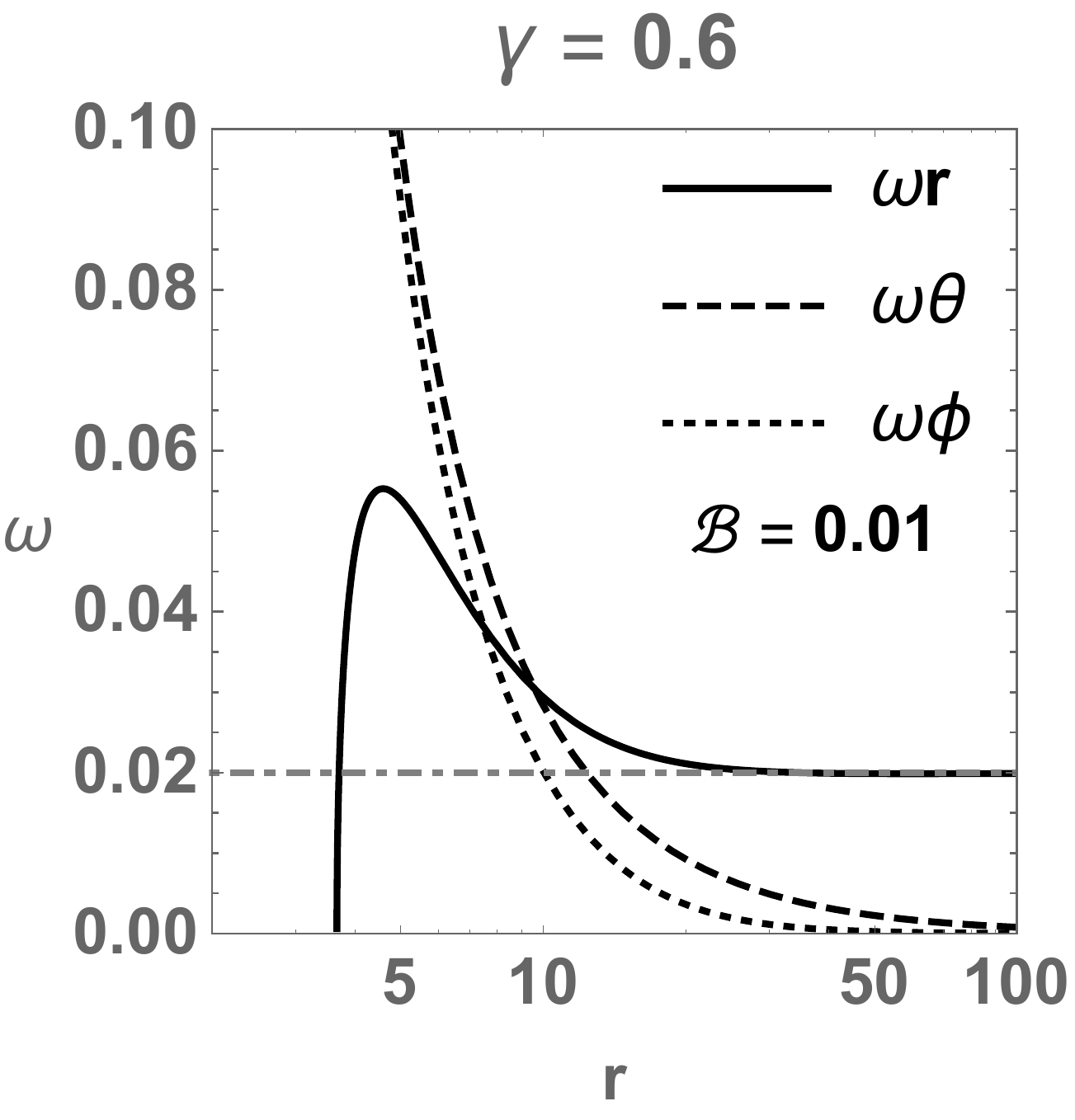}
			\hspace{0.8cm}
			\includegraphics[scale=0.32]{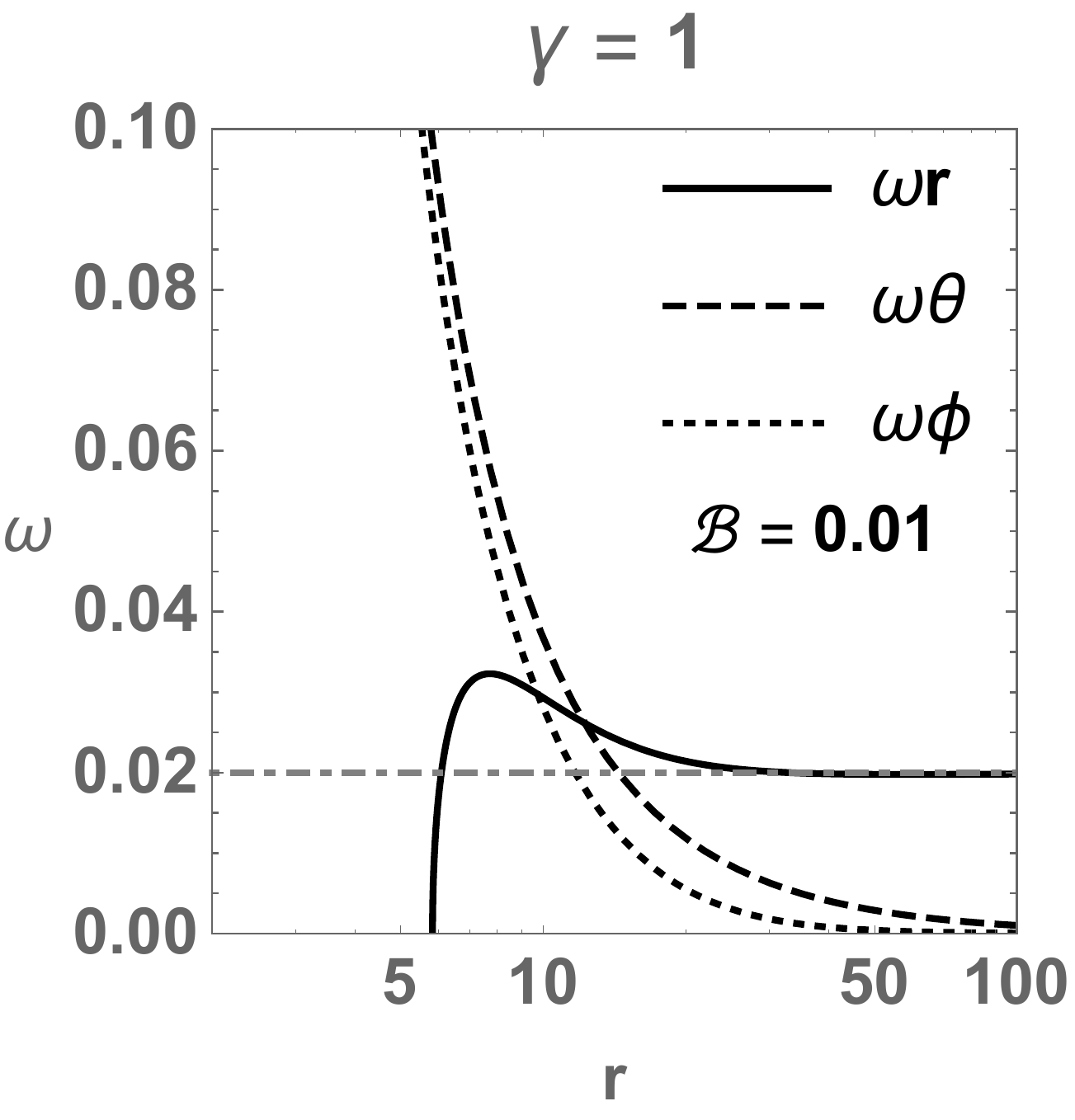}
			\hspace{0.8cm}
			\includegraphics[scale=0.32]{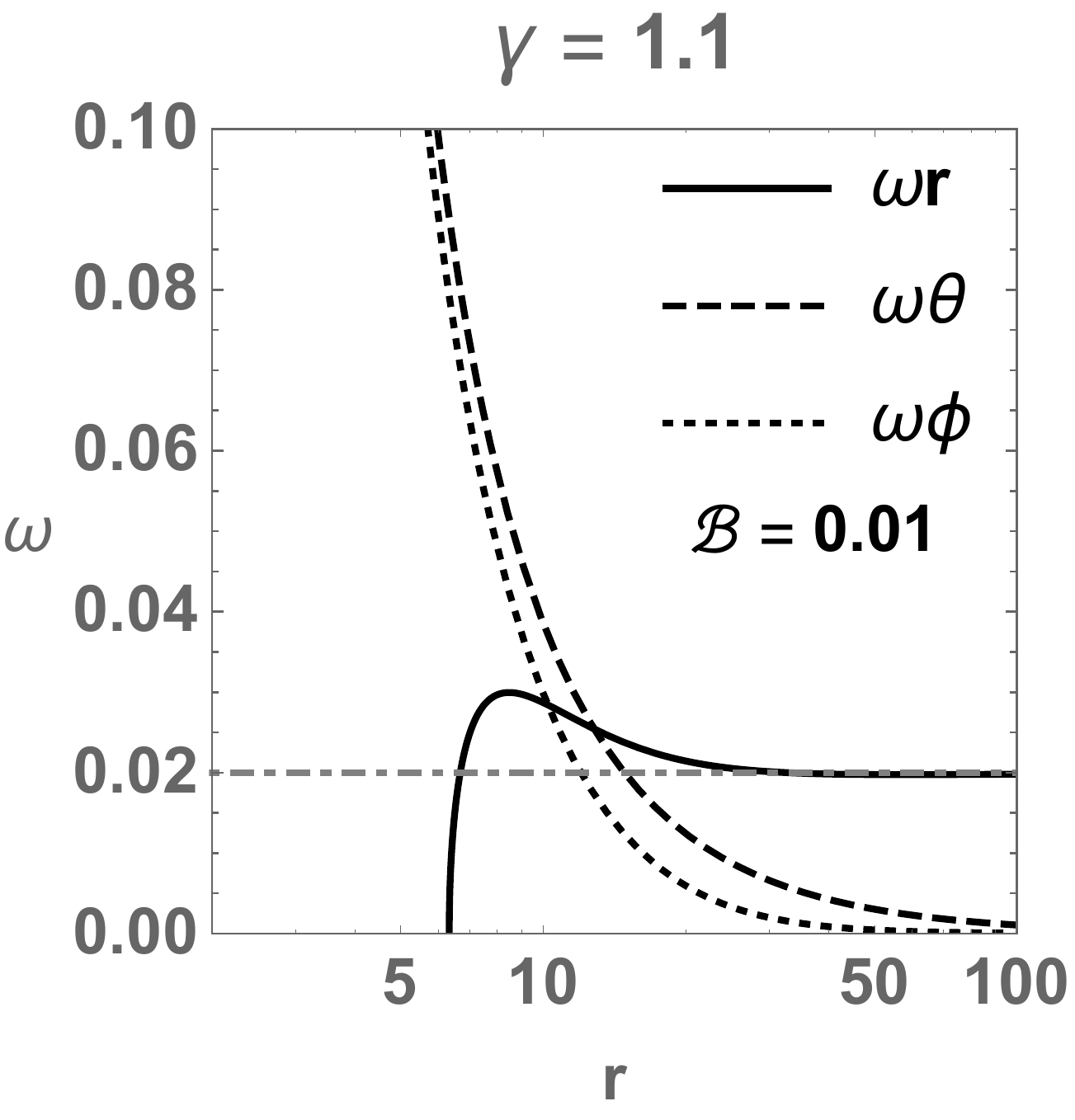}\\
			\includegraphics[scale=0.32]{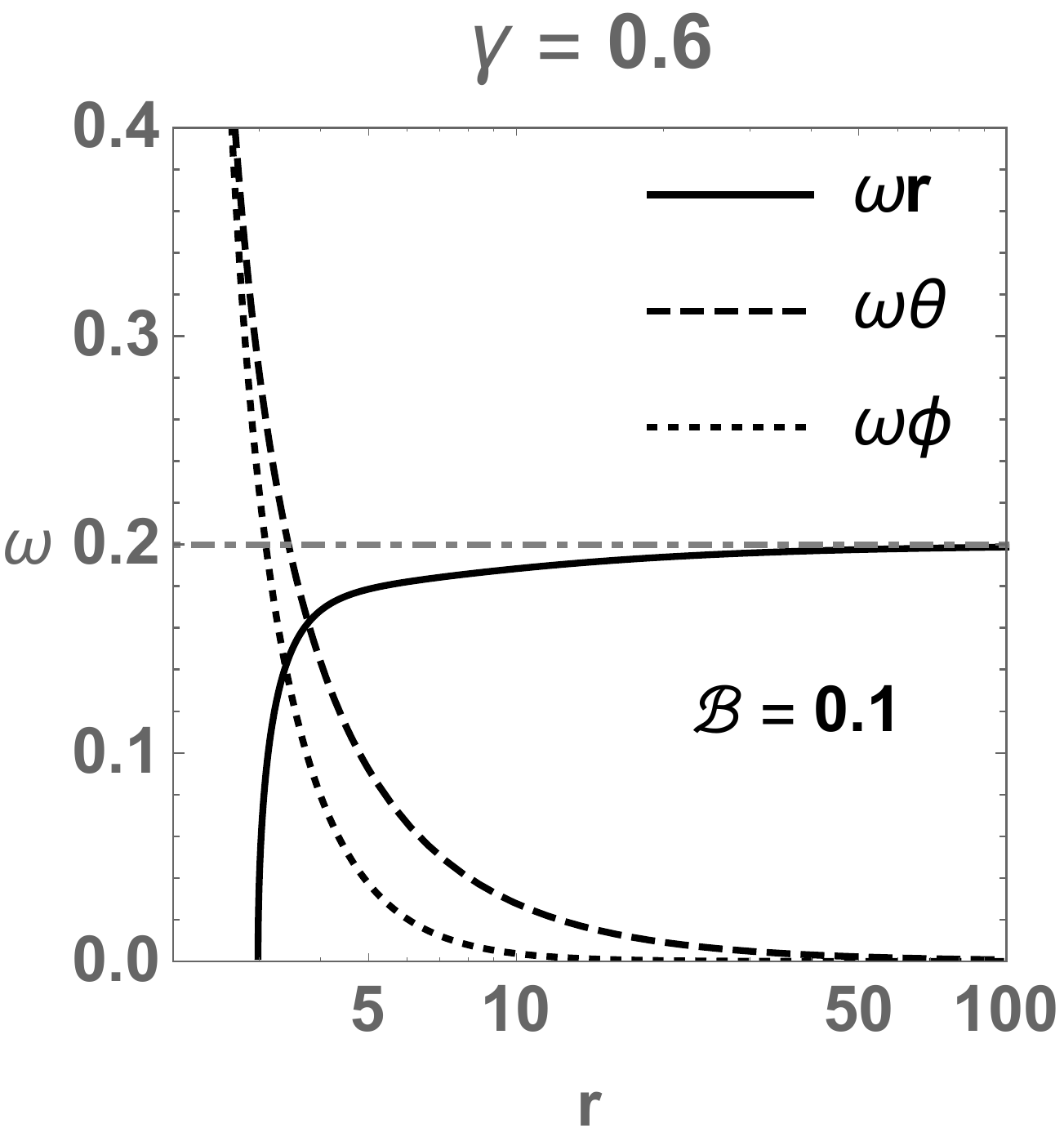}
			\hspace{0.8cm}
			\includegraphics[scale=0.32]{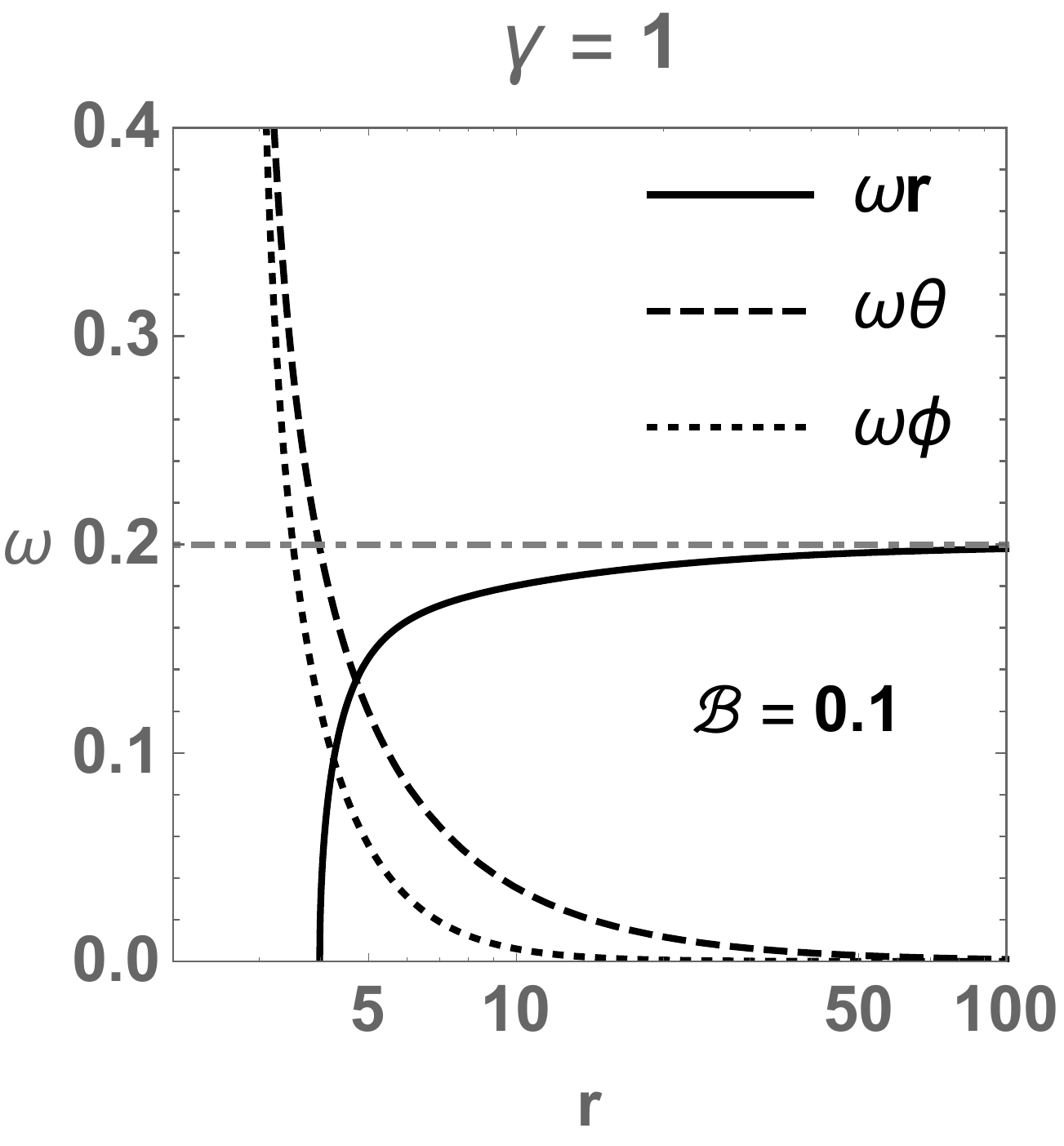}
			\hspace{0.8cm}
			\includegraphics[scale=0.32]{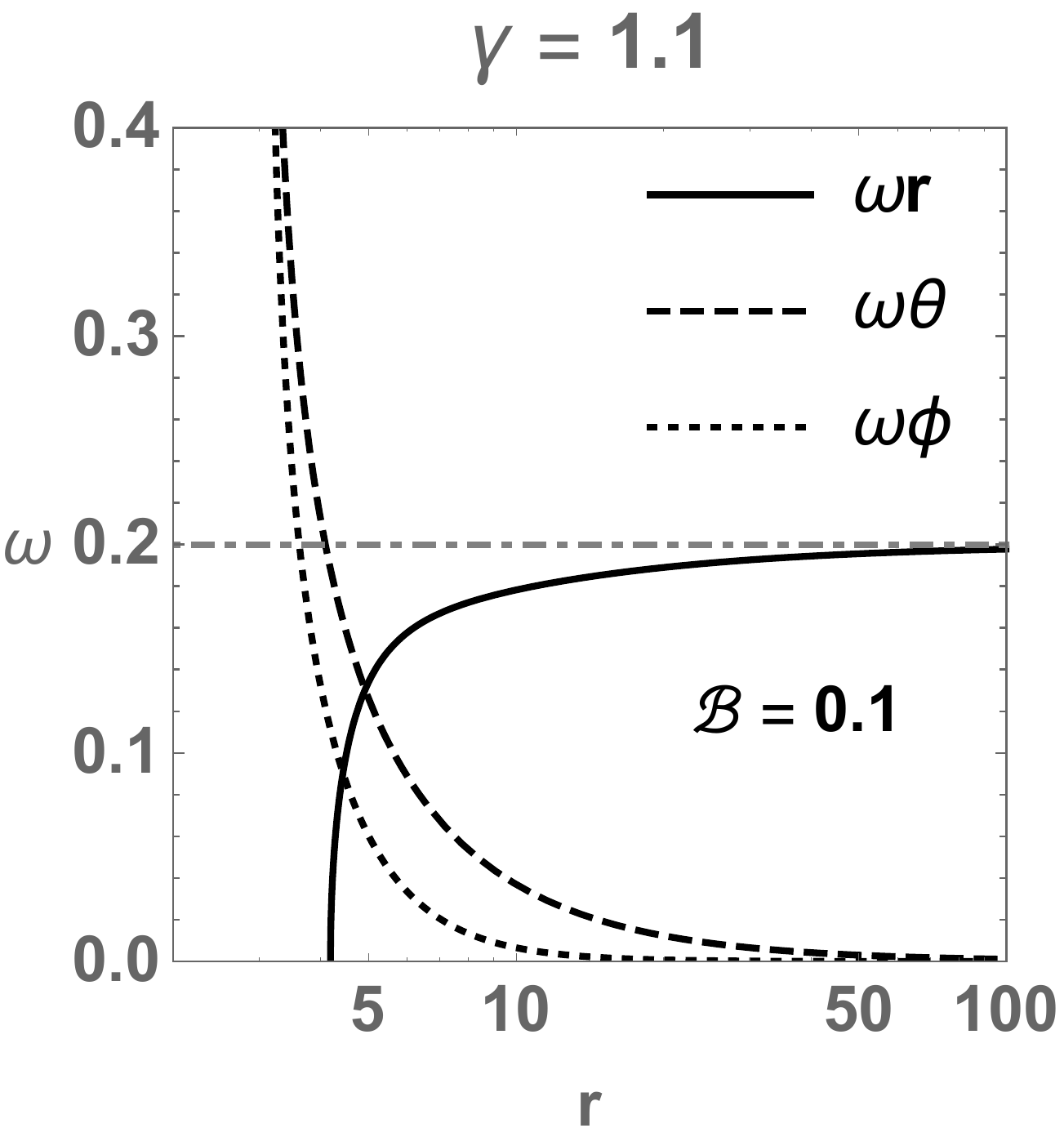}	
			\caption{Plots of $\omega_r$, $\omega_\theta$ and $\omega_\phi$ as functions of $r$ for different values of $\gamma$ and $\mathcal{B}$. The gray dot-dashed line corresponds to the Larmor frequency $\omega_L$ which has a constant value. We set $m=1$.
				\label{figure4}}
		\end{figure*}
	\end{center}
	\begin{center}
		\begin{figure*}[t]
			\includegraphics[scale=0.33]{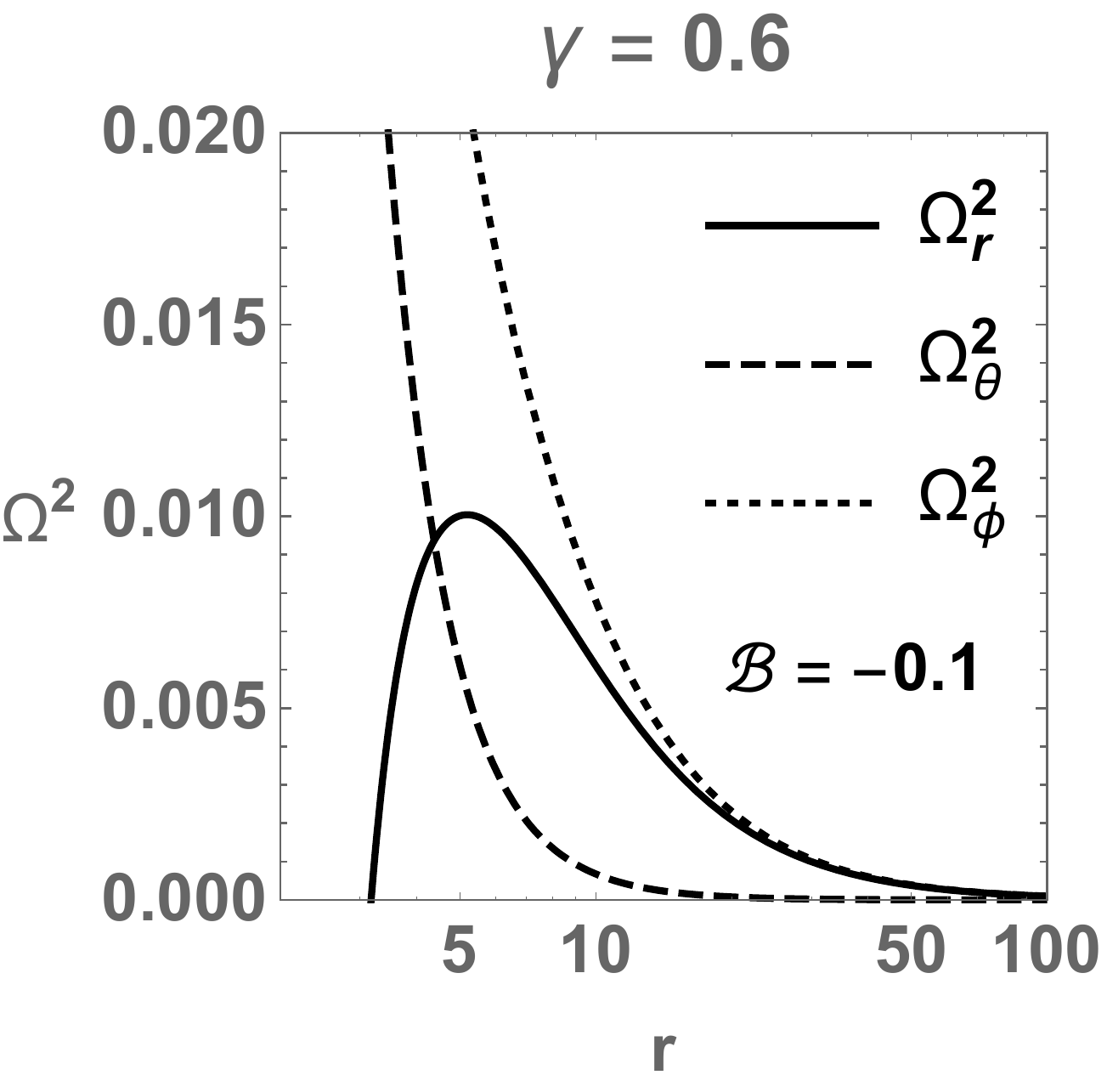}
			\hspace{0.8cm}
			\includegraphics[scale=0.33]{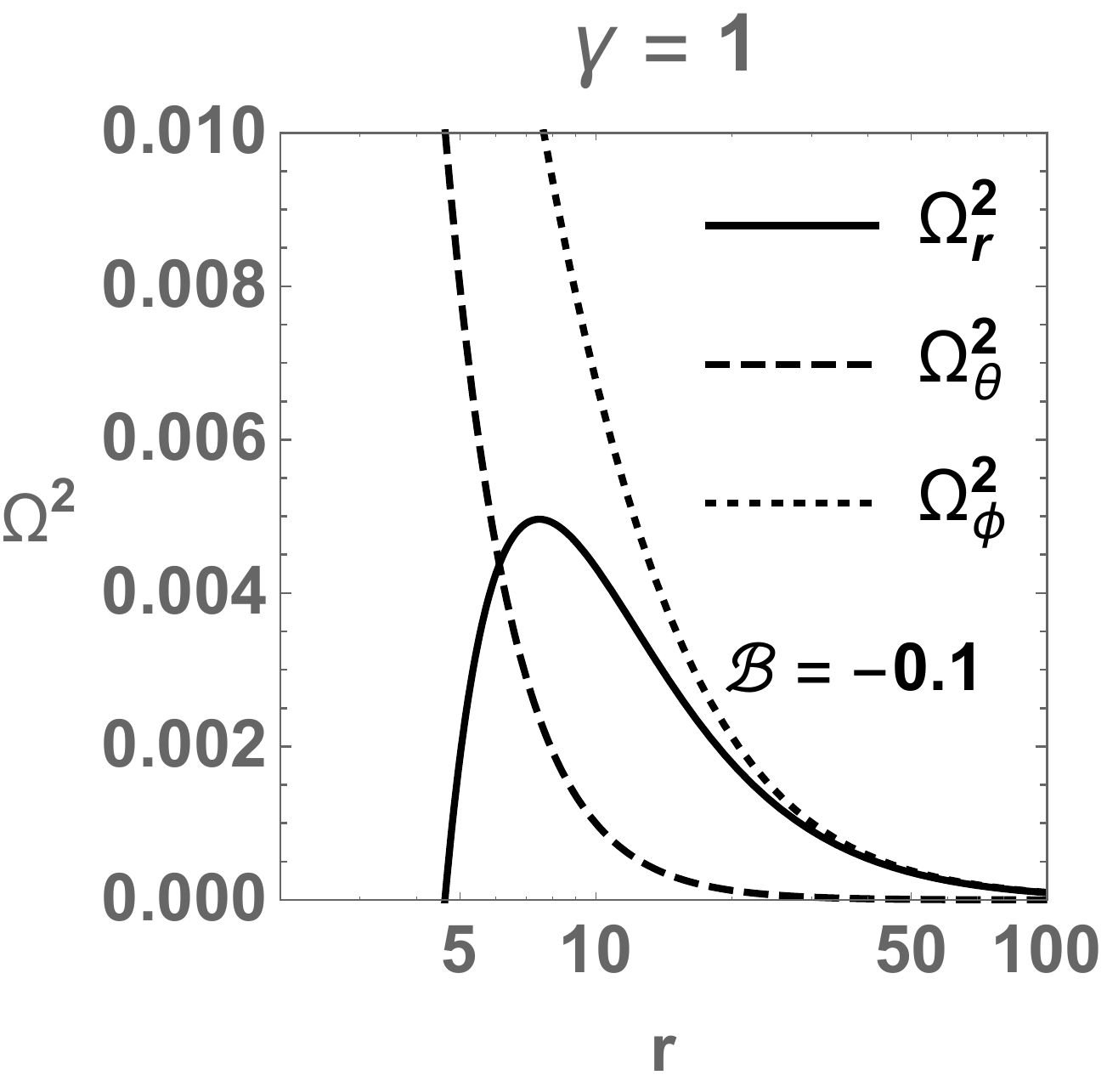}
			\hspace{0.8cm}
			\includegraphics[scale=0.33]{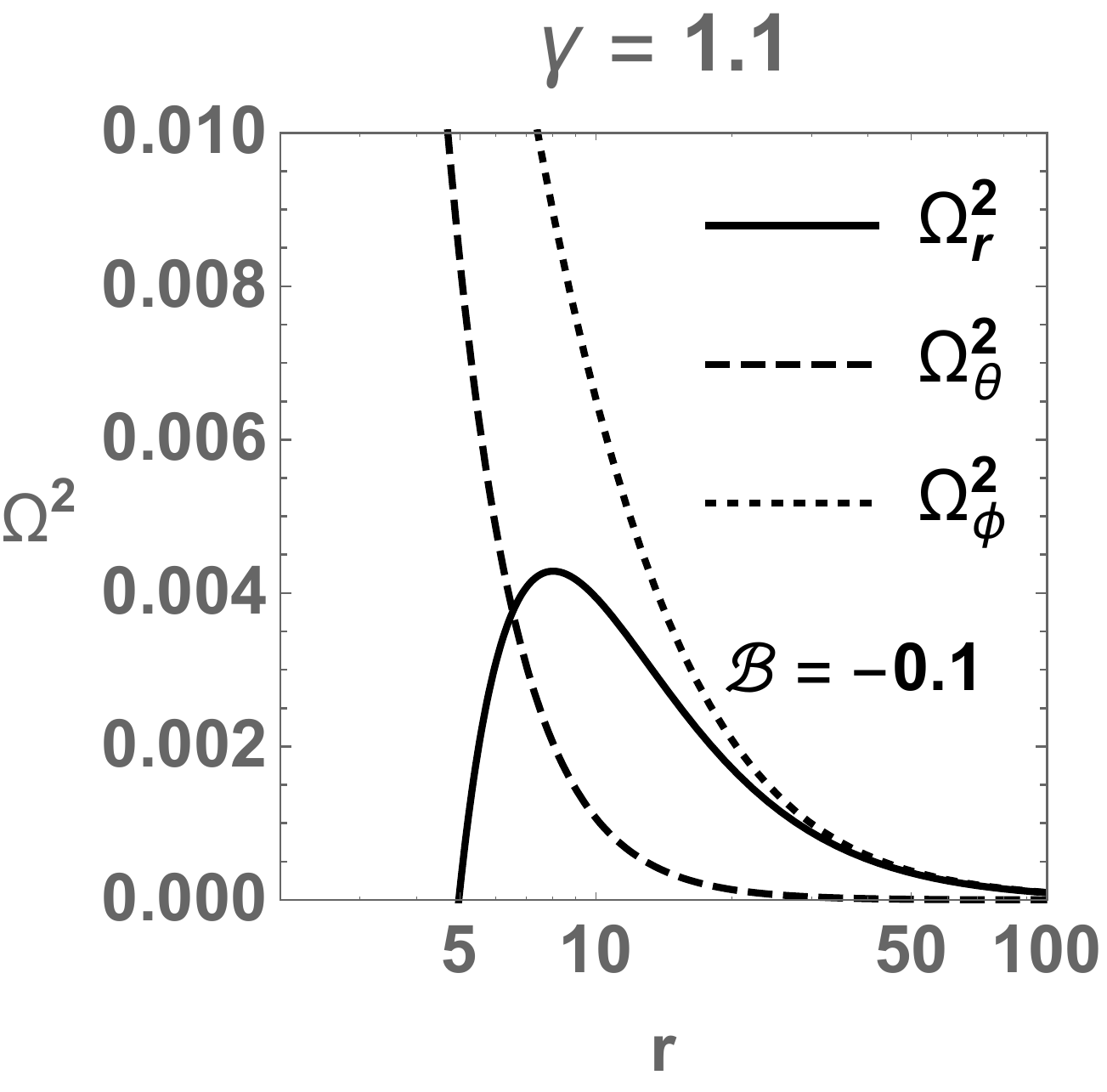}	\\
			\includegraphics[scale=0.33]{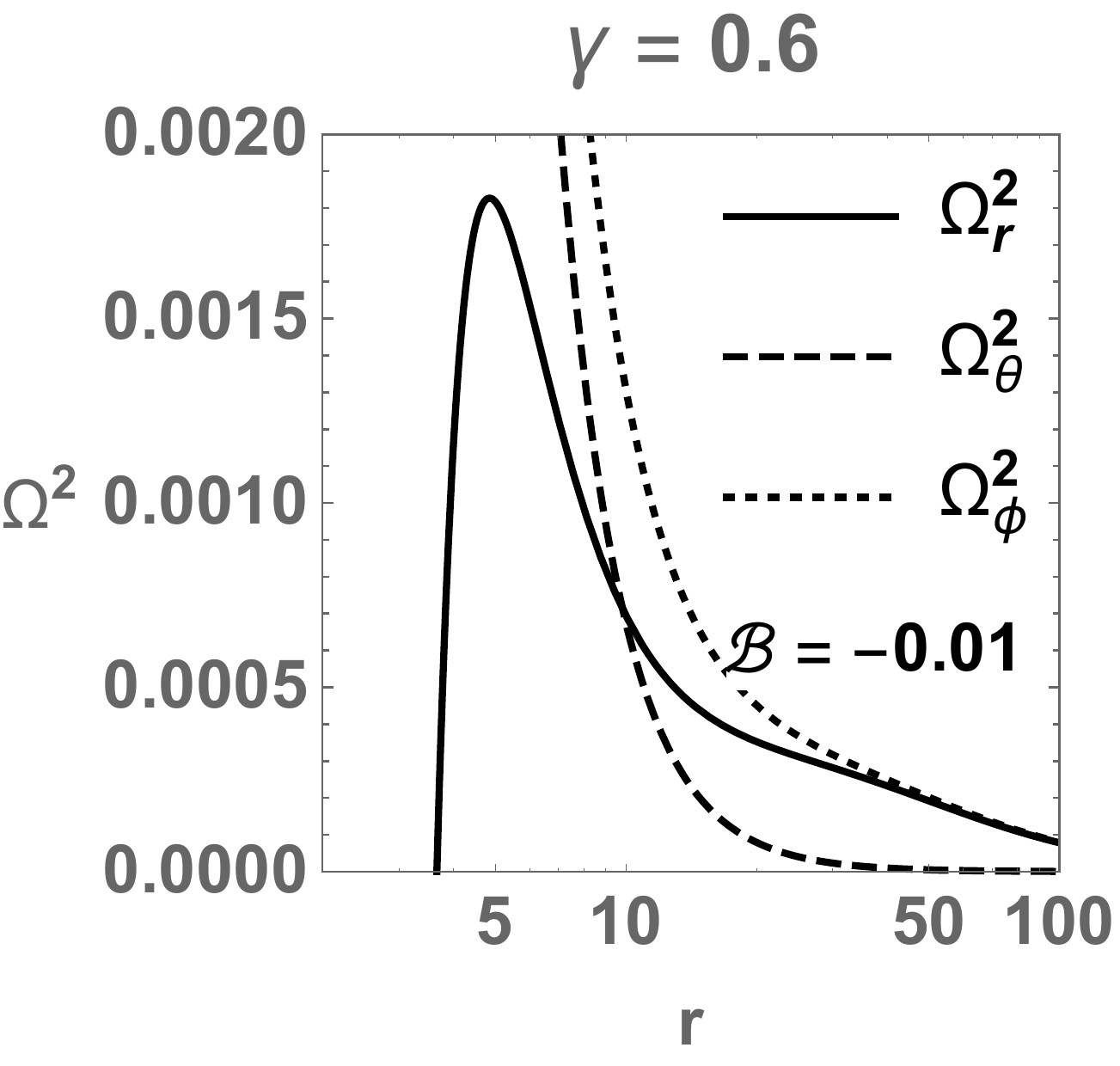}
			\hspace{0.8cm}
			\includegraphics[scale=0.33]{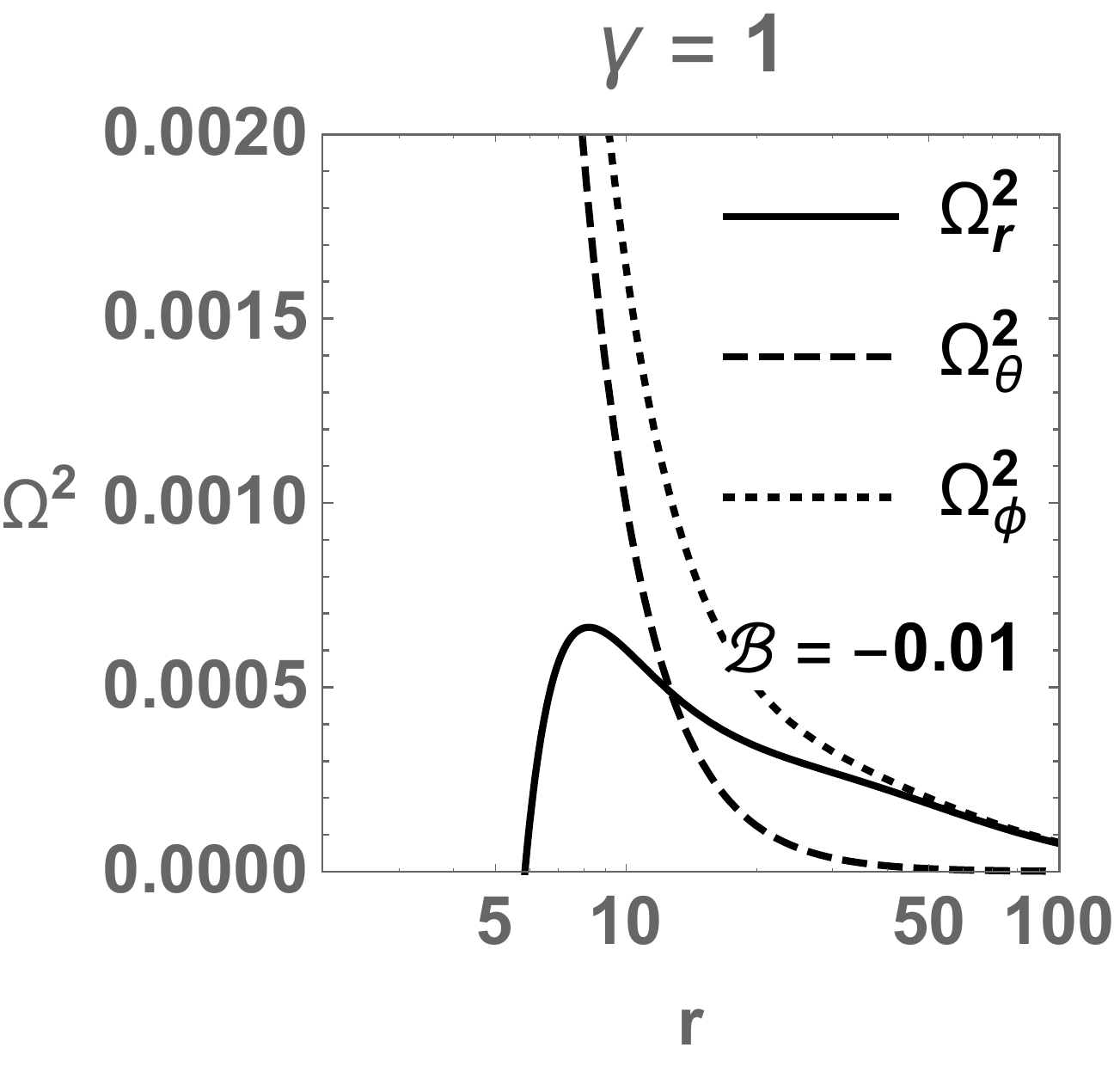}
			\hspace{0.8cm}
			\includegraphics[scale=0.33]{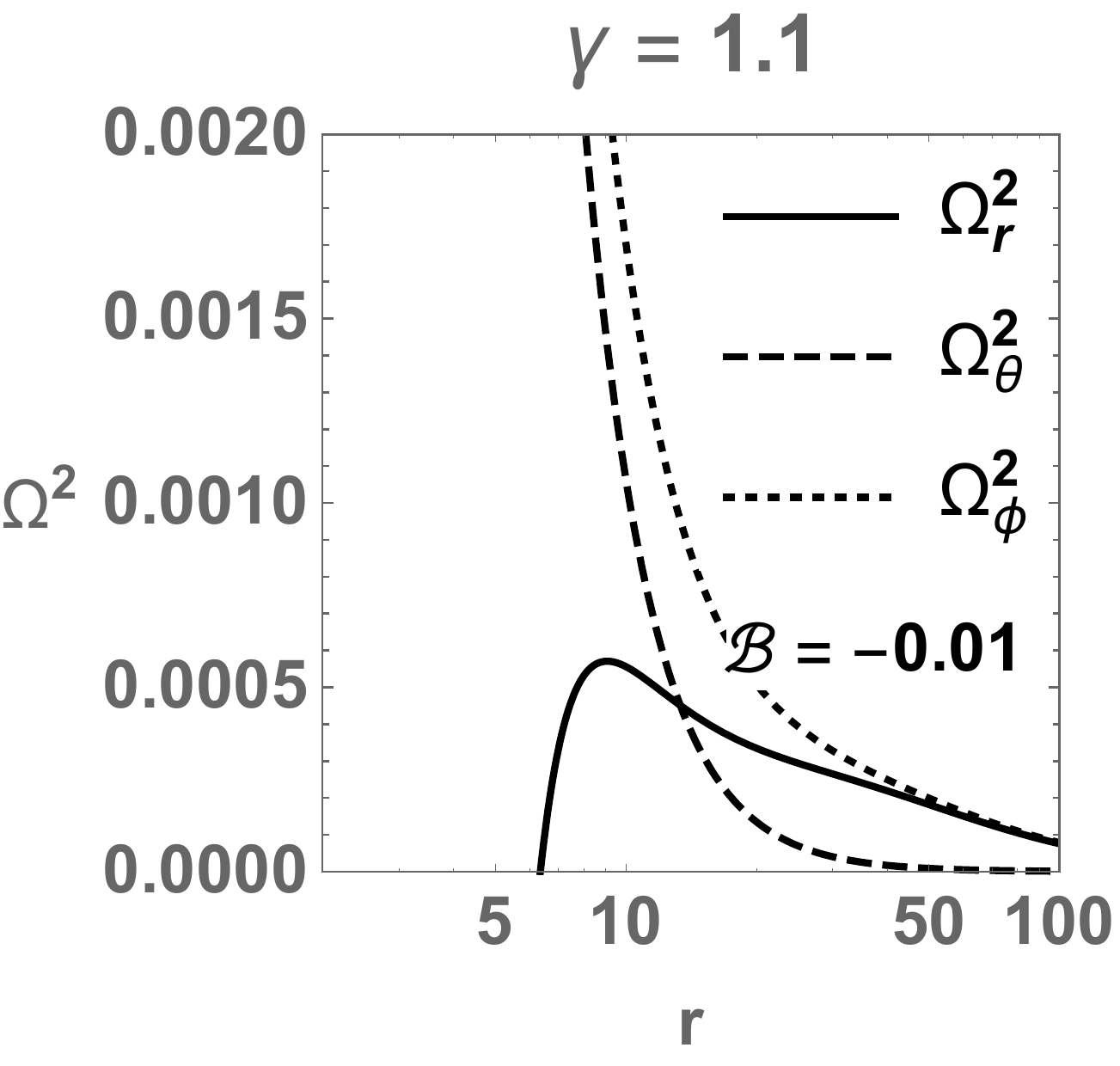}\\
			\includegraphics[scale=0.33]{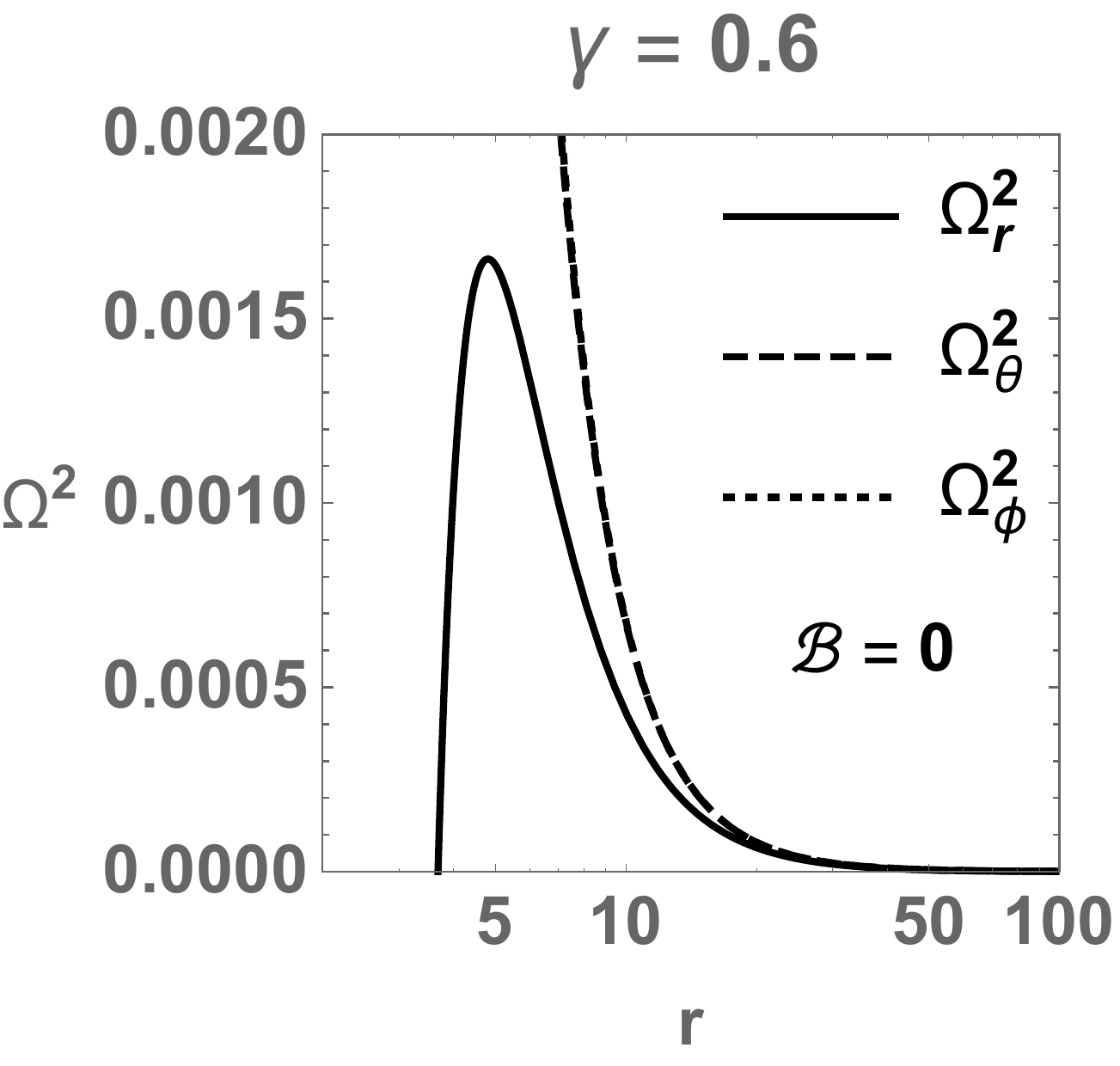}
			\hspace{0.8cm}
			\includegraphics[scale=0.33]{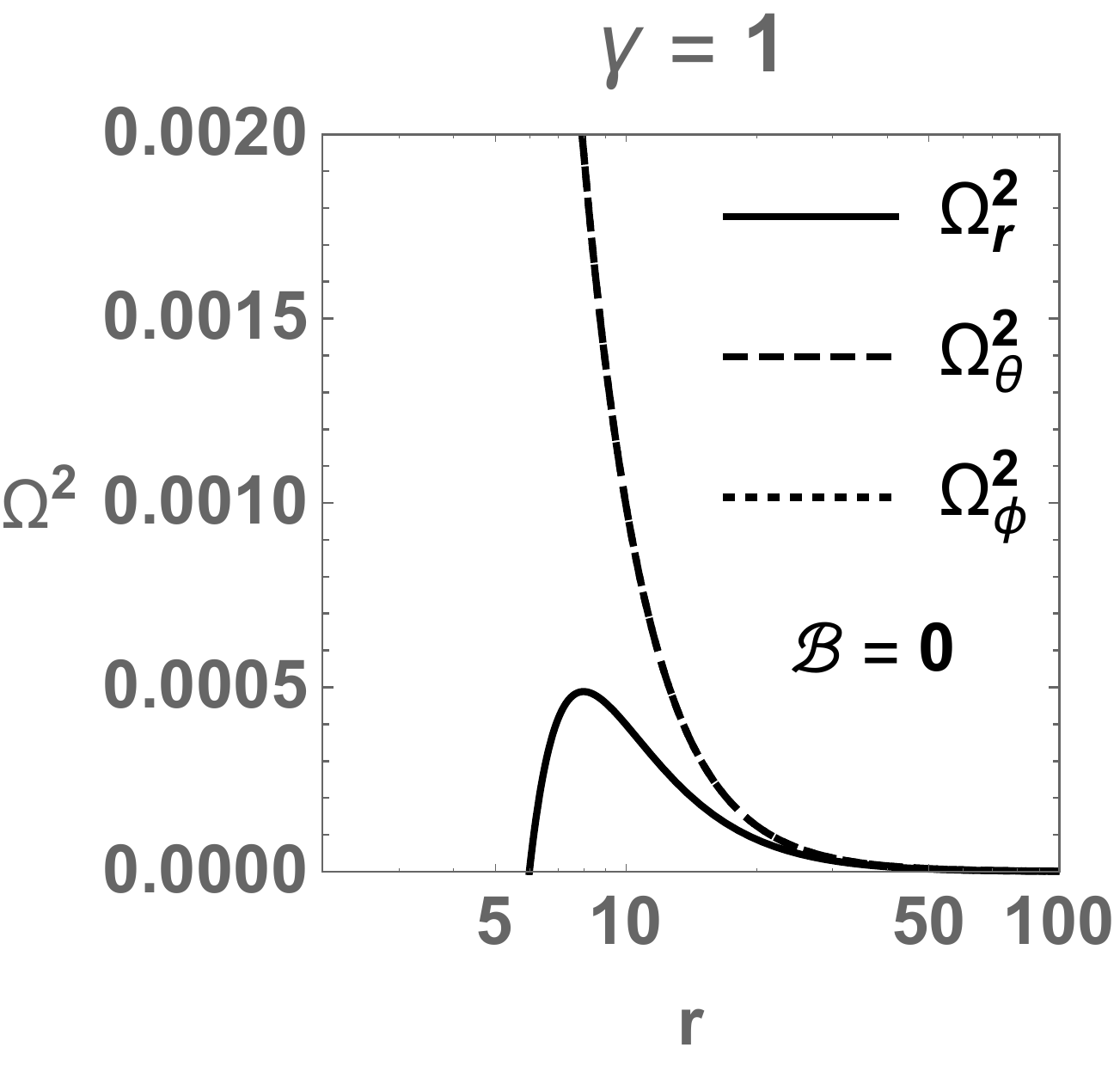}
			\hspace{0.8cm}
			\includegraphics[scale=0.33]{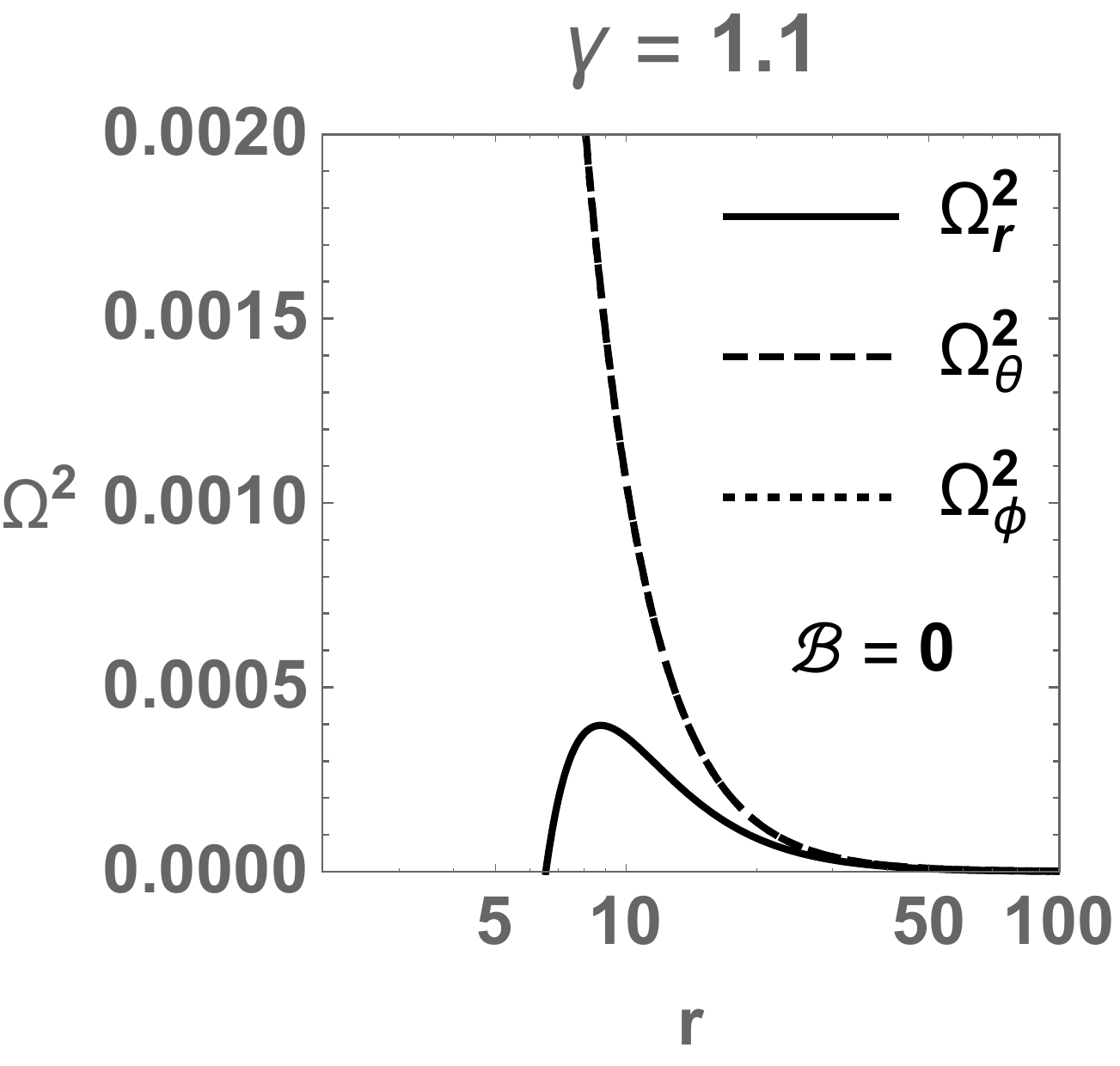}\\
			\includegraphics[scale=0.33]{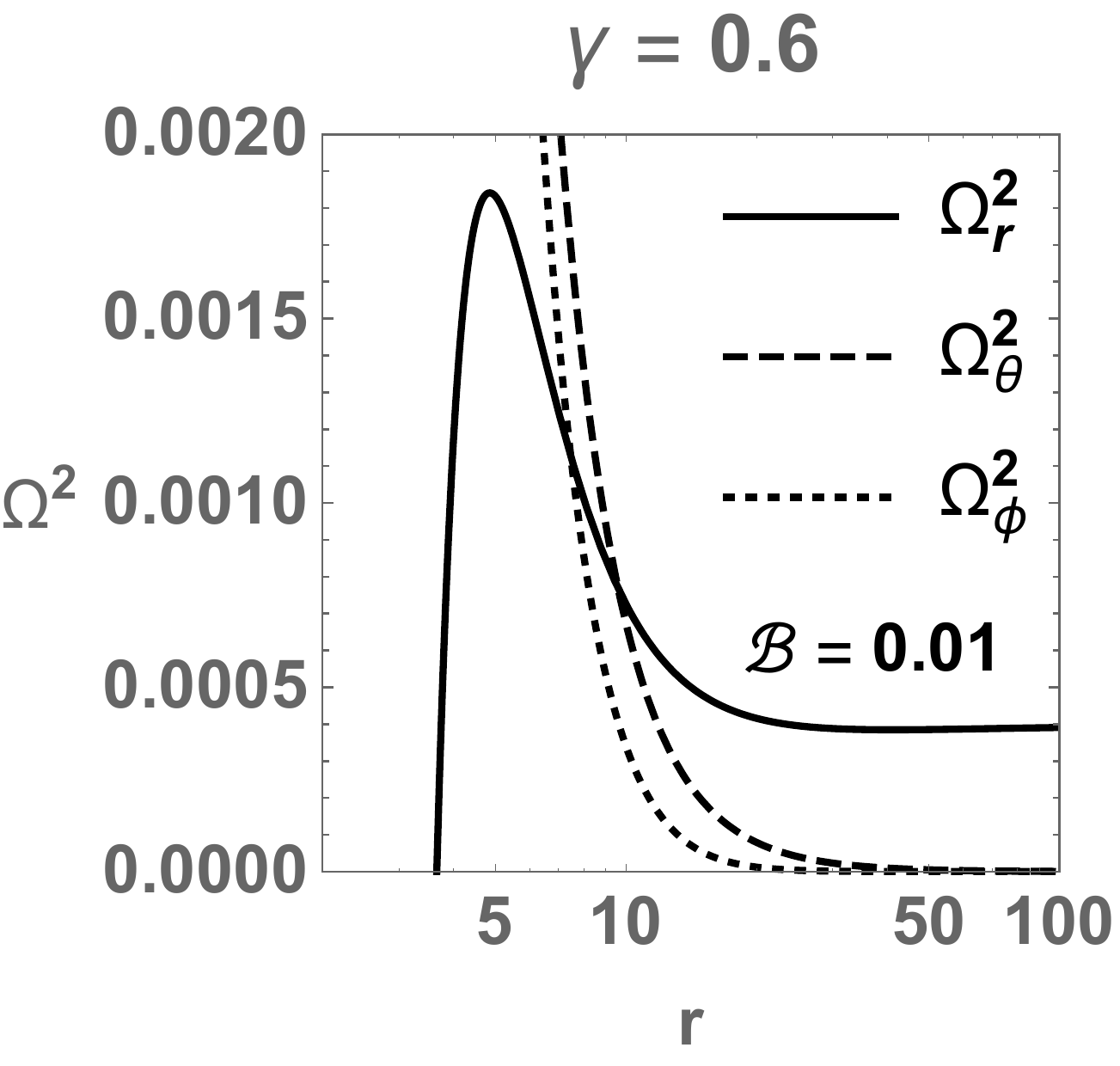}
			\hspace{0.8cm}
			\includegraphics[scale=0.33]{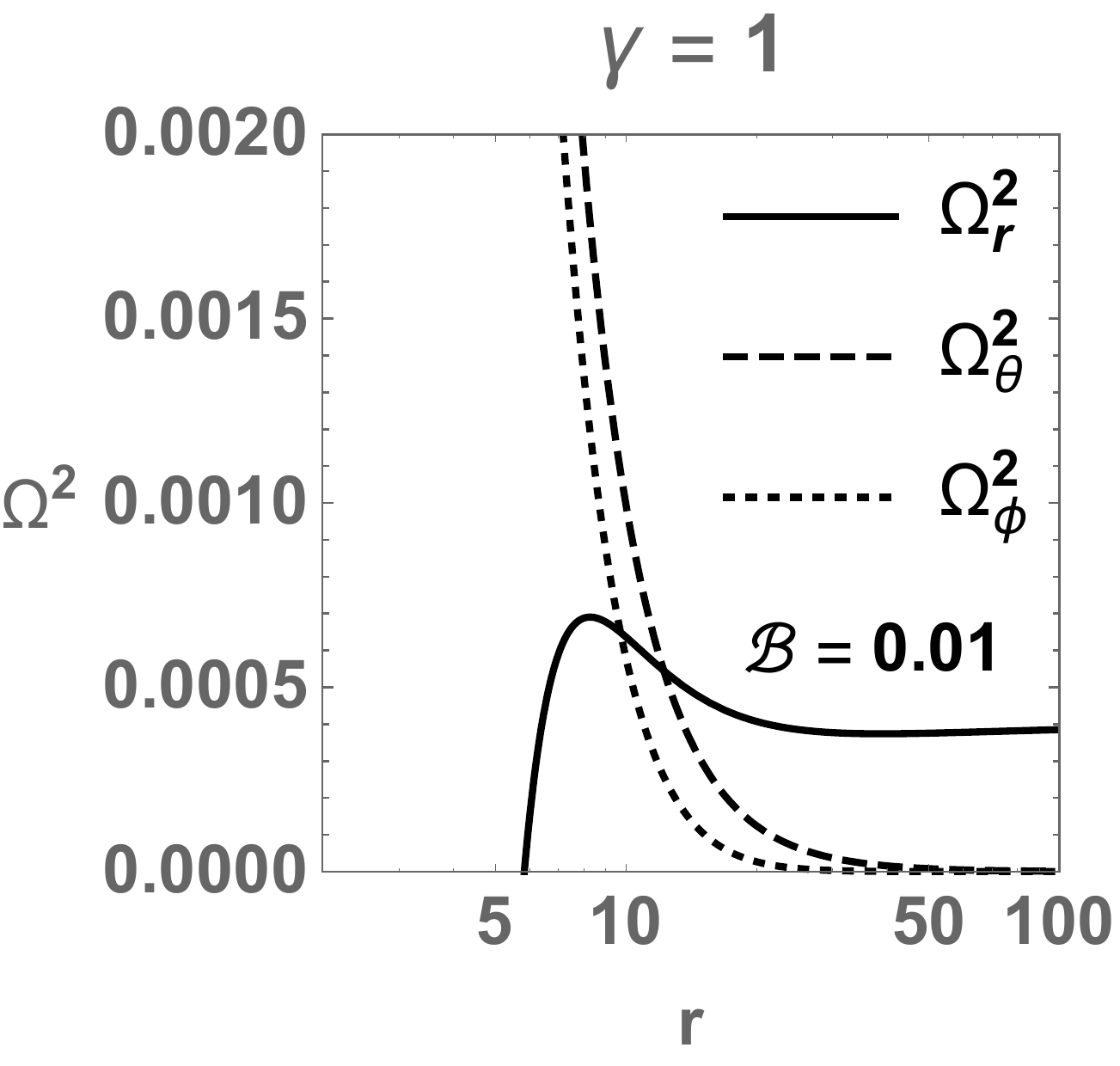}
			\hspace{0.8cm}
			\includegraphics[scale=0.33]{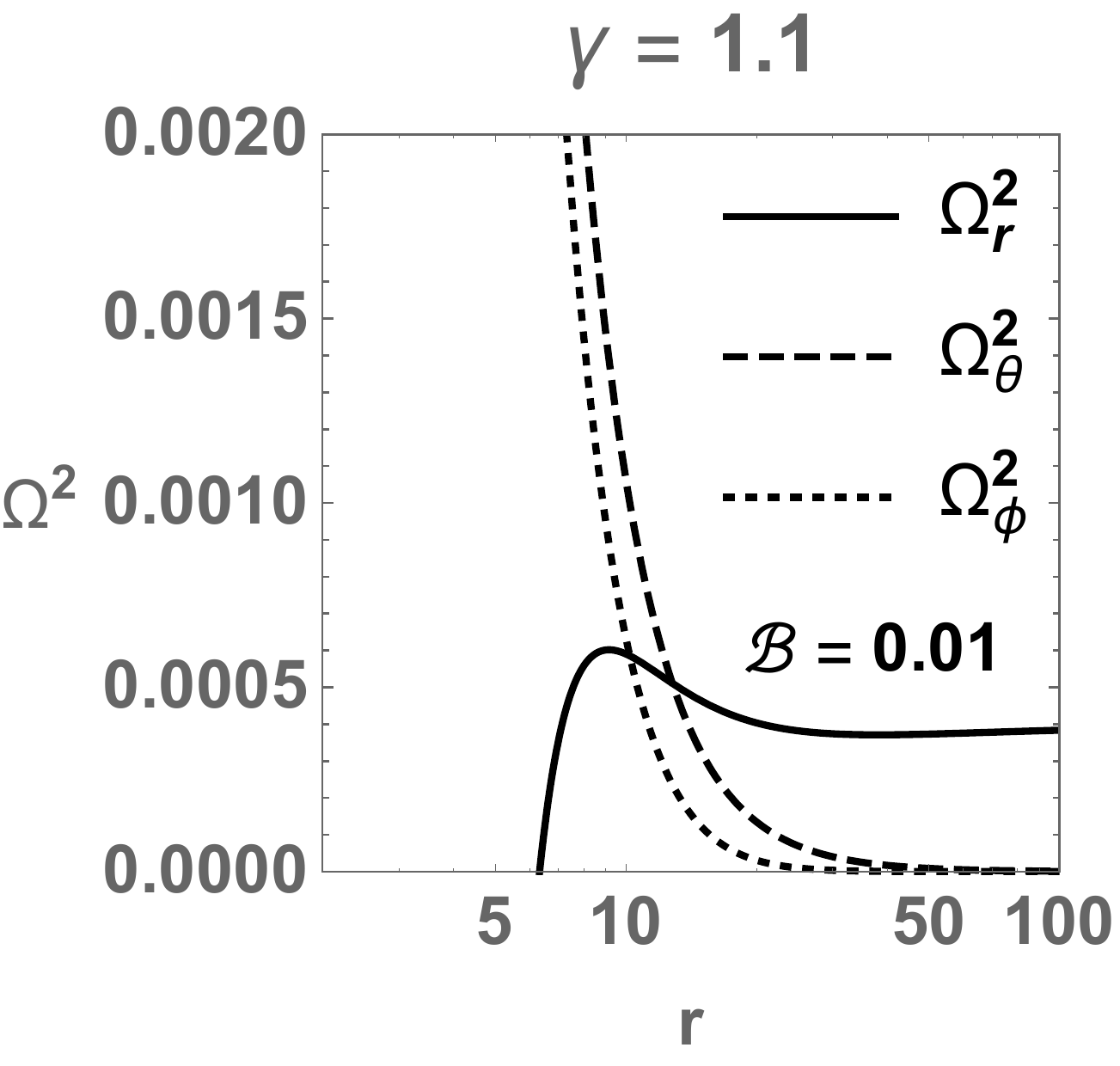}\\
			\includegraphics[scale=0.33]{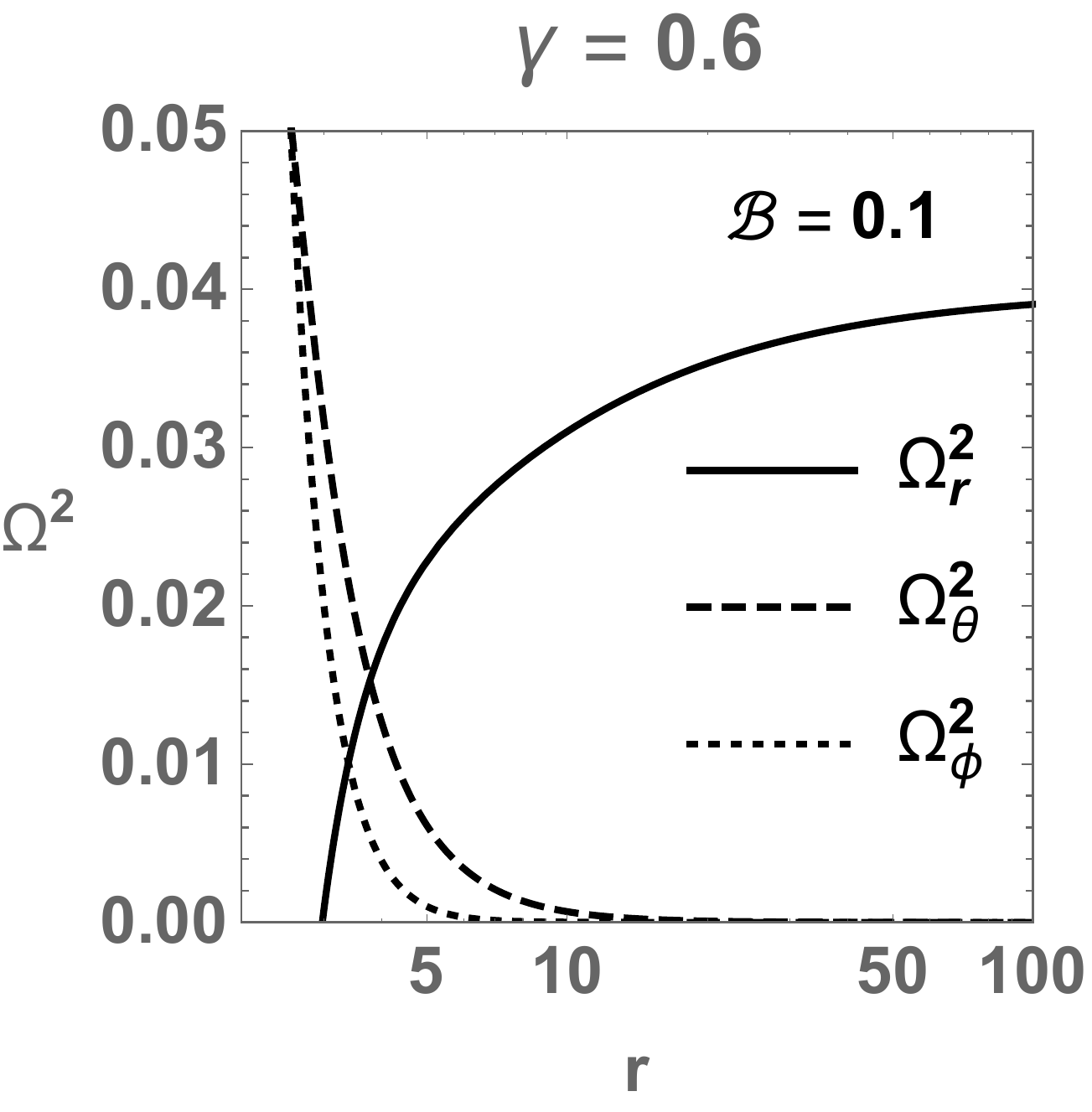}
			\hspace{0.8cm}
			\includegraphics[scale=0.33]{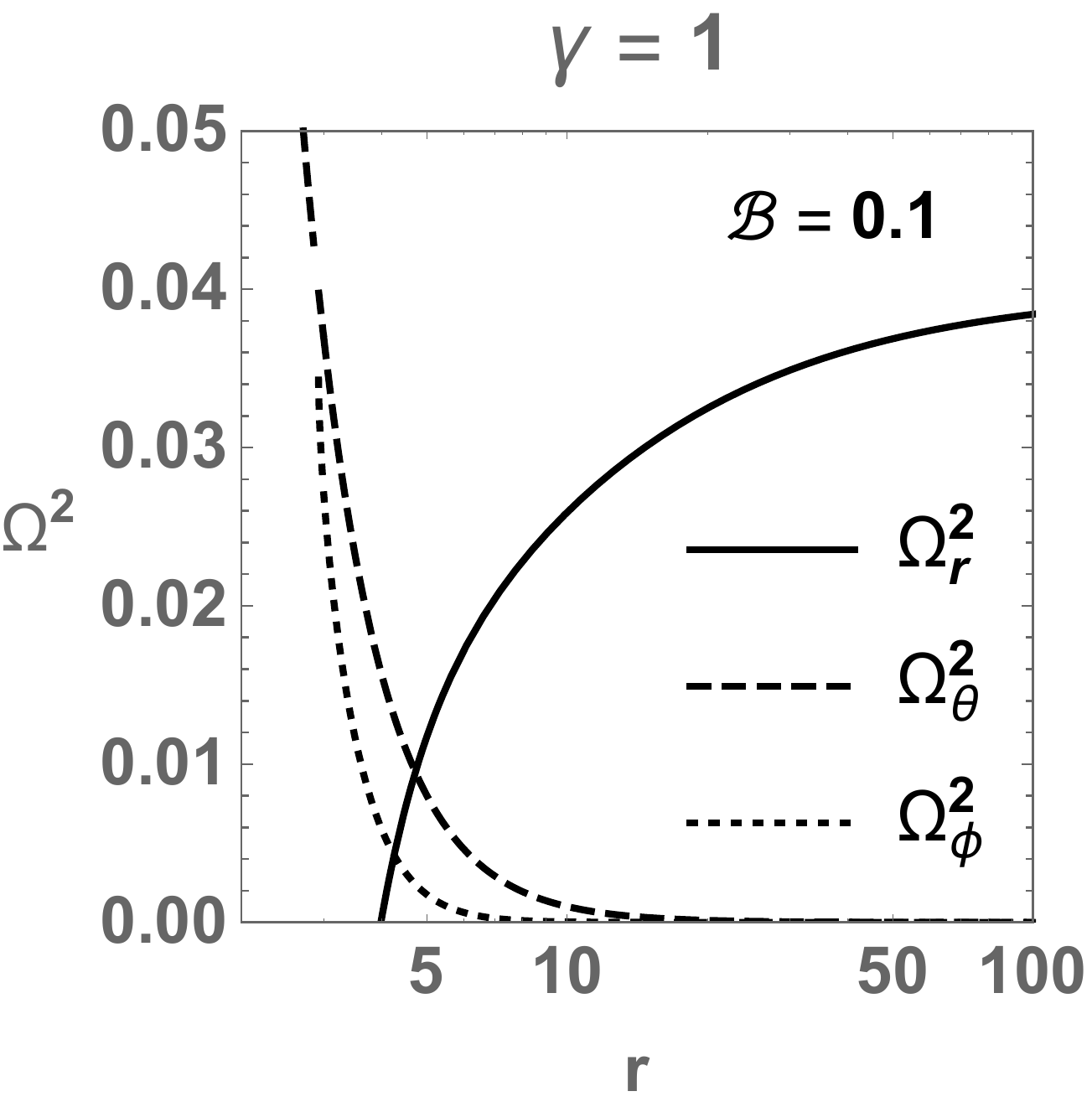}
			\hspace{0.8cm}
			\includegraphics[scale=0.33]{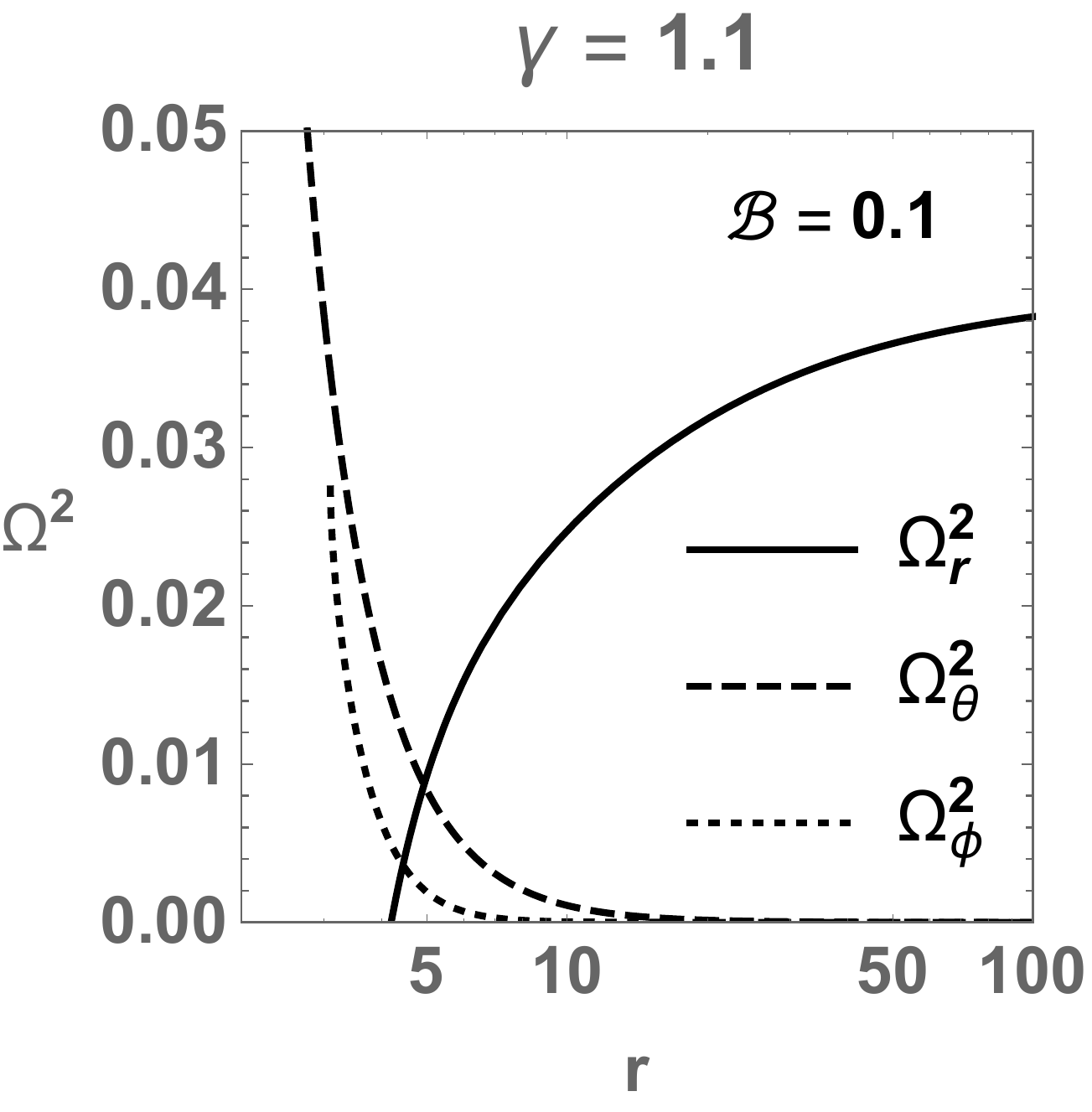}	
			\caption{Plots of the frequencies $\Omega_r$, $\Omega_\theta$ and $\Omega_\phi$ measure by an observer at infinity as a function of $r$ for different values of $\gamma$ and $\mathcal{B}$. We set $m=1$.
				\label{figure5}
			}
		\end{figure*}
	\end{center}
	In the case of small perturbation along the radial direction, i.e. $r= r_0+\delta r$ with $\delta\theta=0$ and in the case of small perturbation in the vertical direction, i.e. $\theta=\theta_0+\delta\theta$ with $\delta r=0$, we can express the functions $\mathcal{R}(r)$ and $\Theta(\theta)$ in powers of $\delta r$ and $\delta \theta$, by expanding near the equilibrium positions $r_0$ and $\theta_0$. In this way, we obtain the equations governing the motion of test charged particles that slightly depart from a circular orbit. 
	The Taylor expansions up to second order in $\delta r$ and $\delta \theta$ are given by~\cite{Toshmatov:2019qih}
	\begin{eqnarray}
	\label{V8}
	\mathcal{R}(r)&=&\mathcal{R}(r_0)+\partial_r\mathcal{R}(r)|_{r_0}\delta r+\frac{1}{2}\partial^2_r\mathcal{R}(r)|_{r_0}\delta r^2+...\\
	\Theta(\theta)&=&\Theta(\theta_0)+\partial_\theta \Theta(\theta)|_{\theta_0}\delta \theta+\frac{1}{2}\partial^2_\theta \Theta(\theta)|_{\theta_0}\delta \theta^2+...
	\end{eqnarray} 
	which, after using Eqs.~(\ref{V7a}) and (\ref{V7b}), allows us to rewrite Eqs.~(\ref{V5}) and (\ref{V6}) as
	\begin{eqnarray}
	\label{V9}
	g_{rr}\delta \dot{r}^2&=&\frac{1}{2}\partial^2_r\mathcal{R}(r)|_{r_0}\delta r^2,\\
	g_{\theta\theta}\delta \dot{\theta}^2&=&\frac{1}{2}\partial^2_\theta \Theta(\theta)|_{\theta_0}\delta \theta^2.
	\end{eqnarray}
	Taking into account that the total energy of the orbit is conserved, one obtains~\cite{Toshmatov:2019qih}.
	\begin{eqnarray}\label{V10a}
	\delta \dot{r}\left[g_{rr}\delta\ddot{r}-\frac{1}{2}\partial^2_r\mathcal{R}(r)|_{r_0}\delta r\right]&=&0,\\ \label{V10b}
	\delta \dot{\theta}\left[g_{\theta\theta}\delta\ddot{\theta}-\frac{1}{2}\partial^2_\theta \Theta(\theta)|_{\theta_0}\delta \theta\right]&=&0.
	\end{eqnarray}
	It is clear that the trivial solutions $\delta\dot{r}=0$, and $\delta\dot{\theta}=0$ correspond to circular orbits. On the other hand, the other solution, given by the quantities inside square brackets of Eqs.~(\ref{V10a}) and (\ref{V10b}), can be expressed in the form of harmonic oscillators as~\cite{Abramowicz:2004tm}
	\begin{eqnarray}
	\label{V11}
	\delta \ddot{r}+\omega^2_r\delta r&=&0,\\
	\delta \ddot{\theta}+\omega^2_\theta\delta \theta&=&0,
	\end{eqnarray}
	where $\omega^2_r$ and $\omega^2_\theta$ are the radial and vertical (latitudinal) frequencies, which are defined by
	\begin{eqnarray}
	\label{V12}
	\omega^2_r&=&-\frac{\partial^2_r\mathcal{R}(r)|_{r_0}}{2g_{rr}}=\frac{\partial^2_r\Pi(r,\theta_0)|_{r_0}}{2g_{rr}},\\
	\omega^2_\theta&=&-\frac{\partial^2_\theta \Theta(\theta)|_{\theta_0}}{2g_{\theta\theta}}=\frac{\partial^2_\theta \Pi(r_0,\theta)|_{\theta_0}}{2g_{\theta\theta}}.
	\end{eqnarray}
	For the $\gamma$-metric immersed in uniform magnetic field, the function $\Pi(r,\theta)$ is given by 
	\begin{widetext}
		\begin{equation}
		\label{V13}
		\begin{aligned}
		\Pi(r,\theta)&=1+\left(1-\frac{2m}{r}\right)^{\gamma-1}\left[\frac{\mathcal{L}}{r\sin\theta}-\mathcal{B}r\sin\theta\left(1-\frac{2m}{r}\right)^{1-\gamma}\right]^2-\left(1-\frac{2m}{r}\right)^{-\gamma}\mathcal{E}^2.
		\end{aligned}
		\end{equation}
		Therefore, the epicyclic frequencies are given by 
		\begin{equation}
		\label{V14}
		\begin{aligned}
		\omega^2_r&=\frac{[r(r-m)^{-2}(r-2m)]^{-\gamma^2}}{r^2(r^2-3mr+2m^2)^2}\big[\mathit{a}(r,\gamma)\mathcal{E}^2+\mathit{b}(r,\gamma)\mathcal{B}^2+\mathit{c}(r,\gamma)\mathcal{L}^2\big]\;,\\
		\omega^2_\theta&=(r-m)^{2(\gamma ^2-1)} [(r-2m) r]^{-\gamma ^2-1} \left[\mathcal{L}^2 \left(r-2m\right)^{2 \gamma }r^{-2\gamma}-(r-2m)^2 r^2 \mathcal{B}^2\right]\;,
		\end{aligned}
		\end{equation}

	where we have defined 
	\begin{equation}
	\label{V15}
	\begin{aligned}
	\mathit{a}(r,\gamma)&=2 \gamma  m r (2 m-r) ((\gamma -1) m+r)\;,\\
	\mathit{b}(r,\gamma)&=r^2 (r-2 m)^2 \left(2 \gamma  (\gamma +1) m^2-2 (\gamma +1) m r+r^2\right)\;,\\
	\mathit{c}(r,\gamma)&=(r-2m)^{2\gamma}r^{-2\gamma}\big[2 (\gamma +1) (\gamma +2) m^2-6 (\gamma +1) m r+3 r^2\big],
	\end{aligned}
	\end{equation}
	\end{widetext}
	and the values for $\mathcal{E}^2=V_{\text{eff}}(r,\pi/2)$ and $\mathcal{L}=\mathcal{L}_{\text{E}+}$ are given by Eqs.~(\ref{IV.7}) and~(\ref{IV.10}), respectively.
	There is a third fundamental frequency given by oscillations about the azimuthal angle $\phi$ which can be obtained from Eq.~(\ref{IVC1}). This is given by
	\begin{equation}
	\label{V15a}
	\omega^2_\phi=\left[\mathcal{L}r^{-1-\gamma}\left(r-2m\right)^{\gamma-1}-\mathcal{B}\right]^2.
	\end{equation}
	
	In the case of vanishing magnetic field, we obtain again the same results as in Ref.~\cite{Toshmatov:2019qih}.
	On the other hand, and differently from the neutral case, when $\mathcal{B}\neq 0$ we also have the so-called Larmor angular frequency $\omega_{L}$, which is associated with the uniform magnetic field itself, and is given by the relation~\cite{Kolos:2015iva}
	\begin{equation}
	\label{V16}
	\omega_L=\frac{qB}{m_0}=2|\mathcal{B}|.
	\end{equation}
	
	It is important to point out that the Larmor frequency $\omega_L$ does not dependent on the radial coordinate and therefore it becomes important at large distances, where the magnetic field dominates over the gravitational field. 
	
	In Fig.~\ref{figure4}, we show the behavior of $\omega_r$, $\omega_\theta$ and $\omega_\phi$ as functions of $r$ for different values of $\gamma$ and $\mathcal{B}$. For completeness we also show the corresponding value of the Larmor frequency $\omega_L$.
	When the $\gamma$-space-time is not immersed in a uniform magnetic field (i.e. when $\mathcal{B}=0$), $\omega_\phi$ and $\omega_\theta$ only coincide when $\gamma=1$ (i.e. Schwarzschild). Moreover, as expected, they asymptotically tend to zero as $r$ increases, similar to the Schwarzschild case~\cite{Kolos:2015iva} (see the third row in Fig.~\ref{figure4}).  
	In the same figure, it is possible to see that the radial frequency $\omega_r$ has a maximum value, which decreases as $\gamma$ increases. Moreover, we can see that $\omega_r$ vanishes at the ISCO. On the other hand, when $\mathcal{B}\neq0$, Fig.~\ref{figure4} shows different behaviors. It is worth noting that for $\mathcal{B}<0$ both $\omega_r$ and $\omega_\phi$ tend asymptotically to the Larmor frequency $\omega_L$, while $\omega_\theta$ tends to 0. Nevertheless, when $\mathcal{B}>0$, only the radial frequency $\omega_r$ tends to $\omega_L$ as $r$ increases.

	\subsection{Frequencies measured by distant observers} 
	
	The epicyclic frequencies derived in the previous section in Eqs.~(\ref{V14}) and (\ref{V15a}) are measured with respect to the proper time of a comoving observer. Therefore, to obtain the frequencies measured by an observer at infinity, it is necessary to take into account the redshift factor $d\tau/dt$. Hence, we have that 
	\begin{equation}
	\label{VB1}
	\Omega_i=\omega_i \frac{d\tau}{dt},
	\end{equation} 
	where $i=r$, $\theta$ or $\phi$. For the $\gamma$-metric, the redshift factor can be obtained from Eq.~(\ref{IV.4}) and is given by 
	\begin{equation}
	\label{VB2}
	\frac{d\tau}{dt}=\frac{r^{-\gamma}(r-2m)^\gamma}{\mathcal{E}}.
	\end{equation}
	Therefore, the frequencies in Eq.~(\ref{VB1}), are given by
	\begin{equation}
	\label{VB3}
	\Omega_i=\omega_i\frac{r^{-\gamma}(r-2m)^\gamma}{\mathcal{E}}.
	\end{equation}
	It is important to point out that $\mathcal{E}=\mathcal{E}(r)$ in Eq.~(\ref{VB3}) is the specific energy at the circular orbit given by $\mathcal{E}=V_{\text{eff}}(r,\pi/2,\mathcal{L}_{E+})$. Therefore, we obtain 
	\begin{widetext}
		\begin{equation}
		\label{VB4}
		\begin{aligned}
		\Omega^2_r&=\frac{r^{-\gamma(\gamma+2)}(r-m)^{2\gamma^2}(r-2m)^{\gamma(2-\gamma)}}{r^2(r^2-3mr+2m^2)}\left[\mathit{a}(r,\gamma)+\frac{\mathit{b}(r,\gamma)\mathcal{B}^2+\mathit{c}(r,\gamma)\mathcal{L}^2}{\mathcal{E}^2}\right]\;,\\
		\Omega^2_\theta&=\frac{(r-m)^{2(\gamma^2-1)}(r-2m)^{-(\gamma-1)^2}r^{-(\gamma+1)^2}}{\mathcal{E}^2} \left[\mathcal{L}^2 \left(r-2m\right)^{2 \gamma }r^{-2\gamma}-(r-2m)^2 r^2 \mathcal{B}^2\right]\;,\\
		\Omega^2_\phi&=\left[\frac{\mathcal{L}r^{-(1+2\gamma)}(r-2m)^{2\gamma-1}-\mathcal{B}r^{-\gamma}(r-2m)^\gamma}{\mathcal{E}}\right]^2.
		\end{aligned}
		\end{equation}
	\end{widetext}
	
	In Fig.~\ref{figure5} we show the behavior of $\Omega^2_r$, $\Omega^2_\theta$ and $\Omega^2_\phi$ as functions of $r$ for different values of $\gamma$ and $\mathcal{B}$. From the figure, we see that the latitudinal $\Omega_\theta$ and azimuthal $\Omega_\phi$ frequencies measured by a distant observer vanish as $r$ increases for all the values of $\gamma$ and $\mathcal{B}$. In particular, when $\mathcal{B}=0$ and $\gamma=1$, these frequencies coincide ($\Omega_\theta=\Omega_\phi$). 
	Furthermore, for positive (negative) values of $\mathcal{B}$, $\Omega_\phi$ is always smaller (greater) than $\Omega_\theta$. 


	\section{Conclusions}
	
	Observations of stellar mass and super-massive black hole candidates rely on the measurement of the spectrum of light emitted by their accretion disks. In particular X-ray reflection spectroscopy appears to be a promising tool to probe the nature of the geometry in the vicinity of black hole candidates
	\cite{Bambi:2016sac,Cao:2017kdq,Tripathi:2018lhx}.
	The recent observation of the `shadow' of the super-massive black hole candidate at the core of the galaxy M87 
	\cite{Akiyama:2019cqa,Akiyama:2019brx,Akiyama:2019sww} 
	also suggests that the possibility to test experimentally the geometry in the vicinity of such astrophysical compact objects may soon be at hand. 

	In both cases, it is of paramount importance to know the behavior of the motion of test particles in the gas of the accretion disk, since the light emitted from the accretion disk is the only medium through which the measurements are made. 
	The behaviors of test particles and, as a consequence, the properties of accretion disks, are influenced by several factors, besides the geometry, of which the most relevant are the presence of additional matter fields
	\cite{Joshi:2013dva,Boshkayev:2018sbj,Ilyas:2016nio},
	magnetic fields
	\cite{Wald:1974np,Petterson:1975sg} 
	and the particle's spin
	\cite{Abramowicz:1979,Semerak:1999qc,Toshmatov:2019bda}.
	Therefore it is very important to know how such factors may alter the observational features of the accretion disks, in order to exclude the possibility of degeneracies such as, for example, the measurement of a black hole spin which could be also due to a non-spinning source with quadrupole moment\cite{Bambi:2013hza}.
	
	The behavior of test particles in the Kerr and Schwarzschild geometries has been studied in detail
	\cite{Thorne:1974ve,Page:1974he,Abramowicz:1988sp,Abramowicz:1996ap,Johannsen:2013asa,Gimeno-Soler:2018pjd}. 
	On the other hand, the same behavior in other exact solutions of Einstein's vacuum field equations has not been thoroughly studied so far. One reasonable hypothesis is that if departures from the black hole geometry do exist they would manifest in the presence of quadrupole moment in the source of the gravitational field. In this respect, the $\gamma$-metric represents the ideal test-bed to investigate departures from the behavior expected by black holes.
	
	Following a program initiated by some of us, we attempted here to characterize the behavior of charged test particles on orbits slightly departing from the ISCO of the $\gamma$-metric immersed in an external magnetic field. We have shown that the presence of a non-vanishing deformation parameter affects the motion of charged particles and changes the epicyclic frequencies.
	
	As more and more precise measurements of the properties of super-massive black hole candidates become available, these models will allow testing the hypothesis that the geometry outside such objects is well described by the Kerr metric.

	\begin{acknowledgments}
			C.A.B.G wishes to dedicate this work to Prof. Alirio Calderón-Molina as a tribute to his memory. The authors want to thank M.~Kološ for helpful discussion. The work of C.A.B.G. and C.B. was supported by the Innovation Program of the Shanghai Municipal Education Commission, Grant No.~2019-01-07-00-07-E00035, and the National Natural Science Foundation of China (NSFC), Grant No.~11973019. C.A.B.G. also acknowledges support from the China Scholarship Council (CSC), grant No.~2017GXZ019022 and the Nazarbayev University for hospitality. This research is supported by Grants No. VA-FA-F-2-008, No. MRB-AN-2019-29 and No. YFA-Ftech-2018-8 of the Uzbekistan Ministry for Innovative Development. This research is partially supported by an Erasmus+ exchange grant between SU and NUUz. A.A. is supported by a postdoc fund through PIFI of the Chinese Academy of Sciences. D.M. acknowledges support by Nazarbayev University Faculty Development Competitive Research Grant No. 090118FD5348 and by the Ministry of Education and Science of the Republic of Kazakhstan's target program IRN: BR05236454.
	\end{acknowledgments}


\end{document}